\documentclass{aa}  
\usepackage{amsmath}
\usepackage[dvipsnames]{xcolor}
\usepackage{booktabs}
\usepackage[colorlinks=true, allcolors=blue]{hyperref}
\usepackage{stfloats}
\usepackage{graphicx}
\usepackage{subcaption}
\usepackage{longtable}
\usepackage{siunitx}
\usepackage{tabularray}
\usepackage{colortbl}
\usepackage{txfonts}
\begin{document} 

   \title{A catalogue of candidate milli-parsec separation massive black hole binaries from long term optical photometric monitoring}

   \author{V. Foustoul
          \inst{1},
          N. A. Webb\inst{1},
          R. Mignon-Risse \inst{2,3},
          E. Kammoun\inst{4,5,6},
          M. Volonteri\inst{7}
          \and
          C. A. Dong-Páez\inst{7}
          }

   \institute{IRAP, Université de Toulouse, CNRS, CNES, 9 avenue du Colonel Roche, 31028, Toulouse, France\\
   \email{vincent.foustoul@irap.omp.eu}
   \and
    Department of Physics, Norwegian University of Science and Technology, NO-7491 Trondheim, Norway
    \and
    Université Paris Cité, CNRS, CNES, Astroparticule et Cosmologie, F-75013 Paris, France
    \and 
    Cahill Center for Astrophysics, California Institute of Technology, 1216 East California Boulevard, Pasadena, CA 91125, USA
    \and
    Dipartimento di Matematica e Fisica, Univsersit\`{a} Roma Tre, via della Vasca Navale 84, I-00146 Rome, Italy
    \and
    INAF--Osservatorio Astrofisico di Arcetri, Largo Enrico Fermi 5, I-5025 Firenze, Italy
   \and
   Institut d’Astrophysique de Paris, Sorbonne Université, CNRS, UMR 7095, 98 bis bd Arago, 75014 Paris, France}
   \date{}

  \abstract
   {The role of mergers in the evolution of massive black holes is still unclear, and their dynamical evolution, from the formation of pairs to binaries and the final coalescence, carries large physical uncertainties. The identification of the elusive population of close massive binary black holes (MBBHs) is crucial to understand the importance of mergers in the formation and evolution of supermassive black holes.} 
   {It has been proposed that MBBHs may display periodic optical/ultra-violet variability. Optical surveys provide photometric measurements of a large variety of objects, over decades and searching for periodicities coming from galaxies in their long-term optical/UV lightcurves may help identify new MBBH candidates.}
   {Using the Catalina Real-Time Transient Survey (CRTS) and Zwicky Transient Facility (ZTF) data, we studied the long-term periodicity of variable sources in the centre of galaxies identified using the galaxy catalogue Glade+.}
   {We report 36 MBBHs candidates, with sinusoidal variability with amplitudes between 0.1 and 0.8 magnitudes over 3-5 cycles, through fitting 15 years of data. The periodicities are also detected when adding a red noise contribution to the sine model. Moreover, the periodicities are corroborated through Generalized Lomb Scargle (GLS) periodograms analysis, providing supplementary evidence for the observed modulation. We also indicate 58 objects, that were previously proposed to be MBBH candidates from analysis of CRTS data only. Adding ZTF data clearly shows that the previously claimed modulation is due to red noise. We also created a catalogue of 221 weaker candidates which require further observations over the coming years to help validate their nature. Based on our 36 MBBHs candidates, we expect $\sim$20 MBBHs at $z<1$, which is commensurate with simulations. Further observations will help confirm these results.}
   {}
   \authorrunning{Foustoul et al.}
\titlerunning{Candidate milli-parsec separation massive black hole binaries}
   \maketitle
   
\section{Introduction} \label{sec:introduction}

The evolution and growth of massive black holes is a key unsolved problem in galaxy formation, and it has significant repercussions in the field of gravitational wave astronomy. The recent plausible detection of gravitational waves at nanoHz frequency by Pulsar Timing Arrays \citep{EPTA_detection, nanograv, PTA_GW_background, 2023RAA....23g5024X} hints at the ability of supermassive black holes (SMBHs) with mass $>10^8\, \mathrm{M_{\odot}}$ and relatively low redshift ($z<1-2$) to form tight binaries on the way to coalescence through emission of gravitational waves. The formation of massive binary black holes (MBBHs) is expected in the galaxy evolution framework, when two galaxies, each hosting a SMBH, merge \citep[e.g.,][]{MBH_path_to_coalescence_2, 2002MNRAS.331..935Y, model_hierarchique_mergers}. After two galaxies hosting SMBHs merge, the black holes sink towards the centre of mass of the galaxies, and if the dynamical decay is efficient they eventually merge. In this path to coalescence, black holes lose angular momentum from different interactions, in three major stages \citep{MBH_path_to_coalescence}. Firstly, the SMBHs are brought closer by dynamical friction with the gaseous and stellar environment in the galaxy remnant. Secondly, the loss of angular momentum is achieved through the interaction between black holes and stars, in a so-called slingshot interaction, and/or through viscous torques in a circumbinary disk \citep[e.g.,][]{2002ApJ...567L...9A}. The dynamics of MBBHs have been recently summarized in \citet{2023LRR....26....2A}. Finally, when the black holes are close enough together, the merging is dominated by the loss of energy through gravitational wave (GW) emission. MBBHs of mass $M \approx 10^{4-7}\, \mathrm{M_{\odot}}$ are expected to be loud mHz gravitational wave sources, to be detected by the first space-based gravitational wave detector LISA \citep[Laser Interferometer Space Antenna;][]{LISA} up to a redshift $z \approx 20$ \citep{Lisa_freq}. The more massive MBBHs ($M > 10^8 \, \mathrm{M_{\odot}}$) are expected to emit gravitational waves in the nHz range and should be detectable with the Pulsar Timing Arrays \citep{PTA_1990, iPTA, EPTA}. 

\begin{figure}[!h]
    \centering
    \includegraphics[width=\hsize]{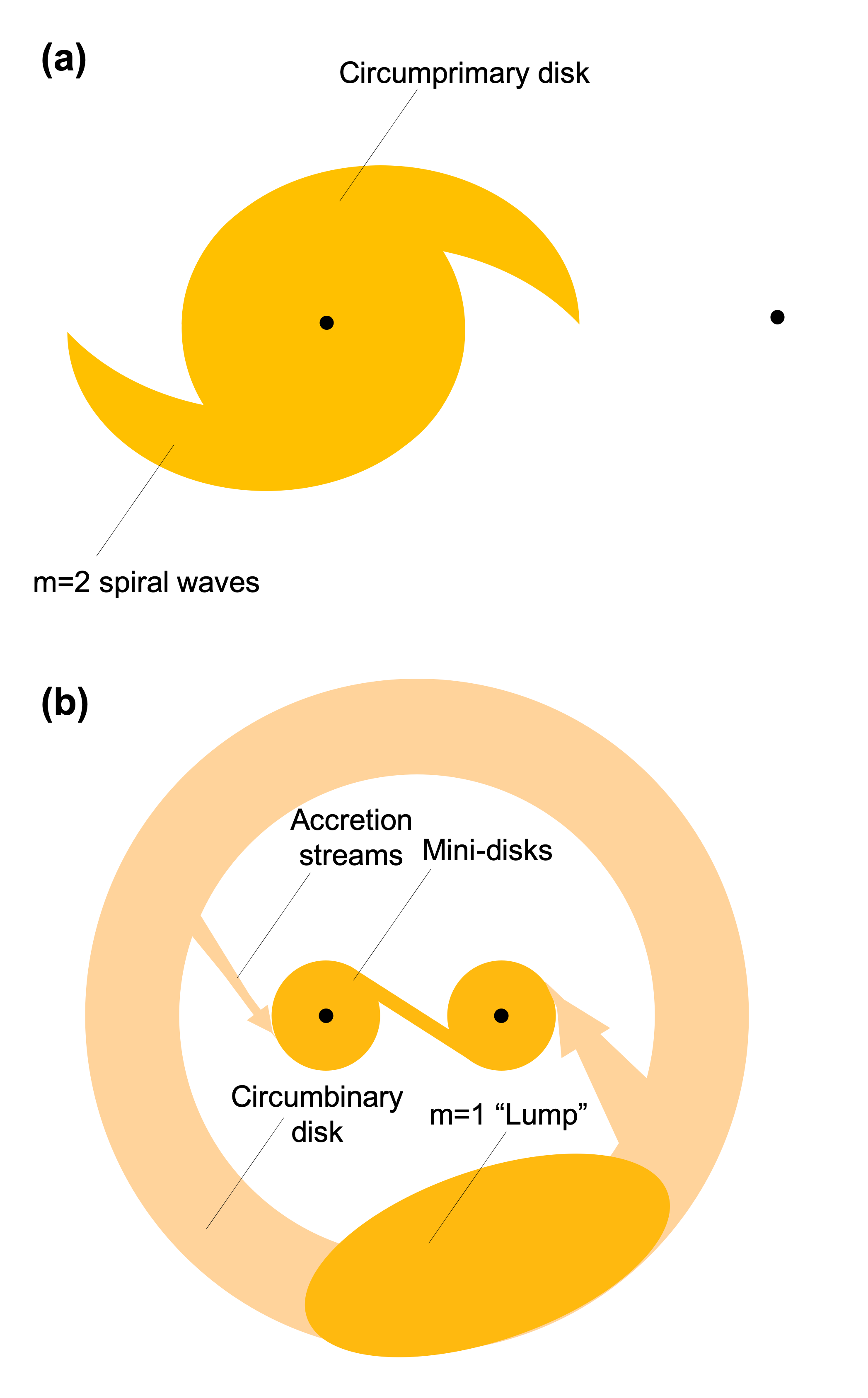}
    \caption{Expected geometry of MBBH in the presence of a circumprimary disk only (panel a) and in the presence of a circumbinary disk (CBD, panel b). In that case, the system is composed of three main features: a CBD, accretion streams and mini-disks.}
    \label{fig:MBBH}
\end{figure}

The geometry of MBBHs has been simulated to determine the observational signatures of such systems (e.g.  \citealt{SMBBH2}, \citealt{simulation_MBH}). Depending on the orbital separation and mass ratio of the binary, two different configurations can occur. If the infalling gas has enough angular momentum, it has been shown that gravitational torques pull and expel gas, creating a cavity of radius $r \approx 2a(1+e)$, with $a$ and $e$ respectively the semi-major axis and eccentricity of the orbit (e.g. \citealt{simulation_MBH}). If a circumbinary disk (CBD) is already present, its inner edge is located at distance $r_\mathrm{CBD} \sim 2a(1+e)$. Shocks at the inner edge of the disk make the gas plunge into the cavity, forming accretion streams and mini-disks around each black hole \citep{SMBBH2,SMBBH3}. An overdensity is also expected to form in the circumbinary disk (e.g. \citealt{Simu_BBH_lump_shi12}, \citealt{lump}). However, if the infalling gas does not have enough angular momentum, the CBD will not form and an accretion disk will form around one or each black hole \citep{single_accretion_disk}. In this case, the disk will be truncated at the radius $r\sim 1/3$ of the orbital separation \citep{tidal_torques_close_binary_systems}. Moreover, \cite{m2_modes_accretion_disk} showed that tidal interactions will generate $m=2$ spiral waves in the disk. A schematic of  both configurations for MBBHs is shown in Fig.~\ref{fig:MBBH}. 

Synergies between future observatories like Athena \citep[Advanced Telescope for High ENergy Astrophysics; ][]{Athena} and LISA are expected to provide constraints on the evolutionary paths leading to supermassive black holes and improve the detection of MBBHs. LISA will be able to detect mergers of massive black holes and with Athena it may be possible to detect the MBBHs before merger. Indeed, as sub-pc MBBHs in the LISA mass range approach merger, the X-ray modulation produced by the relativistic Doppler boosting of the mini-disks should be detectable. Following the coalescence of the massive black holes, the X-ray Integral Field Unit \citep[X-IFU, ][]{X_ifu_new_athena} on board of Athena should provide important information on the MBBH environment, notably the density, temperature and chemical enrichment, probing the history of the MBBH evolution and formation. However, it may be possible to make some headway in understanding the evolution of supermassive black holes using current data and observatories. Many confirmed MBBHs and dual AGN at kiloparsec separation have been discovered already \citep[e.g.][]{NGC6240,Mrk463,dual_AGN,LBQS,NGC7727, Mrk739, 2024ApJ...972..185} but close sub-parsec MBBHs are yet to be confirmed. Possibly the best close MBBH candidate is OJ 287 \citep{OJ287}. OJ 287 is a BL Lacertae object located at a redshift $z=0.306$. It shows periodic outbursts every $\approx 12$ years in its optical and X-ray lightcurve \citep{OJ287}, possibly due to interactions between the two black holes and their immediate environments. However, the MBBH nature causing the observed flares in OJ 287 has recently been contested, as the model predicted a flare in 2022 that has not yet been observed \citep[e.g.][]{OJ_287_doubt}. Other close sub-pc MBBH candidates have been proposed such as the galaxy Mrk 533, which shows a Z-shaped radio jet, interpreted by two orbiting massive black holes. Also, \cite{spectroscopic_mbbh_candidates} proposed 3 quasars as possible sub-pc MBBHs, identifying systematic and monotonic changes in their H$\beta$ broad emission line profile. Variability due to accretion flows in the optical lightcurves of close MBBHs is also expected \citep{variability_accretion}. It has been shown that AGN are more likely to appear in galaxy mergers, as the merger triggers inward falling gas and thus accretion onto the galactic nuclei \cite[e.g.][]{merger_agn_trigger}. Indeed, the fraction of AGN increases in galaxy mergers in comparison to non-mergers \citep[e.g.][]{AGN_mergers,agn_fraction_mergers,agn_mergers_2,agn_mergers_3}. Furthermore, observational studies suggest that AGN activity peaks in the post-merging phase of galaxy mergers, as the star formation rate (SFR) and SMBH accretion rate increase in post-mergers galaxies \citep[e.g.][]{post_merger_galaxies_properties,galaxy_pairs_agn}. Therefore, detecting AGN activity may help to probe past galaxy mergers and could trace the presence of a MBBH.

We organize the paper as follows. In Section~\ref{Methods} we present the sample selection methodology and describe the different techniques used to search for the periodicities in the lightcurves. We show the results of our investigation in Section~\ref{Results}. In Section~\ref{discussion}, we compare our results with previous simulations, discuss the origin of the proposed periodicities, analyse X-ray spectra and fluxes of the suggested candidates and present the caveats of our work. Finally, we draw our conclusions in Section~\ref{conlusion}.   

\section{Methods}\label{Methods}
\subsection{Sample selection and Data processing}
In 2015, \cite{graham} undertook a systematic search for close supermassive black hole binaries in quasars of the Catalina Real-Time Transient Survey (CRTS) and reported a list of 111 possible candidates. However, quasars are subject to statistical variations called red noise.  As discussed in \cite{False_periodicities}, quasars initially proposed as MBBH candidates due to their optical sinusoidal modulation, can be better modelled by a red noise process. This is due to the fact that quasar lightcurves can display seemingly periodic variations due to red noise and, as proposed by \cite{False_periodicities}, it is essential to observe at least 5 periods to infer the presence of a periodic behaviour. We revisit the \cite{graham} sample of candidates and study sources from \cite{ZTF_variable_catalog} catalogue.

\subsubsection{Graham et al. sample}
We have taken all the candidates from the \cite{graham} sample (magnitudes between 15 and 20) and included more recent data from the Zwicky Transient Factory (ZTF) to extend the baseline and determine whether the sinusoidal oscillations are sustained over a sufficiently long time period. CRTS observations were made using the Catalina Sky Survey (CSS) Schmidt telescope, between 2006 and 2016 in the northern sky, covering around 33000 square degrees, magnitudes were measured in the V-band with a limiting magnitude between 20 and 21 \citep{CRTS}. ZTF is a 48 inch Schmidt telescope located at the Palomar observatory, it has a 47 square degree field of view and covers $\sim$ 30000 square degrees in the northern sky. ZTF observations started in March 2018 and are still ongoing. They are done on a two-nightly cadence. Magnitudes are measured in g, r and i bands with a limiting magnitude of $\sim$ 20.5 \citep{ZTF_performance}. 

\subsubsection{Chen et al. sample}
\begin{figure}[!h]
    \includegraphics[width=\hsize]{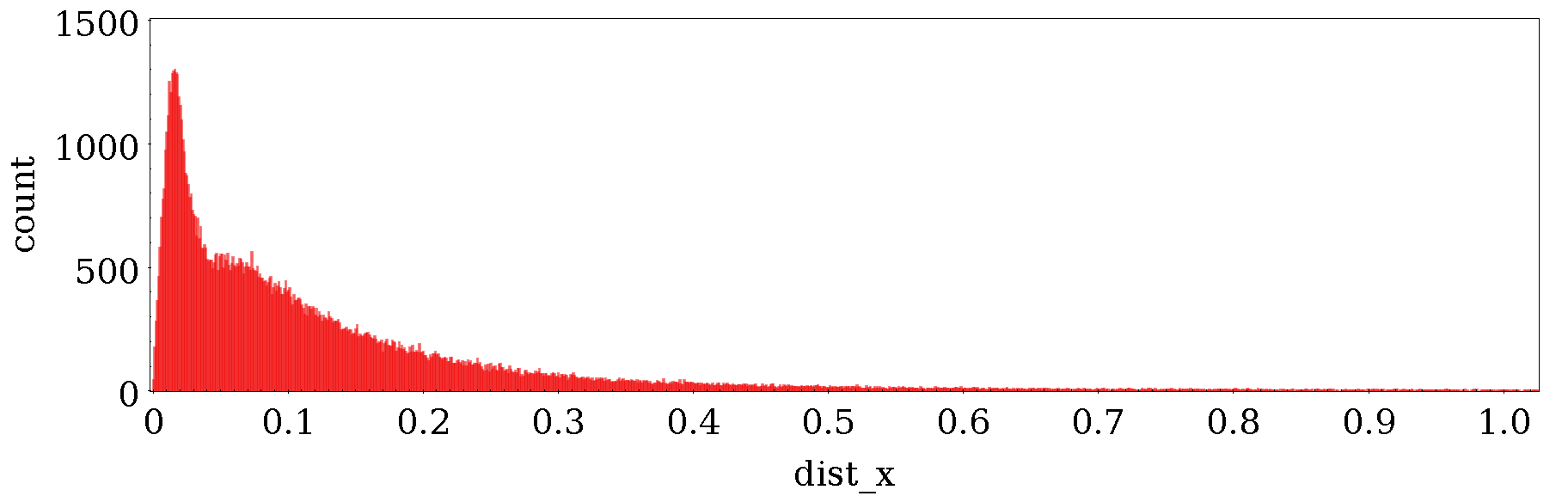}
    \caption{Histogram representing the number of detected sources by ZTF in terms of the distance separation in arc seconds between the Glade+ and ZTF positions for sources in \cite{ZTF_variable_catalog}.}
    \label{fig:histo}
\end{figure}

In 2020, \cite{ZTF_variable_catalog} published a catalogue of periodic variable sources using the ZTF Data Release 2 (DR2) \citep{ZTF_performance}, and includes sources with magnitudes down to r $\sim$ 20.6. This catalogue contains 781 602 confirmed variable sources and 1 381 527 objects which are possibly variable. Most of these objects are stars, and since we are interested in MBBH candidates, we cross-correlated this catalogue with a galaxy catalogue, Glade+ \citep{glade+}. Glade+ includes $\sim$ 23 million objects, separated into 22.5 million galaxies and $\sim$ 750 000 quasars. Glade+ is complete to a luminosity distance of 44 Mpc \citep{glade+}. So that we obtain statistically reliable results, we investigated sources in \cite{ZTF_variable_catalog} with at least 400 ZTF measurements in a single band. In order to download ZTF data, we used the Infrared Science Archive (IRSA) API to query the best match between the Glade+ position of the selected objects (Right Ascension (R.A.) and Declination (Decl.)) and the ZTF data, reducing the odds of miss-matching sources, in a 0.4 arc seconds radius \citep{ZTF}. We used a maximum radius of 0.4 arc second in order to include all good matches. Figure~\ref{fig:histo} shows the distribution of the distance between the matched ZTF and Glade+ sources in our sample. We can clearly see that the number of sources remains similar above 0.4 arc seconds as the majority of matches are between 0 and 0.4 arc seconds.   

In ZTF data, observations affected by clouds and/or the moon are flagged with a catflag of \num{32 768}. As suggested by the ZTF public data release, after downloading ZTF data, we removed such bad observations by only keeping data points with a catflag less than \num{32 768}. 

\subsection{Sine model}
As a first test, we fitted only the ZTF light curves for sources from \cite{ZTF_variable_catalog}, to select those that appear to have sinusoidal modulation, using a sine function as follows: 

\begin{equation}
    y=A \times \sin \left( 2\pi \left(\frac{t}{P} \right)+t_{0} \right)+B
    \label{sine}
\end{equation}

\noindent Where $A$, $P$, $t_{0}$ and $B$ are respectively the amplitude, the period, the phase and the background of the model. Whilst it is clear that not all MBBH will show strictly sinusoidal modulation, it is a first approach to search for modulation from such systems, as using more complex methods to search in such a large dataset is computationally prohibitive. We computed the reduced chi-squared value for each object and retained sources with a reduced chi-squared less than 1.3 which corresponds to a 4$\sigma$ confidence interval that the sinusoidal modulation is a good description of the data. For these sources we added CRTS observations when available to evaluate whether or not the variability was sustained in both epochs. CRTS observations were retrieved from the Catalina Surveys Data Release 2 (CSDR) website \footnote{\url{http://nesssi.cacr.caltech.edu/DataRelease/CSDR2.html}}, using the sources R.A. and Decl. to query the objects lightcurves. The sine model is a simplified model as it does not account for quasar's intrinsic variability (see Sec.~\ref{DRW}), and assumes accretion disks to be stable apart from the binary modulation. However, we used this model as a tool to identify possible MBBHs candidates and more complex models are discussed below, considering disk instabilities might be needed to fully explain the observed variability.

\subsection{Data intercalibration}
In order to compare ZTF and CRTS observations of \cite{graham} and \cite{ZTF_variable_catalog} sources, it is mandatory to perform intercalibration as they have been taken with different instruments in different filters. We extended CRTS lightcurves with ZTF data taken in the g filter. In order to take into account wavelength differences between CRTS and ZTF-g filters, we intercalibrated CRTS and ZTF-g observations using the Python package {\it PyCALI} \citep{PyCALI}, taking ZTF data as reference, in the same way as \cite{ztf_crts_search}. Intercalibrating ZTF and CRTS data together takes into account colour variability and wavelength dependent effects. This allows us to compare both data set in order to see whether or not they show the same variability. Moreover, in order to increase the signal to noise we grouped data points into bins spanning a maximum of 4 days. The new magnitude and errors in each bin are computed using Eq.~\ref{Eq:weighted_mean} and Eq.~\ref{Eq:erreurs} respectively.

\begin{equation}
    \mu=\frac{\sum_{i}^{N}x_{i}/\sigma^{2}_{i}}{\sum_{i}^{N}1/\sigma^{2}_{i}}
    \label{Eq:weighted_mean}
\end{equation}

\begin{equation}
    \sigma^{2}(\mu)=\frac{1}{\sum_{i}^{N} \sigma_{i}^{2}}
    \label{Eq:erreurs}
\end{equation}

$\mu$, $x_{i}$, $\sigma(\mu)$, $\sigma_{i}$ and $N$ correspond to the new calculated magnitude on the binned data point, the magnitude of each observed data points, the new calculated error on the binned data point, the magnitude error of each observed data point in the bin interval and the number of observations in the considered bin, respectively. Also, to limit stochastic effects, we discarded sources with less than 25 CRTS observations after binning. 

\subsection{Fitting technique}\label{sec: nested_sampling}
To find the best fit sinusoid to the intercalibrated lightcurves, we used a nested sampling Monte Carlo algorithm MLFriends \citep[]{Ultranest_ML,Ultranest_methods} using the Ultranest\footnote{\url{https://johannesbuchner.github.io/UltraNest/}} package \citep{Ultranest}  to derive the posterior probability distribution of the model parameters. Ultranest computes the posterior distribution probability by integrating the marginal likelihood Z (Eq.~\ref{eq:ultranest}) over the entire parameter space, where L(D|$\theta$) is the likelihood function and $\pi(\theta)d\theta$ is the prior probability density of the parameters $\theta$ \citep{Ultranest_Nested_sampling}. 

\begin{equation}
    Z=\int L(D| \theta) \times \pi(\theta)d\theta
    \label{eq:ultranest}
\end{equation}

The nested sampling technique consists of zooming towards best-fit models in the whole parameter space by regularly increasing the likelihood threshold. For further information on nested sampling, please refer to \cite{Ultranest_Nested_sampling}. Ultranest allows us to calculate the posterior distribution of the parameters and their standard deviations for the sinusoid, considering a set of priors, see Table~\ref{Tab:Priors_table}. Considering the sampling of the data and the gaps in the lightcurve, we used a flat period prior between 500 and 2500 days to obtain a reliable number of cycles. We describe the other priors used for the fitting in Table~\ref{Tab:Priors_table}. To ensure that the parameter space is correctly sampled we used 400 live points during the fit. After calculating the posterior distribution for each lightcurve of the final sample, which contains variable CRTS and ZTF observations, we selected sources that show a similar period for both data sets in the \cite{graham} and \cite{ZTF_variable_catalog} samples.

{\renewcommand{\arraystretch}{1.1}
\begin{table*}[t]
\caption{Prior distribution for the sine function used to fit the data.}
\begin{tabular*}{1\textwidth}{@{\extracolsep{\fill}}ccc}
\hline
Parameter & Description & Prior \\ \hline
A & Amplitude of the sine modulation (Mag) & Uniform[0.1, 100] \\
B & Background (Mag) & Uniform[5,25] \\
P & Period of the variability (days) & Uniform[500,2500] \\
t$_{0}$ & Phase of the sine function & Uniform[0,2$\pi$] \\
\hline
\end{tabular*}
\label{Tab:Priors_table}
\end{table*} 
}

\subsection{Damped random walk (DRW)}\label{DRW}
AGN exhibit variability in all wavebands. Quasars, which are supermassive black holes actively accreting matter, display intrinsic optical variability called red noise. Distinguishing between red noise and variability caused by a binary system is a hard task and \cite{False_periodicities} showed that apparently sinusoidal variations might be caused by red noise. The damped random walk (DRW) model, a first order continuous autoregressive model or CAR(1) process, is used to represent the optical stochastic variability in quasars \citep{DRW}.

\noindent The DRW is defined by an auto-covariance function of the form :

\begin{equation}
    ACV(t)=\frac{c\tau}{2}\times \exp(\frac{-t}{\tau})
    \label{ACV_DRW}
\end{equation}

\noindent Where $\tau$ and $c$ corresponds to the characteristic time-scale and total variance of the DRW \citep{False_periodicities}. In order to introduce the DRW model, we used the Gaussian process (GP) python package celerite \citep{celerite}. A GP model is made of a mean function $\mu_{\theta}(x)$ and a covariance or "kernel" function $k_{\alpha}(x_{n},x_{m})$. The log-likelihood of a GP model $L(\theta, \alpha)$ is defined with the following equation :

\begin{equation}
    \ln L(\theta, \alpha) = -\frac{1}{2}r_{\theta}^{t} K_{\alpha}^{-1} r_{\theta} -\frac{1}{2}\ln\, det K_{\alpha} -\frac{N}{2}\ln (2\pi)
\end{equation}

\noindent Where $r_{\theta} = y - \mu_{\theta}(t)$ is the vector of residuals and K is the covariance matrix with elements given by $K_{nm} = \sigma_{n}^{2}\delta_{nm} + k(t_{n},t_{m})$ with $\delta_{nm}$ the Kronecker delta and $k(t_{n},t_{m})$ the covariance function. We used the covariance function $k(|t_{n} - t_{m}|)=a \exp (-\nu_{bend}\times|t_{n} - t_{m}|)$, which is of similar form to the DRW auto-covariance function and where a is the variance of the DRW. The power spectral density (PSD) of the DRW model has the form of a bending power-law $\nu^{\gamma}$, with $\nu$ the frequency and $\gamma$ being 0 and smoothly bending to -2 above the bending frequency $\nu_{bend}$. In order to test our lightcurves with the DRW model, we fitted the DRW kernel to the lightcurves of our candidates using the nested sampling package Ultranest (see Section \ref{sec: nested_sampling}). Randomly sampling $a$ and $\nu_{bend}$ from the best-fitting parameter distribution derived with the nested sampling fitting, we simulated 10,000 lightcurves for each source using \citep{lightcurves_simulations} algorithm. We sampled the simulated lightcurves at the same rate than the binary candidates and calculated their $\chi^{2}_{\nu}$ to evaluate their goodness of fit. We compared the calculated $\chi^{2}_{\nu}$ for the DRW and the sine model, which gives a probability of the preferred model for each object in our sample.

\subsection{Generalised Lomb-Scargle periodogram}\label{sec: GLS}
In order to further confirm our tentative periodicities, we computed the Generalised Lomb-Scargle periodogram \citep[GLS; ][]{GLS}, using the frequency range corresponding to the period prior described in \ref{Tab:Priors_table}, for each source retained from \cite{graham} and \cite{ZTF_variable_catalog} samples. The GLS is a period finding technique for irregularly sampled lightcurves. The highest peaks in the GLS periodogram correspond, in general, to the most likely periods. In order to assess the significance of a given peak, we followed the methodology described in \cite{GLS_significance} and computed the {\it global p-value} of the highest peak in each calculated GLS, i.e, the probability to detect a peak of this power at any given frequency. This method consists of simulating 10 000 lightcurves following a DRW power spectral density (PSD) (see Section~\ref{DRW}). The bending frequency of the PSD is defined by $\nu_{bend}$ = 1/2$\pi\tau$ (see Eq.~\ref{ACV_DRW}). Using $\tau$ calculated in the DRW, we ensured that the whole PSD was properly sampled. The simulated lightcurves are generated using the same method as described in Section \ref{DRW} and are sampled at the same rate as the tested lightcurves. Firstly, for a given source, we computed the probability of the highest peak in its GLS periodogram, by computing the probability a peak of this power can be reproduced in the simulated lightcurve at this specific frequency. This probability is called the single p-value. However, spurious periodicities due to red noise can occur at every frequency. So, to evaluate the global significance of a given candidate, we computed the single p-value of the highest peak in each of the 10 000 simulated lightcurves and determined the global significance, called the global p-value, of the candidate by calculating how many time the simulations produce a single p-value probability lower than the peak candidate.

\subsection{A more realistic model}\label{red_noise_sine_model}
The pure sinusoidal model is a simple model that we used to find potential MBBH candidates. However, AGN are variable objects due to instabilities in the accretion disk. Thus, in the case of a MBBH, the sinusoidal variability should be modulated by red noise. We tested our candidates with a more realistic model, composed of a DRW GP kernel (as described in Section \ref{DRW}) to account for the red noise modulation added to a sine function. We fitted the lightcurves for our proposed candidates with the DRW model using the nested sampling algorithm (see Section \ref{sec: nested_sampling}). The prior distributions used for the sine function are described in Table~\ref{Tab:Priors_table}. To assess whether this model is preferred over the DRW model described in Section~\ref{DRW}, we compute the Bayes ratio. The Bayes ratio R is defined as the ratio between the marginal likelihoods, or evidence, of the two models :  
\begin{equation}
    R=\frac{Z(DRW+SINE)}{Z(DRW)}
\end{equation}
In order to interpret the computed Bayes ratio, we adopt the widely used Jeffreys scale \citep{Bayes_factor_scale}. Values of the Bayes ratio R in the range 3.2 $<$ R $<$ 10 and R $>$ 10 are, respectively, substantial or strong evidence for the presence of sinusoidal variability in a DRW model. R$<$1 indicates that the DRW model is preferred over the DRW plus sine modulation model.

\subsection{Gravitational decay}
The orbital decay due to gravitational wave emission is linked to the inspiral motion of the binary and it is determined analytically through Post-Newtonian approximations. The gravitational wave emission of the inspiral motion is related to the separation of the binary. Indeed, the time to merge $t_\mathrm{m}$ of the binary, which depends on the energy loss through gravitational wave emission, is associated with the binary separation according to a power-law function, $r_{12} \propto t_\mathrm{m}^{1/4}$ \citep[in geometric units $\mathrm{G}=\mathrm{c}=1$,][]{time_to_merge} : 

\begin{equation}
    t_\mathrm{m}=\frac{5r_{12}^{4}(t)}{256\nu M^{3}}
    \label{eq:t_to_merge}
\end{equation}

\noindent Where, $M$ is the total mass of the binary and $\nu$ is the symmetric mass ratio parameter $\nu=M_{1}M_{2}/M^{2}$, with $M_{1}$ and $M_{2}$ the masses of, respectively, the primary, secondary black holes in the binary. However, the orbital separation is also linked to the orbital period, $P \propto r_{12}^{3/2}$. Thus, it is possible for a given period, to compute, in units of M, the expected binary separation and the orbital decay due to gravitational wave emission, assuming a given mass ratio. However, for mili-parsec separation MBBHs, the change of the period due to the emission of gravitational waves is thought to be longer than the time span of the observations used here. 
We will discuss this further in Section~\ref{sec: phys interp}.

\section{Results} \label{Results}
We investigated the \cite{graham} sample of MBBH candidates. Out of 111 possible candidates,  89 sources also have ZTF observations. Fitting intercalibrated CRTS and ZTF data for all 89 objects, we found 26 objects that showed similar variability in the CRTS and ZTF data. 

\begin{figure}[!h]
   \includegraphics[width=\hsize]{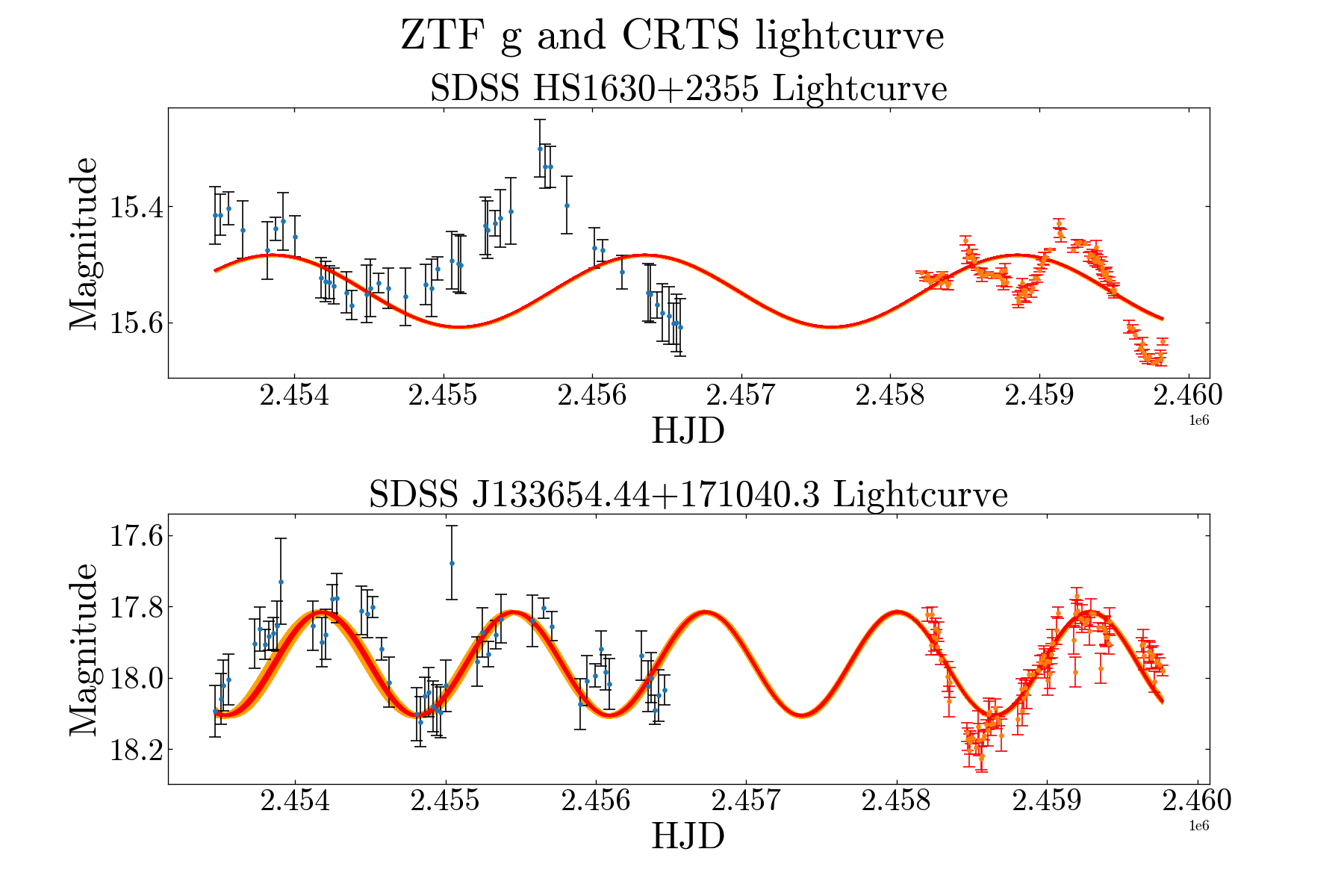}
    \caption{Top : CRTS (black) and ZTF-g (red) intercalibrated lightcurve for a rejected candidate from \cite{graham} sample. Bottom : CRTS (black) and ZTF-g (red) intercalibrated lightcurve for an accepted candidate from \cite{graham} sample.}
    \label{fig:Good_bad_candid}
\end{figure}

\begin{figure}[!h]
    \includegraphics[width=\hsize]{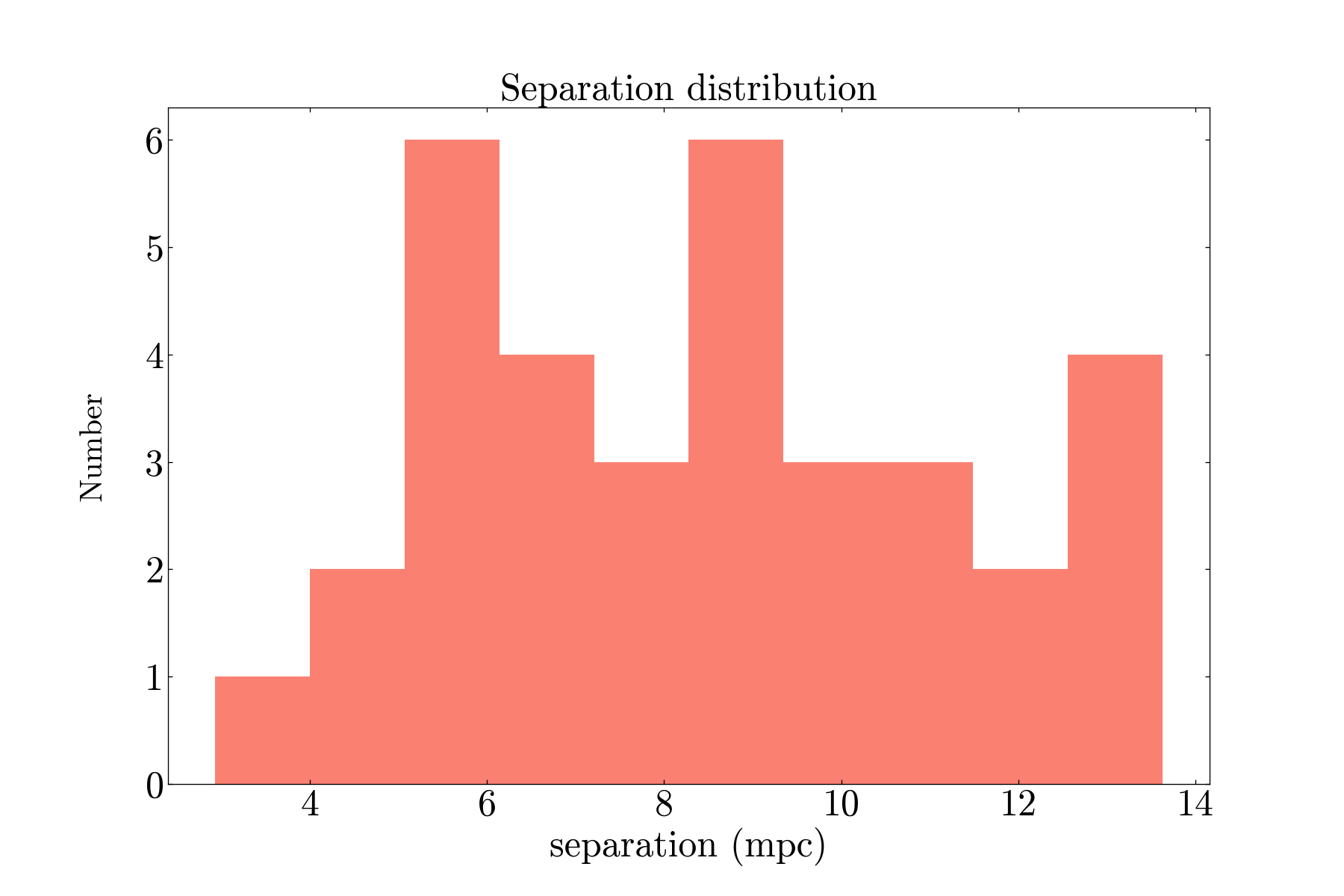}
    \caption{Calculated separation distribution considering circular orbits and equal mass black holes in the binaries }
    \label{fig:separation}
\end{figure}

We rejected 58 candidates, as the ZTF variability was clearly different from the CRTS variability (see for example Fig.~\ref{fig:Good_bad_candid}, top panel). Finally, for 5 sources there was insufficient ZTF data to draw reliable conclusions as there was less than half a period of ZTF observations. After cross-matching \cite{ZTF_variable_catalog} with Glade+ \citep{glade+}, we identified \num{21 862} variable objects, referenced either as galaxies or quasars. We fitted the ZTF data for those \num{21 862} objects with a sine function and selected sources with a $\chi^{2}_{\nu}$ < 1.3, reducing it to 1010 galaxies. We extended their ZTF lightcurves with CRTS observations and fitted the intercalibrated data altogether. Out of the initial sample of 1010 galaxies, 10 showed consistent variability between CRTS and ZTF observations. We display in Appendix~\ref{appendix:Appendix_A_lightcurves} the CRTS and ZTF lightcurves of all the 36 objects, CRTS and ZTF observations are respectively black and red dots. Note however, that in the joint lightcurves, the $\chi_{\nu}^{2}$ remains quite high for some fits due to some scatter in the data. \cite{graham} identified between 2 and 3 cycles per proposed candidate, for the 36 objects we identify between $\sim$3 and 5 cycles for each candidate by extending the lightcurves using both CRTS and ZTF data. This is close to the number of periods proposed in \cite{False_periodicities} to infer periodic behaviour. We give the SDSS spectral redshift in Table~\ref{Tab:Table1}. This allows us to estimate the expected separation of the binary candidates, using redshift corrected periods, supposing equal masses for the two black holes and using the mass estimates from \cite{Masse_candidates} (see Fig.~\ref{fig:separation}).  Masses estimated in \cite{Masse_candidates} are done using continuum luminosity and H$\beta$, Mg(II) and C(IV) width measurements in SDSS optical spectra. The results are presented in Table~\ref{Tab:Table1}.

\begin{figure}[!h]
  \includegraphics[width=\hsize]{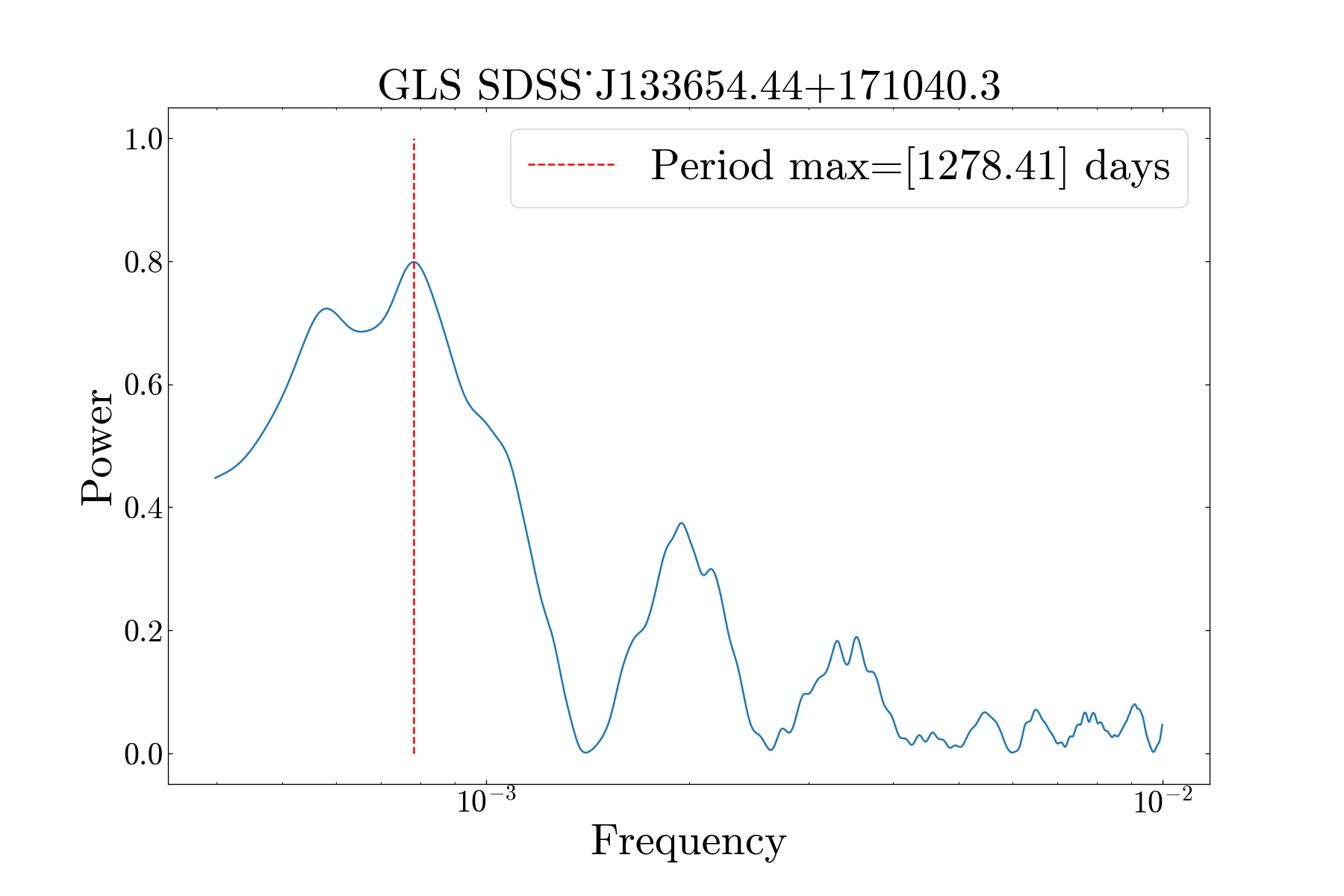}
    \caption{GLS periodogram of the source SDSS J133654.44+171040.3, the period indicated from the strongest peak is in agreement with the fitted period by Ultranest}
    \label{fig:Lomb_Scargle}
\end{figure}

The highest peaks in the GLS periodogram for the 36 sources give the same periods, within the error bars, as those inferred through fitting the data (see Table~\ref{Tab:Table1}). We show in Figure~\ref{fig:Lomb_Scargle} an example of the GLS periodogram for the quasar SDSS J133654.44+171040.36. 

\begin{figure}[!h]   
    \includegraphics[width=\hsize]{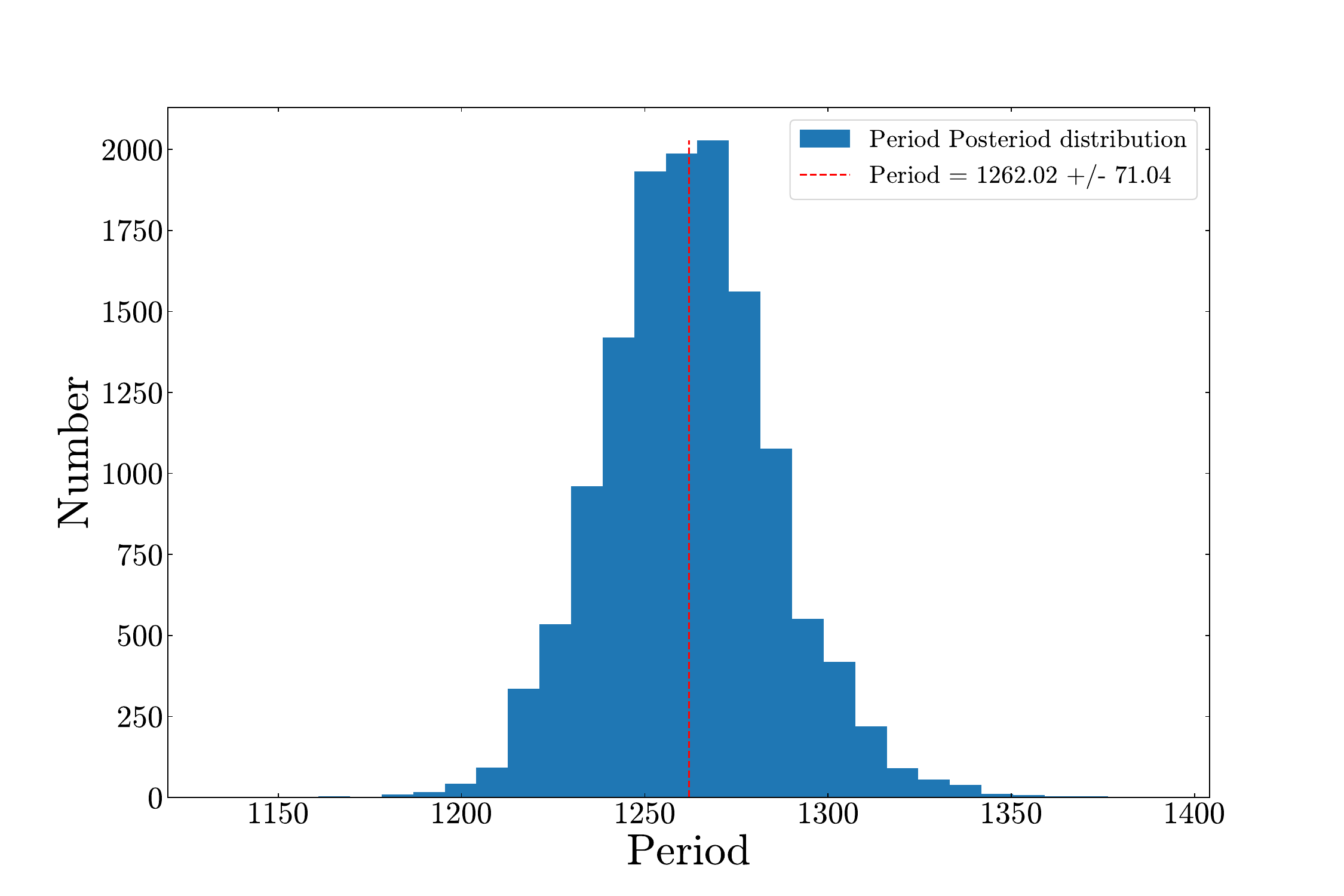}
    \caption{Period posterior distribution for the source SDSS J133654.44+171040.3 using a red noise plus sine modulation model}
    \label{fig:Period_dist}
\end{figure}

After modelling the lightcurves with the damped random walk, as described in Section~\ref{DRW}, we compared the results to those obtained with the sinusoidal fit. The characteristic time scale $\tau$ found for our sources is similar to the value inferred for quasars as the mean $\tau$ in the sample of 36 candidates is $\sim$ 125 days \citep[e.g.][]{tau_qso_1, tau_qso_2}. For the 36 sources, the sine model gives a better fit as the $\chi_{\nu}^{2}$ remains higher in the 10$^{4}$ simulated lightcurves. Using the more realistic model, incorporating red noise into the sinusoidal modulation, as described in Section~\ref{red_noise_sine_model}, the posterior period distributions resulting from the nested sampling computation peak at similar periods but with larger uncertainties than the sine only model for most of our sources (see Table~\ref{Tab:Table1}). We show in Fig.~\ref{fig:Period_dist} the posterior period distribution for a typical candidate. The Bayes ratio R calculated to compare this model with the pure DRW model gives a low value (R<1) for 30 sources, preferring the DRW only model. 5 sources have a Bayes ratio 1<R<3.2 indicating that the two models are equivalent. Finally, 1 object, J141425+171811, shows strong evidence for the DRW and sine model, as its Bayes ratio R is larger than 7. However, as discussed in Sec.~\ref{discussion}, the Bayes ratio analysis might be inconclusive as it remains low, even in the presence of sinusoidal modulation, because of the sparse, noisy data.

\begin{figure}[!h]   
    \includegraphics[width=\hsize]{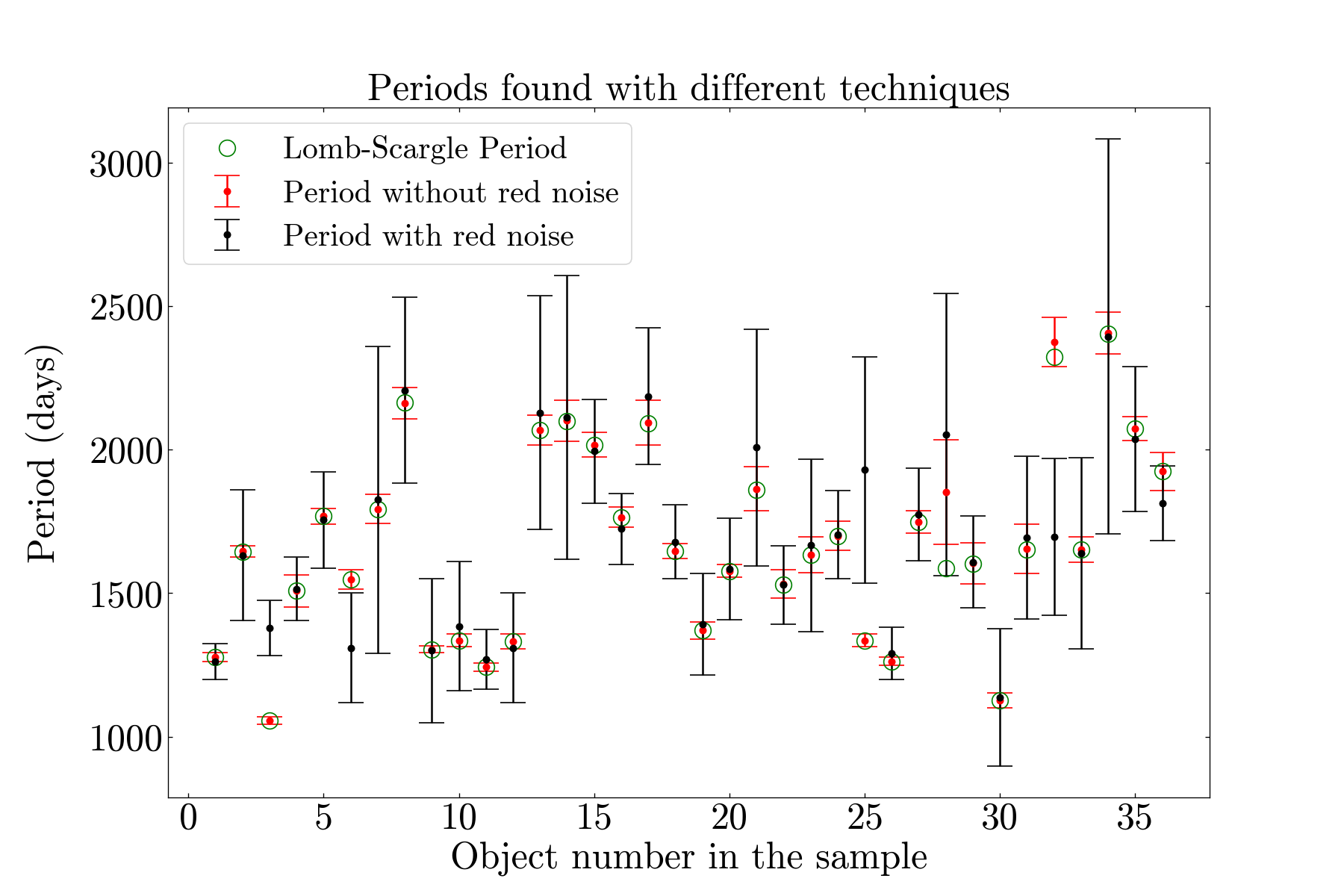}
    \caption{Periods and uncertainties found for the candidates in Table~\ref{Tab:Table1} with the different techniques applied. The object list on the x axis follows the same order as in Table ~\ref{Tab:Table1}. The good agreement of the results from the different techniques can be clearly seen.}
    \label{fig:Period_diff_tech}
\end{figure}

The periods found for each object with the pure sine model, the Lomb-Scargle periodogram and the sine modulated red noise model are comparable and are in general consistent within the error bars (see Fig.~\ref{fig:Period_diff_tech}). For the 36 sources, we computed the significance of the highest peaks in their GLS periodogram using the method described in Section~\ref{sec: GLS}. The significance of the peaks remains quite low for some of the objects, due to the quality of the data and the low number of cycles (3-5). These tests do not completely rule out the red noise origin and additional multi-wavelength evidence is needed to confirm the MBBH possibility.

\begin{figure}[!h]
\includegraphics[width=\hsize]{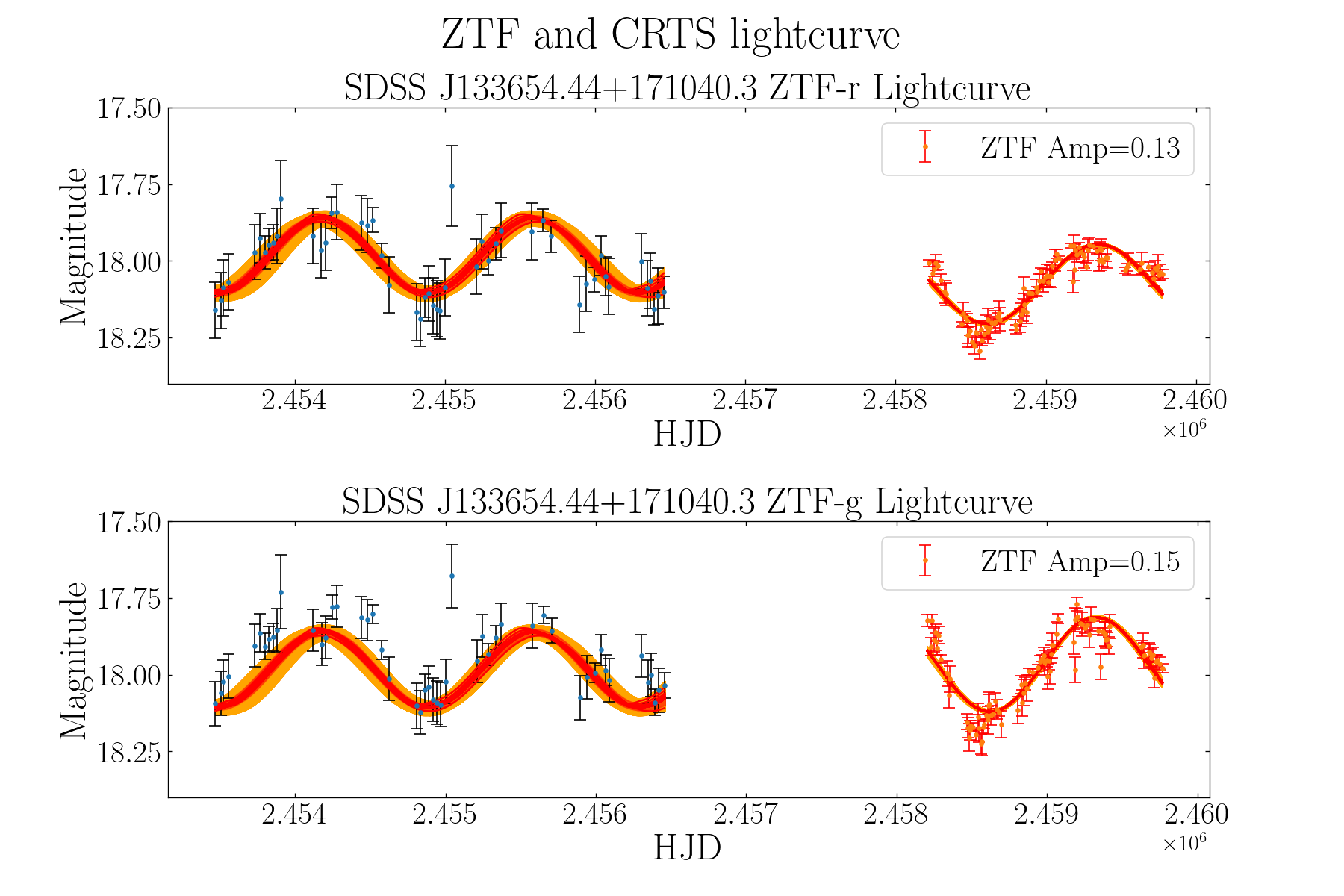}
    \caption{Lightcurve of SDSS J133654.44+171040.3. Top: CRTS data (black) and ZTF data (red) taken in the r-band. Bottom: CRTS data (black) and ZTF data (red) taken in the g-band.}
    \label{fig:lc_filtres_diff}
\end{figure}

We performed the initial periodicity analysis in the g band. Additionally, we investigated the lightcurves with ZTF data in the r band. We compared the lightcurves from the two wavebands in an additional check for periodicity. We found the same period in both the r and the g band lightcurves for each galaxy. We also compared the amplitudes of the two bands, to constrain the origin of the emission. Figure~\ref{fig:lc_filtres_diff} shows intercalibrated lightcurves of SDSS J133654.44+171040.3, for ZTF data taken in the r-band (top) and g-band (bottom). The periodicities are compatible within 3$\sigma$ but the amplitude of the oscillation in the g-band is slightly larger, values are given in Table~\ref{Tab:Table1}. The increased variability in the blue is expected for MBBHs \citep{Cocchiararo2024}, but can also be observed in single black hole quasars \citep{quasar_variability}.

\section{Discussion} \label{discussion}
\begin{figure}[!h]
    \includegraphics[width=\hsize]{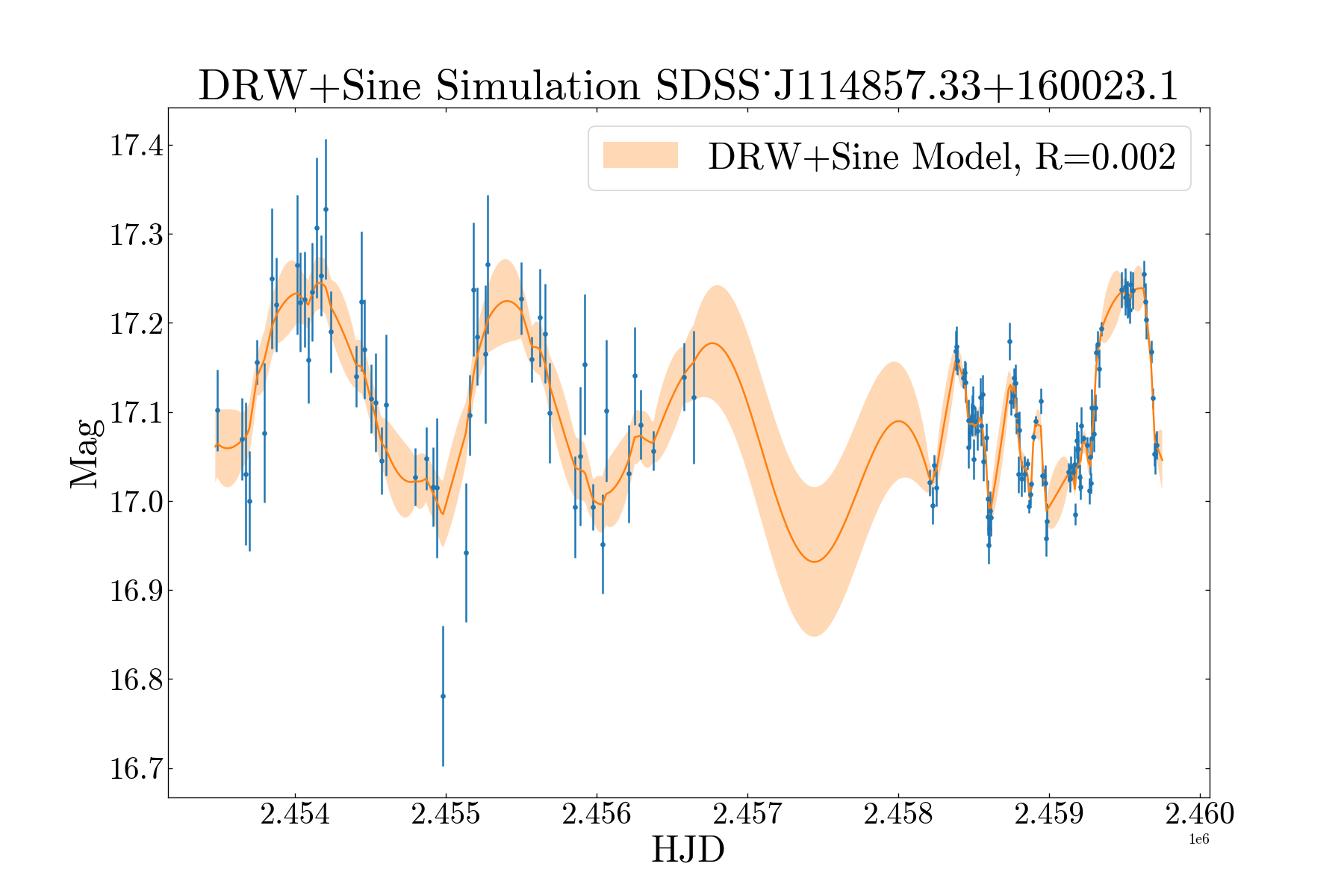}
    \caption{Simulated lightcurve with a DRW and a sine modulation. The lightcurve has the same sampling and error bars as the candidate J083349.55+232809.0. The orange envelope correspond to 1$\sigma$ uncertainty of the best DRW+Sine fitted model}
    \label{fig:DRW_Sine}
\end{figure}

Validating MBBH candidates through lightcurve analysis is a challenging task. From our initial sample of 1099 sources coming from our selected ZTF data and \cite{graham} sample, we identified sinusoidal modulation in the CRTS and the ZTF data sets independently. Of these, 36 show similar periodicities in the CRTS and ZTF data. We show in Appendix~\ref{appendix:Appendix_A_lightcurves} the lightcurves of the identified candidates. However for other objects, when fitted together, the periodic variations identified in individual data sets were different. As an example of such unmatched periodicities, we show in the top panel of Fig.~\ref{fig:Good_bad_candid} the lightcurve of the quasar HS1630+2355. HS1630+2355 is a quasar coming from the \cite{graham} sample to which we added ZTF data. We can see that CRTS data exhibit apparently periodic sinusoidal variations which do not match ZTF data. According to Eq.~\ref{eq:t_to_merge} and if we consider the variability to probe the orbital period of the binary, the change of period due to the emission of the gravitational waves is too small for this separation scale to explain the observed variation in periodicity.
Thus, we assign this difference in identified periodicities to red noise and discard the binary scenario to explain the variability. Out of the 89 sources for which there are ZTF observations, we discard 58 of them as the identified variability observed in CRTS data is not seen in ZTF observations. In the final sample of MBBH candidates, all of our candidates have data spanning between 3 and 5 cycles, necessary to help rule out red noise behaviour. We determine similar periods through the Lomb-Scargle analysis and the DRW+Sine model, further reinforcing the likelihood of the presence of periodic modulation, which also lends support to the idea that these galaxies are home to close binary black holes. However, in most of our sources, the Bayes ratio indicates that the DRW model is preferred or is equivalent to the DRW plus a sinusoidal modulation. In order to estimate the Bayes ratio when a sine modulation is present, we simulated lightcurves with a sine modulation added to a DRW model for each candidate and computed the Bayes ratio. The parameters of the DRW and the sinusoidal modulation (e.g. a, $\nu_{bend}$, A, B, P and t$_{0}$) are set to the best values derived from the fitting of the DRW+Sine model to each candidates. The amplitude of the periodic modulation added in the simulated lightcurves is $\sim$ 0.1-0.2 magnitudes. The variance of the DRW process modelled in the simulations remains lower than the amplitude of the sinusoidal modulation by one order of magnitude for each candidates. The simulated lightcurves have the same sampling and error bars as the candidates. Despite the fact that we introduced a sinusoidal modulation, we retrieve simulated lightcurves with lower Bayes ratios than the candidate MBBH Bayes ratios (see for example Fig.~\ref{fig:DRW_Sine}, R is lower than the lowest R calculated for our proposed candidates). R in the simulated lightcurves goes as low as $\sim$ 0.001. This is likely due to the paucity and low signal to noise ratio of the data; therefore, completing the lightcurves with additional data, taken with ZTF or the upcoming Vera C. Rubin Observatory, will be crucial to confirm which model is more probable.

In addition, we checked the SDSS optical spectra of the candidates looking for broadening or asymmetries in the optical lines which could indicate the presence of a binary. However, no definitive evidence for binarity was visible because of the separation of the candidates and the limited spectral resolution.

\subsection{Comparison with simulations}
\begin{figure*}[!h]
    \centering
    \begin{subfigure}[t]{0.45\textwidth}
        \centering
        \includegraphics[height=1.9in]{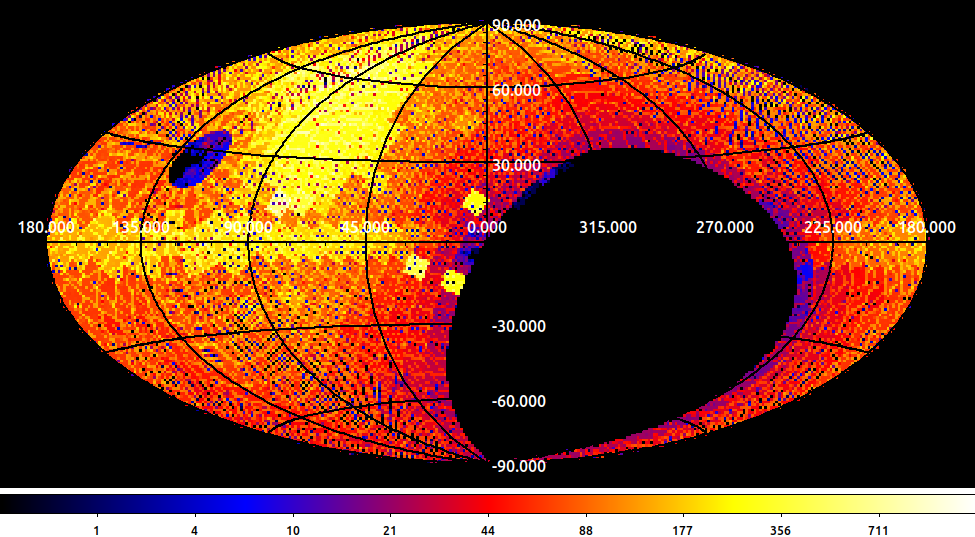}
    \end{subfigure}
    \hspace{0.5cm}  
    \begin{subfigure}[t]{0.45\textwidth}
        \centering
        \includegraphics[height=1.9in]{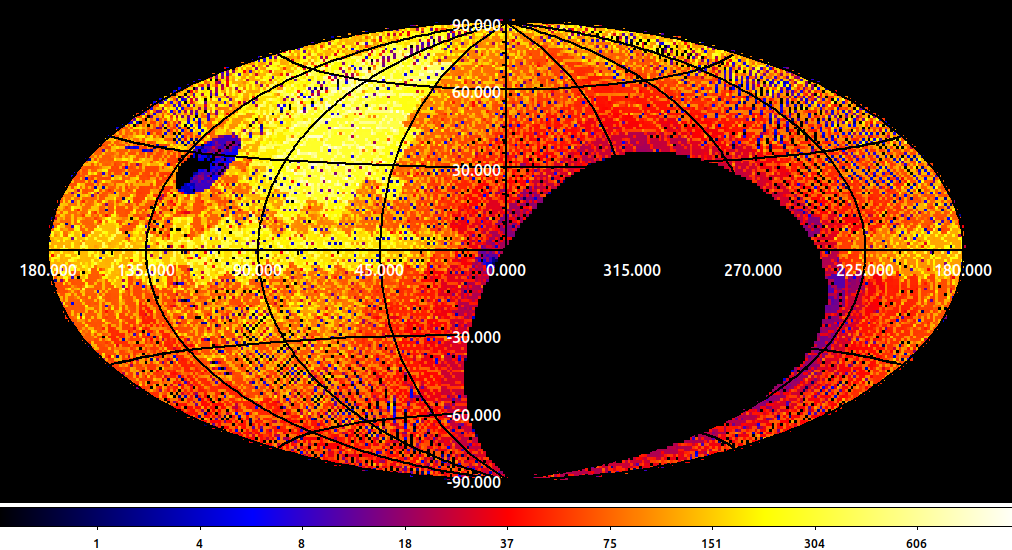}
    \end{subfigure}
    \caption{Number of observations per CCD-quadrant in ZTF DR2 in galactic coordinates centred at l,b=0,0. Left and right figures are respectively number of observations made in r and g filters.}
    \label{fig:n_obs}
\end{figure*}

\begin{figure}[!h]
    \includegraphics[width=\hsize]{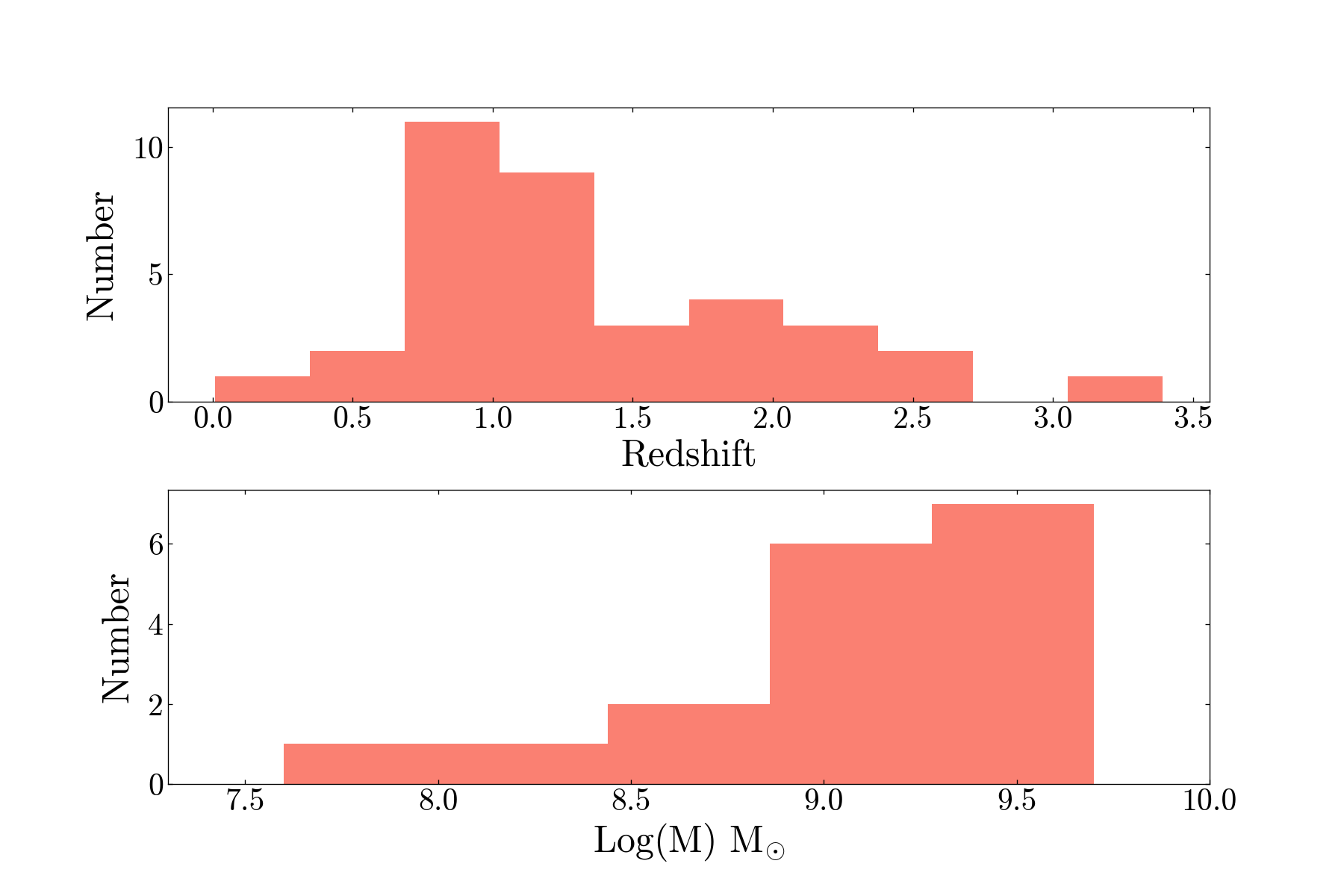}
    \caption{Top : Histogram representing the redshift distribution of proposed MBBHs candidates. Bottom : Histogram representing the Log(M) distribution of proposed MBBHs candidates}
    \label{fig:histo_mass_candidates}
\end{figure}

Some of the candidates in our sample are from the \cite{ZTF_variable_catalog} catalogue, which gives periodic sources found in ZTF DR2. DR2 only covers a small part of the sky (see Fig.~\ref{fig:n_obs}), which creates a selection bias. As a result, the candidates that we retrieved from this catalogue are grouped in the area with the most observations. We calculated the expected number of sub-parsec MBBHs candidates at redshift $z<1$ in the Universe, extrapolating from our candidates. The MBBHs candidates we propose have been found with joint CRTS and ZTF observations and the ZTF and CRTS survey cover around 30000 square degrees on the sky \citep{CRTS}. Thus, extrapolating to the whole sky area and considering a homogeneous distribution, we could expect to find 19 sub-parsec supermassive binary quasars at a redshift $z<1$. \cite{MBBHs_low_redshift_expectations} found that they are relatively rare objects, with only around \num{10} objects at $z<0.7$ but growing by a factor of $\approx 5-10$ for a redshift $z<1$. Our result is slightly lower, as our study might not be complete due to the magnitude limitations of the surveys. Further, our sample seems representative of the whole sample as the total mass of the objects in our sample is quite high, with only 2 objects with a mass lower than $10^{8} \, \mathrm{M_{\odot}}$ and 20 objects with a mass higher than $10^{9} \, \mathrm{M_{\odot}}$ (See Fig.~\ref{fig:histo_mass_candidates}). This is expected as the detection lifetime of MBBHs increases with the binary mass \citep{residence_time_SMBH}. However, our number of candidates is higher than expected from \cite{2019MNRAS.485.1579K} as they compute the CRTS survey to be able to detect between 0.2 and 5 binaries. Therefore, our list of candidates might contain a significant number of false positives. 

\subsection{Physical interpretation}\label{sec: phys interp}
We can see that the identified periods are the same in r and g filters and that the fitted amplitudes differ slightly but are the same within the error for each candidate, see  Table~\ref{Tab:Table1}. These fitted amplitudes are systematically higher in the g band (the average amplitude in the blue filter is 0.167 and in the red filter it is 0.145), suggesting that the amplitude of the variability is higher in the g band. Stronger modulation in the blue band is suggestive of an accretion disk origin \citep[e.g.][]{Cocchiararo2024}. However, a similar increase of the variability in the blue can be seen in regular quasars \citep{quasar_variability}. Different origins for the observed periodicity are plausible, such as a blob of gas orbiting in the CBD, referred to as the \lq lump{\rq} (\citealt{BBH_emission_lump_farris15}, \citealt{tang_late_2018}\footnote{In \cite{tang_late_2018}, the assumed gas temperature is ${\sim}10$ times higher than in other theoretical studies (e.g. \citealt{bbh_sed_notch_roedig14}, \citealt{bbh_emission_dascoli18}), resulting in X-ray emission rather than optical for the CBD.}, \citealt{lump_modulation_MR24}, of plausible instability origin: \citealt{lump}); a periodic accretion flow due to the secondary black hole being closer to the inner edge of the circumbinary disk \citep{accretion_flow_varability}; quasi-periodic oscillations (QPOs) due to the beat frequency between the orbital frequency and the frequency of the blob of gas in the CBD (e.g. \citealt{QPO}, \citealt{lump_modulation_MR24}) or Doppler boosting emission from the orbiting mini-disks \citep{doppler_boosting}. Indeed, the relativistic effects due to the orbiting black holes, would periodically boost or dim the emitted light of mini-disks moving towards or away from us. 

\begin{figure}[!h]
\includegraphics[width=\hsize]{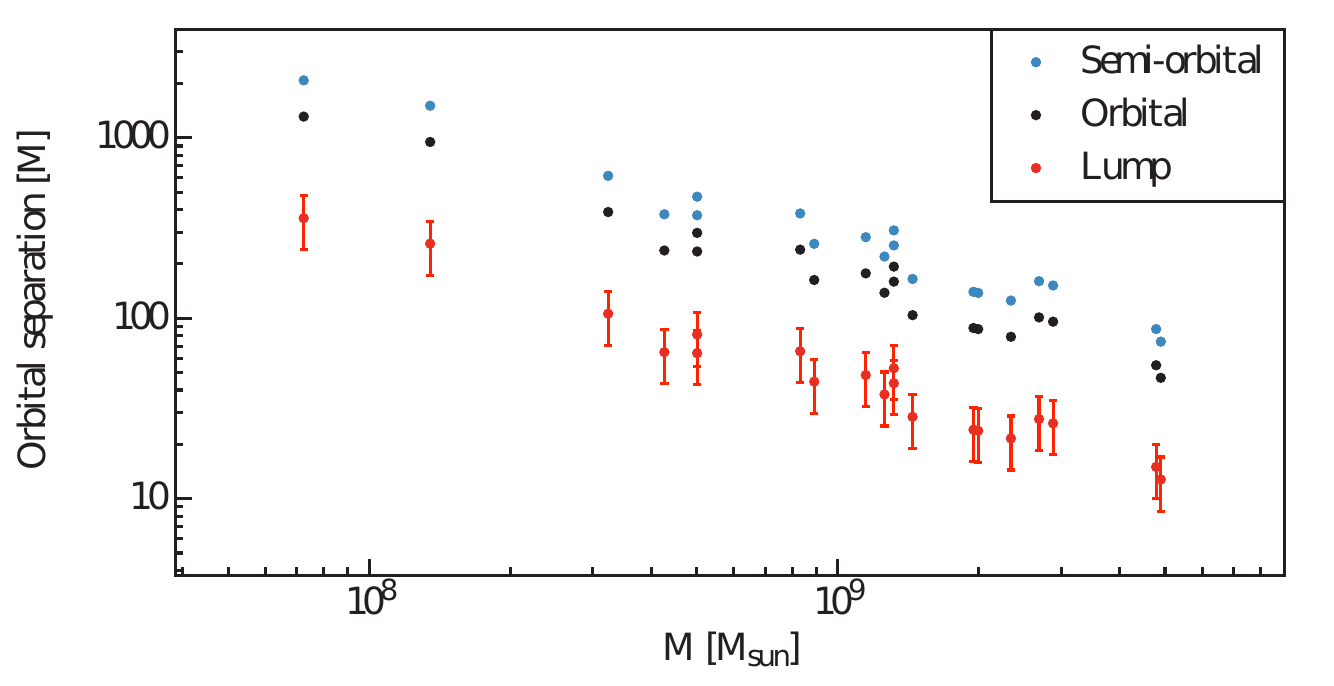}
    \caption{Orbital separation as a function of $M$, depending on the origin of the modulation (semi-orbital, orbital, or due to the lump's orbit). 
    The orbital separation is given in units of the total MBBH mass $M$, under the geometric unit system $\mathrm{G}=\mathrm{c}=1$.
    For the lump-origin points, the error bars are obtained by considering a lump period between $4\, P$ and $10\, P$.}
    \label{fig:M_r12}
\end{figure}

In Fig.~\ref{fig:M_r12} we show the theoretical orbital separation, under the geometric unit system $\mathrm{G}=\mathrm{c}=1$\footnote{In this unit system, spatial scales in physical units, denoted $d'$, are simply given by $d'=d \, (M/1\mathrm{M_\odot}) \times 1.477$~km, where $d$ is the spatial scale in geometric units.}.
This unit system allows for direct comparison between different BBH sources, since spatial scales are then normalized by the gravitational radius and can be used to estimate the region of gravitational influence of the BBH. The separation is derived from the observed variability period, corrected for the redshift, and assuming Keplerian orbits and equal-mass binaries (the mass ratio has a negligible influence on the period though, in these units), depending on the physical timescale on which the variability occurs: the orbital period $P$, the semi-orbital period $P/2$, or the lump (blob) period ${\sim}4-10\, P$ (e.g. \citealt{Shi12}, \citealt{Noble12}, \citealt{Westernacher22}, \citealt{lump}) which gives the error bars.
This representation helps us to see if the orbital separation is consistent with the MBBH geometry associated with the assumed variability origin (see Fig.~\ref{fig:MBBH}).
Indeed, the larger the separation in the BBH, the less likely it is to possess a CBD, as it would require a larger angular momentum budget for the infalling gas building the CBD, in which case the variability might not be attributed to the lump.
We can reasonably assume that BBHs with an orbital separation larger than $1000$~M are unlikely to host a CBD.
On the one hand, variability occurring on the orbital or semi-orbital period can occur both in circumprimary disks (e.g. via accretion modulation produced by the $m=2$ spiral arms) and CBD configurations (e.g. via Doppler boosting of the mini-disks, \citealt{doppler_boosting}).
The circumprimary disk scenario would be more likely for systems with an orbital separation ranging $100-1000$~M.
On the other hand, since the lump period is larger than the orbital period, the derived orbital separation is consistently smaller.
As a consequence, except for two sources (J152157.02+181018.6 and J141425.92+171811.2), we get an orbital separation $\lesssim 100$~M which is compatible with a CBD configuration and therefore with the lump origin.
Hence, interestingly, neither the CPD nor CBD geometry can be ruled-out here.

Let us briefly discuss the implications for the time to merger.
If the variability is at the orbital or semi-orbital period, all systems are more than $10^5$ days away from merger.
In that case, the decrease of the period due to gravitational wave emission is indeed too small to be perceived, as could be expected.
However, if the variability is at the lump period, the MBBH could be as close as $600$~days before merger and five of our candidates have a time to merger smaller than $10$ years: J160730.33+144904.3, J134855.92-032141, J081133.43+065558.1, J083349.55+232809.0 and J104941.01+085548.4. For these systems, the impact of the inspiral motion onto the lump-related modulation period may not be negligible and depends also on the time at which BBH-CBD decoupling occurs \citep{SimuBBH_inspiral_farris15}. From a theoretical point of view, the survival of the lump-related modulation close to merger is still uncertain and calls for dedicated studies. It is however very unlikely that five sources have such a short time to merger, so it is highly likely that most, if not all, show modulation due to the binary orbit or semi-orbit. To confirm the origin, further data are necessary. 

\begin{figure}[!h]
\includegraphics[width=\hsize]{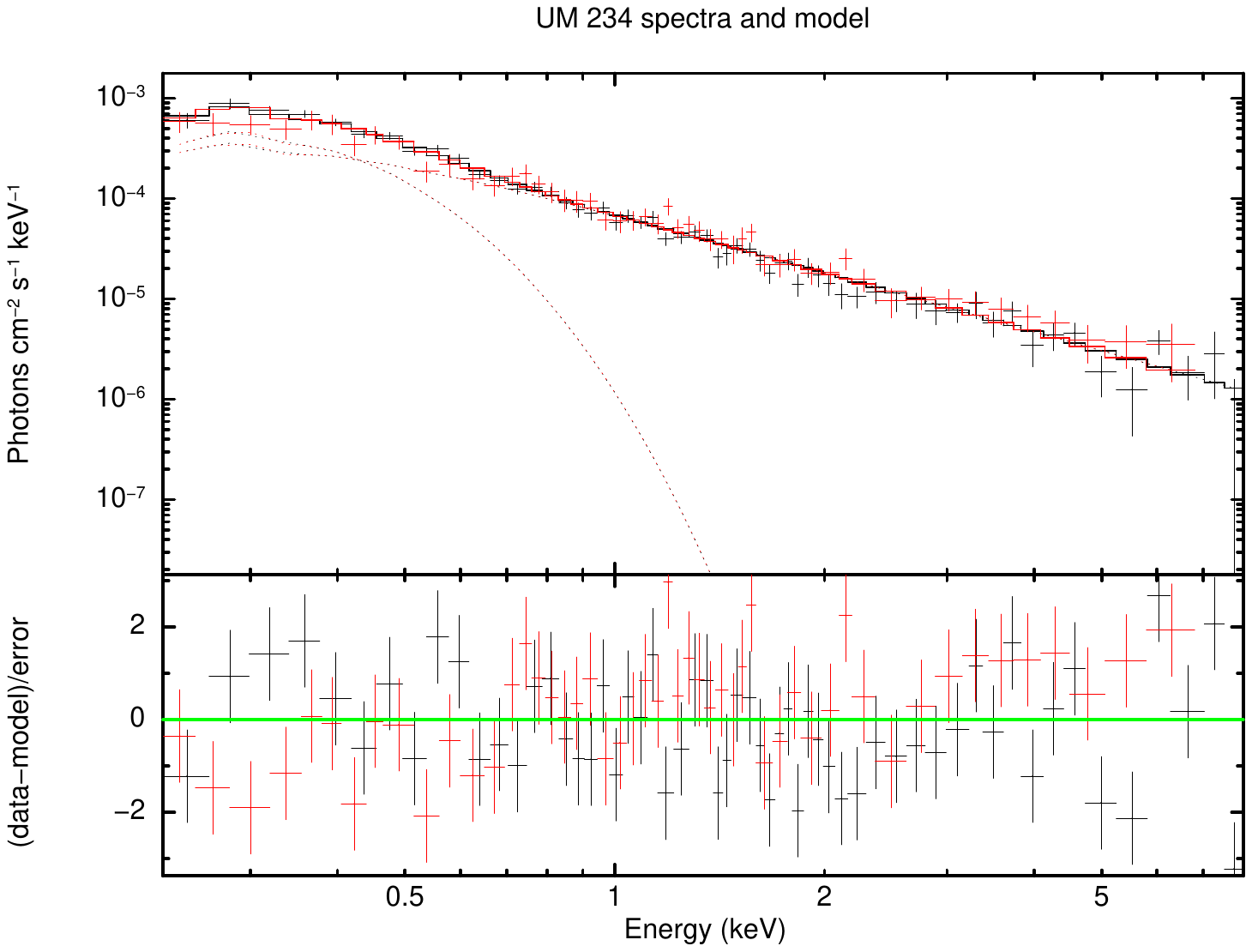}
\caption{Top : UM 234 X PN (black) and MOS2 (red) spectra modelled with an absorbed power-law and blackbody. Bottom : residuals of the applied model}
\label{Fig:UM234_X}
\end{figure}

Other diagnostics can be used to rule out some of the aforementioned scenarii. If the detected emission is due to Doppler boosting of the mini-disks, the variability is expected to be stronger in the UV and the X-rays \citep{Multi_messenger_SMBBH}. Multi-wavelength observations would therefore be useful to help constrain the variability, yet difficult considering the timescales of the detected periodicities. Two sources in our sample, J123527.36+392824.0 and UM 234, have been observed once in X-ray with the XMM-Newton observatory. They show a typical AGN spectra, well modelled with an absorbed power-law and blackbody (See Fig~\ref{Fig:UM234_X}), parameters are shown in Table~\ref{Tab:Table2}. If we suppose an expected geometry as described in Fig~\ref{fig:MBBH}, \cite{simulation_MBH} showed that, for a black hole of $10^{6} \, \mathrm{M_{\odot}}$, the soft X-ray emission ($\sim$ 1 keV) is dominated by the CBD and the mini-disks, while the hard X-ray part (>1 keV) is of coronal origin and found to originate from the mini-disks entirely. Considering the redshift and the masses of the two sources we expect their X-ray spectra to be mainly dominated by the coronal emission of the mini-disks. This is indeed what is seen. However, this kind of spectrum is also typical in single type I AGN.

\begin{figure}[!h]
    \includegraphics[width=\hsize]{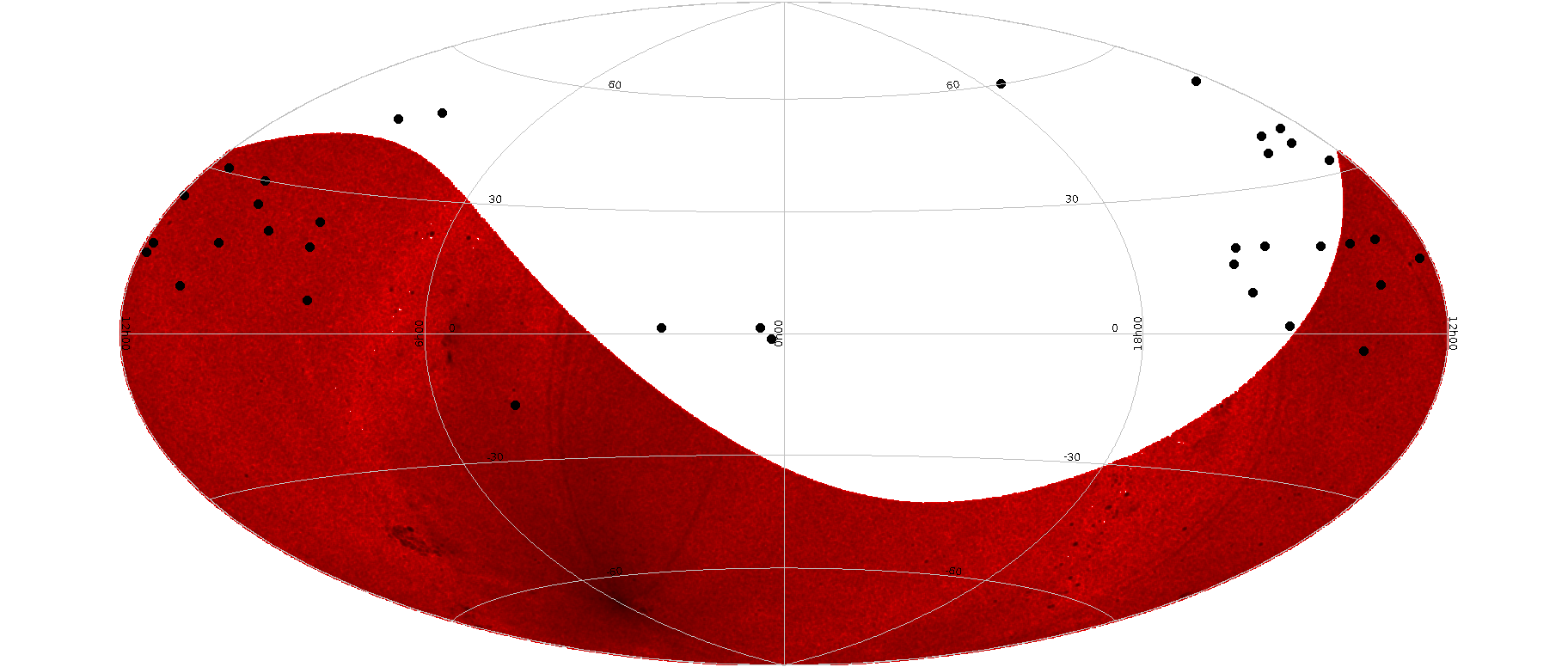}
    \caption{Aitoff projection of the eRASS1 DR1 main catalog, our candidates coordinates are represented with black dots}
    \label{fig:erass_candid}
\end{figure}

Galaxy merger studies show that major galaxy mergers can trigger black hole activity, as  the AGN evolves from a type II to a type I AGN closer to the end of the merger \citep[e.g.][and references therein]{AGN_unification}. Then, detecting type I AGN emission could help trace a past galaxy merger (see Section~\ref{sec:introduction}). The sources in our sample  from \cite{ZTF_variable_catalog} are galaxies, as they were selected by cross-matching with the galaxy catalogue Glade+. Even though only a fraction of our candidates were selected as they are AGN (notably those from the Graham sample), optical spectroscopy from SDSS and X-ray detections using the eRASS DR1 catalogue  \citep{erosita_cat} in the 0.2-2.3 keV band (see Table~\ref{Tab:Table erosita}) indicate that all 36 candidates that we propose as MBBHs are actually quasars. This could provide further evidence for the binary black hole nature. Note, the eROSITA catalogue contains data for half the sky, and 18 out of 36 of our candidates fall within the footprint of this catalogue (see Fig.~\ref{fig:erass_candid}).

In order to determine if these MBBH candidates accrete at fairly high rates, as expected if they are the result of mergers \citep[e.g.][]{post_merger_galaxies_properties, galaxy_pairs_agn}, we tried to estimate the accretion rate as a percentage of the Eddington rate. Out of the 18 candidates in the eRASS DR1 footprint, 7 objects are detected with eROSITA. The median limiting flux for an extragalactic source in the eRASS1 DR1 main catalogue is 6$\times$10$^{-14}$ erg.cm$^{-2}$.s$^{-1}$ \citep{erosita_cat}, so we consider this value as an upper limit for non detected candidates in the catalogue. We show in Table \ref{Tab:Table erosita} the 18 candidates in the part of the sky covered by the eRASS1 catalogue, as well as their redshift, detected flux or upper limit, their predicted flux assuming that they radiate at 20$\%$ of their Eddington luminosity and their computed luminosity. The predicted flux was calculated using the bolometric to X-ray correction proposed in \cite{bolomectric_correction_x_ray}. The detected fluxes or upper limits are compatible with the very high accretion rate of 20$\%$ Eddington, except in the case of J115445+244646, which is the only source that shows a flux significantly lower than the predicted flux, putting an upper limit on its accretion rate around 5$\%$ of its Eddington accretion rate. These high accretion rates indicate a large reservoir of matter available to be accreted, which may originate from a previous merger (see for example Section~\ref{sec:introduction}).

{\renewcommand{\arraystretch}{1.1}
\begin{table*}[!h]
\caption{Candidate MBBHs eROSITA sources}
\begin{tabular*}{1\textwidth}{@{\extracolsep{\fill}}ccccc}
\hline
Object         & Redshift &  \multicolumn{1}{c}{\begin{tabular}[c]{@{}c@{}} Flux \\ ($\times10^{-14}$ ergs/cm²/s)\end{tabular}} & \multicolumn{1}{c}{\begin{tabular}[c]{@{}c@{}}  Predicted flux \\ ($\times10^{-14}$ ergs/cm²/s)\end{tabular}} & \multicolumn{1}{c}{\begin{tabular}[c]{@{}c@{}}   Luminosity \\ ($\times10^{44}$ ergs/s) \end{tabular}} \\ \hline
J114857+160023 & 1.22     & \textless{}6 & 8.09 & \textless{}5.3 \\ 
J133654+171040 & 1.23     & 15.9  & 5.62 & 14.4     \\ 
J100641+253110 & 1.13     & \textless{}6  & 7.43 & \textless{}4.4       \\ 
J115445+244646 & 0.8      & \textless{}6 & 21.1 & \textless{}1.9       \\ \
J102625+295907 & 3.39     & 4.46    & 0.81 & 48.0     \\ 
J113142+304139 & 2.34     & \textless{}6  & 1.45 & \textless{}27.0      \\ 
J092911+203708 & 1.84     & \textless{}6  & 2.91 & \textless{}15.0      \\ 
J043526-164346 & 0.01     & \textless{}6  & & \textless{}0.00014      \\ 
J081133+065558 & 1.27     & \textless{}6  & 6.28 & \textless{}5.9       \\ 
J082926+180020 & 0.81     & 5.61     & 5.93 & 1.8    \\ 
J083349+232809 & 1.15     & 8.31    & 7.0 & 6.4     \\ 
J102255+172155 & 1.06     & 15.0     & 4.04 & 9.4    \\ 
J104941+085548 & 1.18     & \textless{}6  & 6.85 & \textless{}4.9      \\ 
J115141+142156 & 1.00     & \textless{}6  & 6.11 & \textless{}3.2      \\ 
J121457+132024 & 1.49     & \textless{}6  & 4.23 & \textless{}8.7      \\ 
J130040+172758 & 0.86     & 12.6   & 9.09 & 4.7      \\ 
J131909+090814 & 0.88     & 5.7     & 6.01 & 2.2     \\ 
J134855-032141 & 2.1      & \textless{}6  &  2.43 & \textless{}20.0      \\ \hline
\end{tabular*}
\tablefoot{Candidate MBBHs that fall within the footprint of the eRASS1 DR1 main catalogue, their redshift, their detected or upper limit flux in the 0.2-2.3 keV band, their predicted flux supposing that they accrete at 20\% of Eddington and their luminosity corresponding to the detected flux/upper limit. The missing predicted flux is due to the fact that there is no mass measurement for this source in the \cite{Masse_candidates} catalogue.}
\label{Tab:Table erosita}
\end{table*} 
}

We cannot determine if the X-ray flux varies over the orbital period as there is insufficient data in this catalogue version. Future versions will provide further data. If there is a modulation and it is due to periodic Doppler boosting, this would give constraints on the angle at which we are observing the system, as this effect cannot be seen if the object is observed face-on, while the variability amplitude is maximum when the orbit is seen edge-on. 

{\renewcommand{\arraystretch}{1.1}
\begin{table*}[t]
\caption{J123527.36+392824.0 and UM 234 X-ray model parameters}
\begin{tabular*}{1\textwidth}{@{\extracolsep{\fill}}cccccccc}
\hline
object & redshift &  \multicolumn{1}{c}{\begin{tabular}[c]{@{}c@{}}   nH \\ (10$^{22}$ cm$^{-2}$) \end{tabular}}  &  \multicolumn{1}{c}{\begin{tabular}[c]{@{}c@{}}   kT\\ (keV) \end{tabular}} & $\Gamma$ & \multicolumn{1}{c}{\begin{tabular}[c]{@{}c@{}}   $\chi^{2}$ \\ (d.o.f) \end{tabular}} & \multicolumn{1}{c}{\begin{tabular}[c]{@{}c@{}} F$_{0.2-12.0}$\\ ($\times10^{-14}$ ergs/cm²/s) \end{tabular}} & \multicolumn{1}{c}{\begin{tabular}[c]{@{}c@{}}L$_{0.2-12.0}$\\ ($\times10^{44}$ ergs/s) \end{tabular}} \\ \hline
J123527.36+392824.0 & 2.15 & 0.015 & \begin{tabular}{@{}c@{}}
      \\ 
 0.3 $\pm$ 0.1 \\ 
\end{tabular} & \begin{tabular}{@{}c@{}}
1.77  \\ 
1.36 $\pm$ 0.2\\  
\end{tabular}  & \begin{tabular}{@{}c@{}}
110.99 (55)  \\ 
88.66 (53) \\  
\end{tabular}  & \begin{tabular}{@{}c@{}}
$4.7$ \\ 
$11.8$  \\  
\end{tabular}  & \begin{tabular}{@{}c@{}}
$22.3$  \\ 
$20.4$  \\ 
\end{tabular}                                             

\\ \hline
UM 234 & 0.73 & 0.024  & \begin{tabular}{@{}c@{}}
      \\ 
 0.14 $\pm$ 0.02 \\ 
\end{tabular}   & \begin{tabular}{@{}c@{}}
2.27 $\pm$ 0.07  \\ 
1.98 $\pm$ 0.2 \\ 
\end{tabular}  & \begin{tabular}{@{}c@{}}
141.99 (108)  \\ 
108.12 (106) \\  
\end{tabular} & \begin{tabular}{@{}c@{}}
$13.6$   \\ 
$44.9$ \\  
\end{tabular}  & \begin{tabular}{@{}c@{}}
$8.98$   \\ 
$9.66$  \\  
\end{tabular}   \\ 
\hline
\end{tabular*}
\tablefoot{Parameters of J123527.36+392824.0 and UM 234 X-ray spectrum modelled with an absorbed power-law with and without a black body component. The nH column density corresponds to galactic absorption, as the fit shows no intrinsic absorption. X-ray flux and luminosity were computed for the applied model. Uncertainties are given at a 3$\sigma$ significance}
\label{Tab:Table2}
\end{table*} 
}

\subsection{Other possible candidates}
For the sources with too little data, it is even more difficult to differentiate between modulations arising due to binarity or due to red noise. These candidates require many more observations to be considered good MBBH candidates, but we nonetheless create a catalogue of possible candidates, for which further observations are required. This catalogue of possible candidates contains 221 objects. 

\begin{figure}[!h]
    \includegraphics[width=\hsize]{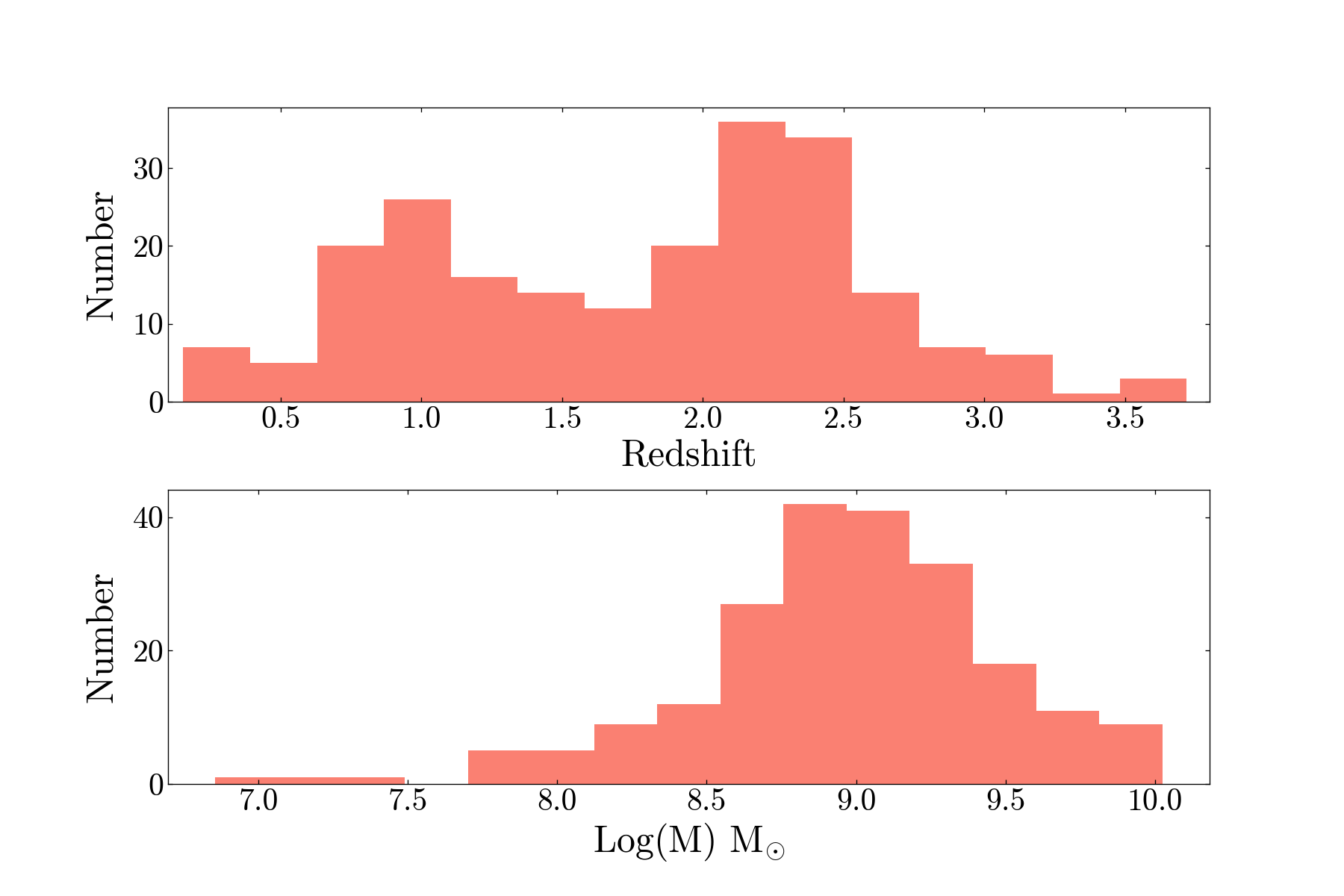}
    \caption{Top : Histogram representing the redshift distribution of possible candidates. Bottom : Histogram representing the Log(M) distribution of possible candidates}
    \label{fig:histo_mass}
\end{figure}

\begin{figure}[!h]
    \includegraphics[width=\hsize]{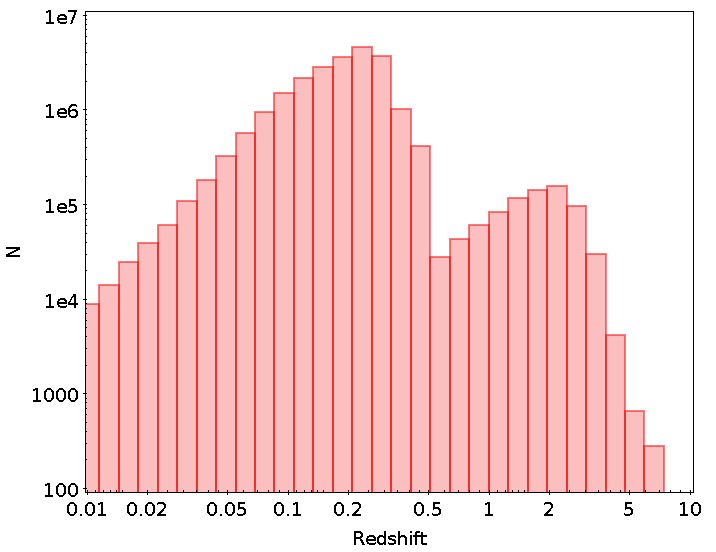}
    \caption{Redshift distribution of galaxies in the Glade+ catalogue.}
    \label{fig:glade_redshift_distrib}
\end{figure}

We show in Fig.~\ref{fig:histo_mass} the Log(M) and redshift distribution of sources in this catalogue. We can see in Fig.~\ref{fig:histo_mass} that the mass distribution of sources in our catalogue is concentrated towards high values with a mean log(M) value of \num{8.95}. This result is expected for a MBBH population as discussed in Section~\ref{discussion}. In the redshift distribution, we can see two peaks, one around a redshift 1 and the other around a redshift 2.3. We can see that the number of candidates increases as the redshift increases between $0.5$ and $1.0$ as predicted by \citep{MBBHs_low_redshift_expectations}. As we probe a larger volume, the chances of detecting more candidates rises. Also, as the number of candidates increases with redshift, it naturally shifts the mass distribution towards high values, as lower mass black holes cannot be detected due to the limiting magnitudes of the surveys. The second peak is likely to emerge from the distribution of galaxies in the Glade+ catalogue (see Fig. ~\ref{fig:glade_redshift_distrib}) which is composed from several different surveys and is therefore inhomogeneous in redshift and sky position.

\subsection{Caveats}
Whilst we have identified some sources that seem to have fairly reliable periods, we can not be sure that these are indeed MBBHs. Indeed, the periods identified may not be due to a binary system but caused by other physical processes, or even instabilities occurring in single BH disks. There is still a chance that the modulation may be due to red noise. Indeed, due to sparse data and low signal to noise, the tests performed are unable to discriminate between a DRW model and a DRW+sinusoidal modulation. Therefore, continued monitoring with current and future surveys, such as the Vera C. Rubin Telescope, will help to confirm these periods. Further, our search focused on finding sinusoidal modulation in optical lightcurves, so we may have missed periodic modulation that differs from a strictly sinusoidal curve, such as that which may be observed from eccentric orbits such as OJ 287 \citep{OJ287}. 

\section{Conclusions}\label{conlusion}
We have identified \num{36} MBBHs candidates through the detection of between 3 and 5 cycles of sinusoidal modulation in their optical lightcurves. These periods are corroborated through GLS analysis. The sparse and noisy data do not allow for the clear statistical preference of the periodic model, regardless of the method used. The amplitude of the variability seems to be stronger in g band compared to r band, as expected for MBBHs. Furthermore, the sources having X-ray observations show typical AGN-like spectra, which is also expected for MBBHs as mergers can trigger AGN activity. We emphasize that the sine model is a simplified model used to identify MBBHs candidates but further investigation with more complex models to account for quasar/AGN variability might be needed to rigorously explain the observed variability. In the binary scenario, the observed periodicity can be explained by different physical origins, such as periodic accretion flow, Doppler boosting modulation of the mini-disks or a blob/lump orbiting in the circumbinary disk. Multi-wavelength follow-up to look for UV/X-ray variability would be a good way to test the binarity of our candidates. Extrapolating from our sample, we determined the number of sub-parsec MBBH candidates at z$<$1, which is consistent with simulations. Also, we refute \num{58} objects previously proposed to be MBBHs  \citep{graham} by using additional ZTF data and demonstrating that the sinusoidal modulation was not sustained over many periods. Finally, we created a catalogue of possible MBBH candidates that show some evidence for periodic modulation, but further observations are required to confirm this. MBBHs at z$<$1 and with orbital periods between a week and a few years are expected to produce gravitational waves in the nano-Hz band accessible to the Pulsar Timing Array (PTA) experiments. We are currently undertaking a targeted search for our z$<$1 candidates in the PTA data. Also, future surveys, such as the upcoming Rubin observatory Legacy Survey of Space and Time \citep[LSST, ][]{LSST}, will be useful to help confirm these candidates. Our method does not account for modulation that deviates from a sinusoidal behaviour such as what has been observed for OJ 287 \citep{OJ287} and our sample is likely to be only a fraction of the total population of MBBHs. 

\begin{acknowledgements}
The authors are grateful to the anonymous referee for careful reading of the paper and the excellent suggestions that helped improve this work. The authors acknowledge funding from the French National Research Agency (grant ANR-21-CE31-0026, project MBH\_waves). NW and VF also acknowledge support from the CNES.

Based on observations obtained with the Samuel Oschin Telescope 48-inch and the 60-inch Telescope at the Palomar
Observatory as part of the Zwicky Transient Facility project. ZTF is supported by the National Science Foundation under Grants
No. AST-1440341 and AST-2034437 and a collaboration including current partners Caltech, IPAC, the Weizmann Institute for
Science, the Oskar Klein Center at Stockholm University, the University of Maryland, Deutsches Elektronen-Synchrotron and
Humboldt University, the TANGO Consortium of Taiwan, the University of Wisconsin at Milwaukee, Trinity College Dublin,
Lawrence Livermore National Laboratories, IN2P3, University of Warwick, Ruhr University Bochum, Northwestern University and
former partners the University of Washington, Los Alamos National Laboratories, and Lawrence Berkeley National Laboratories.
Operations are conducted by COO, IPAC, and UW
Funding for SDSS-III has been provided by the Alfred P. Sloan Foundation, the Participating Institutions, the National Science Foundation, and the U.S. Department of Energy Office of Science. The SDSS-III web site is http://www.sdss3.org/.

SDSS-III is managed by the Astrophysical Research Consortium for the Participating Institutions of the SDSS-III Collaboration including the University of Arizona, the Brazilian Participation Group, Brookhaven National Laboratory, Carnegie Mellon University, University of Florida, the French Participation Group, the German Participation Group, Harvard University, the Instituto de Astrofisica de Canarias, the Michigan State/Notre Dame/JINA Participation Group, Johns Hopkins University, Lawrence Berkeley National Laboratory, Max Planck Institute for Astrophysics, Max Planck Institute for Extraterrestrial Physics, New Mexico State University, New York University, Ohio State University, Pennsylvania State University, University of Portsmouth, Princeton University, the Spanish Participation Group, University of Tokyo, University of Utah, Vanderbilt University, University of Virginia, University of Washington, and Yale University.
RMR acknowledges funding from Centre National d'Etudes Spatiales (CNES) through a postdoctoral fellowship (2021- 2023). RMR has received funding from the European Research Concil (ERC) under the European Union Horizon 2020 research and innovation programme (grant agreement number No. 101002352, PI: M. Linares).
\end{acknowledgements}

\bibliography{ref.bib}

\begin{thebibliography}{89}
\expandafter\ifx\csname natexlab\endcsname\relax\def\natexlab#1{#1}\fi

\bibitem[{{Agazie} {et~al.}(2023){Agazie}, {Anumarlapudi}, {Archibald}, {Baker}, {B{\'e}csy}, {Blecha}, {Bonilla}, {Brazier}, {Brook}, {Burke-Spolaor}, {Burnette}, {Case}, {Casey-Clyde}, {Charisi}, {Chatterjee}, {Chatziioannou}, {Cheeseboro}, {Chen}, {Cohen}, {Cordes}, {Cornish}, {Crawford}, {Cromartie}, {Crowter}, {Cutler}, {D'Orazio}, {Decesar}, {Degan}, {Demorest}, {Deng}, {Dolch}, {Drachler}, {Ferrara}, {Fiore}, {Fonseca}, {Freedman}, {Gardiner}, {Garver-Daniels}, {Gentile}, {Gersbach}, {Glaser}, {Good}, {G{\"u}ltekin}, {Hazboun}, {Hourihane}, {Islo}, {Jennings}, {Johnson}, {Jones}, {Kaiser}, {Kaplan}, {Kelley}, {Kerr}, {Key}, {Laal}, {Lam}, {Lamb}, {Lazio}, {Lewandowska}, {Littenberg}, {Liu}, {Luo}, {Lynch}, {Ma}, {Madison}, {McEwen}, {McKee}, {McLaughlin}, {McMann}, {Meyers}, {Meyers}, {Mingarelli}, {Mitridate}, {Natarajan}, {Ng}, {Nice}, {Ocker}, {Olum}, {Pennucci}, {Perera}, {Petrov}, {Pol}, {Radovan}, {Ransom}, {Ray}, {Romano}, {Runnoe}, {Sardesai}, {Schmiedekamp}, {Schmiedekamp}, {Schmitz},
  {Schult}, {Shapiro-Albert}, {Siemens}, {Simon}, {Siwek}, {Stairs}, {Stinebring}, {Stovall}, {Sun}, {Susobhanan}, {Swiggum}, {Taylor}, {Taylor}, {Turner}, {Unal}, {Vallisneri}, {Vigeland}, {Wachter}, {Wahl}, {Wang}, {Witt}, {Wright}, {Young}, \& {Nanograv Collaboration}}]{nanograv}
{Agazie}, G., {Anumarlapudi}, A., {Archibald}, A.~M., {et~al.} 2023, \apjl, 952, L37

\bibitem[{{Amaro-Seoane} {et~al.}(2023){Amaro-Seoane}, {Andrews}, {Arca Sedda}, {Askar}, {Baghi}, {Balasov}, {Bartos}, {Bavera}, {Bellovary}, {Berry}, {Berti}, {Bianchi}, {Blecha}, {Blondin}, {Bogdanovi{\'c}}, {Boissier}, {Bonetti}, {Bonoli}, {Bortolas}, {Breivik}, {Capelo}, {Caramete}, {Cattorini}, {Charisi}, {Chaty}, {Chen}, {Chru{\'s}li{\'n}ska}, {Chua}, {Church}, {Colpi}, {D'Orazio}, {Danielski}, {Davies}, {Dayal}, {De Rosa}, {Derdzinski}, {Destounis}, {Dotti}, {Dutan}, {Dvorkin}, {Fabj}, {Foglizzo}, {Ford}, {Fouvry}, {Franchini}, {Fragos}, {Fryer}, {Gaspari}, {Gerosa}, {Graziani}, {Groot}, {Habouzit}, {Haggard}, {Haiman}, {Han}, {Istrate}, {Johansson}, {Khan}, {Kimpson}, {Kokkotas}, {Kong}, {Korol}, {Kremer}, {Kupfer}, {Lamberts}, {Larson}, {Lau}, {Liu}, {Lloyd-Ronning}, {Lodato}, {Lupi}, {Ma}, {Maccarone}, {Mandel}, {Mangiagli}, {Mapelli}, {Mathis}, {Mayer}, {McGee}, {McKernan}, {Miller}, {Mota}, {Mumpower}, {Nasim}, {Nelemans}, {Noble}, {Pacucci}, {Panessa}, {Paschalidis}, {Pfister}, {Porquet},
  {Quenby}, {Ricarte}, {R{\"o}pke}, {Regan}, {Rosswog}, {Ruiter}, {Ruiz}, {Runnoe}, {Schneider}, {Schnittman}, {Secunda}, {Sesana}, {Seto}, {Shao}, {Shapiro}, {Sopuerta}, {Stone}, {Suvorov}, {Tamanini}, {Tamfal}, {Tauris}, {Temmink}, {Tomsick}, {Toonen}, {Torres-Orjuela}, {Toscani}, {Tsokaros}, {Unal}, {V{\'a}zquez-Aceves}, {Valiante}, {van Putten}, {van Roestel}, {Vignali}, {Volonteri}, {Wu}, {Younsi}, {Yu}, {Zane}, {Zwick}, {Antonini}, {Baibhav}, {Barausse}, {Bonilla Rivera}, {Branchesi}, {Branduardi-Raymont}, {Burdge}, {Chakraborty}, {Cuadra}, {Dage}, {Davis}, {de Mink}, {Decarli}, {Doneva}, {Escoffier}, {Gandhi}, {Haardt}, {Lousto}, {Nissanke}, {Nordhaus}, {O'Shaughnessy}, {Portegies Zwart}, {Pound}, {Schussler}, {Sergijenko}, {Spallicci}, {Vernieri}, \& {Vigna-G{\'o}mez}}]{2023LRR....26....2A}
{Amaro-Seoane}, P., {Andrews}, J., {Arca Sedda}, M., {et~al.} 2023, Living Reviews in Relativity, 26, 2

\bibitem[{{Amaro-Seoane} {et~al.}(2013){Amaro-Seoane}, {Aoudia}, {Babak}, {Bin{\'e}truy}, {Berti}, {Boh{\'e}}, {Caprini}, {Colpi}, {Cornish}, {Danzmann}, {Dufaux}, {Gair}, {Hinder}, {Jennrich}, {Jetzer}, {Klein}, {Lang}, {Lobo}, {Littenberg}, {McWilliams}, {Nelemans}, {Petiteau}, {Porter}, {Schutz}, {Sesana}, {Stebbins}, {Sumner}, {Vallisneri}, {Vitale}, {Volonteri}, {Ward}, \& {Wardell}}]{Lisa_freq}
{Amaro-Seoane}, P., {Aoudia}, S., {Babak}, S., {et~al.} 2013, GW Notes, 6, 4

\bibitem[{{Amaro-Seoane} {et~al.}(2017){Amaro-Seoane}, {Audley}, {Babak}, {Baker}, {Barausse}, {Bender}, {Berti}, {Binetruy}, {Born}, {Bortoluzzi}, {Camp}, {Caprini}, {Cardoso}, {Colpi}, {Conklin}, {Cornish}, {Cutler}, {Danzmann}, {Dolesi}, {Ferraioli}, {Ferroni}, {Fitzsimons}, {Gair}, {Gesa Bote}, {Giardini}, {Gibert}, {Grimani}, {Halloin}, {Heinzel}, {Hertog}, {Hewitson}, {Holley-Bockelmann}, {Hollington}, {Hueller}, {Inchauspe}, {Jetzer}, {Karnesis}, {Killow}, {Klein}, {Klipstein}, {Korsakova}, {Larson}, {Livas}, {Lloro}, {Man}, {Mance}, {Martino}, {Mateos}, {McKenzie}, {McWilliams}, {Miller}, {Mueller}, {Nardini}, {Nelemans}, {Nofrarias}, {Petiteau}, {Pivato}, {Plagnol}, {Porter}, {Reiche}, {Robertson}, {Robertson}, {Rossi}, {Russano}, {Schutz}, {Sesana}, {Shoemaker}, {Slutsky}, {Sopuerta}, {Sumner}, {Tamanini}, {Thorpe}, {Troebs}, {Vallisneri}, {Vecchio}, {Vetrugno}, {Vitale}, {Volonteri}, {Wanner}, {Ward}, {Wass}, {Weber}, {Ziemer}, \& {Zweifel}}]{LISA}
{Amaro-Seoane}, P., {Audley}, H., {Babak}, S., {et~al.} 2017, arXiv e-prints, arXiv:1702.00786

\bibitem[{{Armitage} \& {Natarajan}(2002)}]{2002ApJ...567L...9A}
{Armitage}, P.~J. \& {Natarajan}, P. 2002, \apjl, 567, L9

\bibitem[{{Artymowicz} \& {Lubow}(1996{\natexlab{a}})}]{variability_accretion}
{Artymowicz}, P. \& {Lubow}, S.~H. 1996{\natexlab{a}}, \apjl, 467, L77

\bibitem[{{Artymowicz} \& {Lubow}(1996{\natexlab{b}})}]{accretion_flow_varability}
{Artymowicz}, P. \& {Lubow}, S.~H. 1996{\natexlab{b}}, \apjl, 467, L77

\bibitem[{{Barret} {et~al.}(2023){Barret}, {Albouys}, {Herder}, {Piro}, {Cappi}, {Huovelin}, {Kelley}, {Mas-Hesse}, {Paltani}, {Rauw}, {Rozanska}, {Svoboda}, {Wilms}, {Yamasaki}, {Audard}, {Bandler}, {Barbera}, {Barcons}, {Bozzo}, {Ceballos}, {Charles}, {Costantini}, {Dauser}, {Decourchelle}, {Duband}, {Duval}, {Fiore}, {Gatti}, {Goldwurm}, {Hartog}, {Jackson}, {Jonker}, {Kilbourne}, {Korpela}, {Macculi}, {Mendez}, {Mitsuda}, {Molendi}, {Pajot}, {Pointecouteau}, {Porter}, {Pratt}, {Pr{\^e}le}, {Ravera}, {Sato}, {Schaye}, {Shinozaki}, {Skup}, {Soucek}, {Thibert}, {Vink}, {Webb}, {Chaoul}, {Raulin}, {Simionescu}, {Torrejon}, {Acero}, {Branduardi-Raymont}, {Ettori}, {Finoguenov}, {Grosso}, {Kaastra}, {Mazzotta}, {Miller}, {Miniutti}, {Nicastro}, {Sciortino}, {Yamaguchi}, {Beaumont}, {Cucchetti}, {D'Andrea}, {Eckart}, {Ferrando}, {Kammoun}, {Lotti}, {Mesnager}, {Natalucci}, {Peille}, {de Plaa}, {Ardellier}, {Argan}, {Bellouard}, {Carron}, {Cavazzuti}, {Fiorini}, {Khosropanah}, {Martin}, {Perry}, {Pinsard},
  {Pradines}, {Rigano}, {Roelfsema}, {Schwander}, {Torrioli}, {Ullom}, {Vera}, {Villegas}, {Zuchniak}, {Brachet}, {Cicero}, {Doriese}, {Durkin}, {Fioretti}, {Geoffray}, {Jacques}, {Kirsch}, {Smith}, {Adams}, {Gloaguen}, {Hoogeveen}, {van der Hulst}, {Kiviranta}, {van der Kuur}, {Ledot}, {van Leeuwen}, {van Loon}, {Lyautey}, {Parot}, {Sakai}, {van Weers}, {Abdoelkariem}, {Adam}, {Adami}, {Aicardi}, {Akamatsu}, {Alonso}, {Amato}, {Andr{\'e}}, {Angelinelli}, {Anon-Cancela}, {Anvar}, {Atienza}, {Attard}, {Auricchio}, {Balado}, {Bancel}, {Barusso}, {Bascu{\~n}an}, {Bernard}, {Berrocal}, {Blin}, {Bonino}, {Bonnet}, {Bonny}, {Boorman}, {Boreux}, {Bounab}, {Boutelier}, {Boyce}, {Brienza}, {Bruijn}, {Bulgarelli}, {Calarco}, {Callanan}, {Campello}, {Camus}, {Canourgues}, {Capobianco}, {Cardiel}, {Castellani}, {Cheatom}, {Chervenak}, {Chiarello}, {Clerc}, {Clerc}, {Cobo}, {Coeur-Joly}, {Coleiro}, {Colonges}, {Corcione}, {Coriat}, {Coynel}, {Cuttaia}, {D'Ai}, {D'anca}, {Dadina}, {Daniel}, {Dauner}, {DeNigris},
  {Dercksen}, {DiPirro}, {Doumayrou}, {Dubbeldam}, {Dupieux}, {Dupourqu{\'e}}, {Durand}, {Eckert}, {Eiriz}, {Ercolani}, {Etcheverry}, {Finkbeiner}, {Fiocchi}, {Fossecave}, {Franssen}, {Frericks}, {Gabici}, {Gant}, {Gao}, {Gastaldello}, \& {Genolet}}]{X_ifu_new_athena}
{Barret}, D., {Albouys}, V., {Herder}, J.-W.~d., {et~al.} 2023, Experimental Astronomy, 55, 373

\bibitem[{{Bate} \& {Bonnell}(1997)}]{single_accretion_disk}
{Bate}, M.~R. \& {Bonnell}, I.~A. 1997, \mnras, 285, 33

\bibitem[{{Begelman} {et~al.}(1980){Begelman}, {Blandford}, \& {Rees}}]{MBH_path_to_coalescence_2}
{Begelman}, M.~C., {Blandford}, R.~D., \& {Rees}, M.~J. 1980, \nat, 287, 307

\bibitem[{{Bellm} {et~al.}(2019){Bellm}, {Kulkarni}, {Graham}, {Dekany}, {Smith}, {Riddle}, {Masci}, {Helou}, {Prince}, {Adams}, {Barbarino}, {Barlow}, {Bauer}, {Beck}, {Belicki}, {Biswas}, {Blagorodnova}, {Bodewits}, {Bolin}, {Brinnel}, {Brooke}, {Bue}, {Bulla}, {Burruss}, {Cenko}, {Chang}, {Connolly}, {Coughlin}, {Cromer}, {Cunningham}, {De}, {Delacroix}, {Desai}, {Duev}, {Eadie}, {Farnham}, {Feeney}, {Feindt}, {Flynn}, {Franckowiak}, {Frederick}, {Fremling}, {Gal-Yam}, {Gezari}, {Giomi}, {Goldstein}, {Golkhou}, {Goobar}, {Groom}, {Hacopians}, {Hale}, {Henning}, {Ho}, {Hover}, {Howell}, {Hung}, {Huppenkothen}, {Imel}, {Ip}, {Ivezi{\'c}}, {Jackson}, {Jones}, {Juric}, {Kasliwal}, {Kaspi}, {Kaye}, {Kelley}, {Kowalski}, {Kramer}, {Kupfer}, {Landry}, {Laher}, {Lee}, {Lin}, {Lin}, {Lunnan}, {Giomi}, {Mahabal}, {Mao}, {Miller}, {Monkewitz}, {Murphy}, {Ngeow}, {Nordin}, {Nugent}, {Ofek}, {Patterson}, {Penprase}, {Porter}, {Rauch}, {Rebbapragada}, {Reiley}, {Rigault}, {Rodriguez}, {van Roestel}, {Rusholme}, {van
  Santen}, {Schulze}, {Shupe}, {Singer}, {Soumagnac}, {Stein}, {Surace}, {Sollerman}, {Szkody}, {Taddia}, {Terek}, {Van Sistine}, {van Velzen}, {Vestrand}, {Walters}, {Ward}, {Ye}, {Yu}, {Yan}, \& {Zolkower}}]{ZTF_performance}
{Bellm}, E.~C., {Kulkarni}, S.~R., {Graham}, M.~J., {et~al.} 2019, \pasp, 131, 018002

\bibitem[{{Bianchi} {et~al.}(2008){Bianchi}, {Chiaberge}, {Piconcelli}, {Guainazzi}, \& {Matt}}]{Mrk463}
{Bianchi}, S., {Chiaberge}, M., {Piconcelli}, E., {Guainazzi}, M., \& {Matt}, G. 2008, \mnras, 386, 105

\bibitem[{{Bowen} {et~al.}(2019){Bowen}, {Mewes}, {Noble}, {Avara}, {Campanelli}, \& {Krolik}}]{QPO}
{Bowen}, D.~B., {Mewes}, V., {Noble}, S.~C., {et~al.} 2019, \apj, 879, 76

\bibitem[{{Buchner}(2016)}]{Ultranest_ML}
{Buchner}, J. 2016, Statistics and Computing, 26, 383

\bibitem[{{Buchner}(2019)}]{Ultranest_Nested_sampling}
{Buchner}, J. 2019, \pasp, 131, 108005

\bibitem[{{Buchner}(2021{\natexlab{a}})}]{Ultranest_methods}
{Buchner}, J. 2021{\natexlab{a}}, Statistics Surveys, arXiv:2101.09675

\bibitem[{{Buchner}(2021{\natexlab{b}})}]{Ultranest}
{Buchner}, J. 2021{\natexlab{b}}, The Journal of Open Source Software, 6, 3001

\bibitem[{{Burke} {et~al.}(2021){Burke}, {Shen}, {Blaes}, {Gammie}, {Horne}, {Jiang}, {Liu}, {McHardy}, {Morgan}, {Scaringi}, \& {Yang}}]{tau_qso_2}
{Burke}, C.~J., {Shen}, Y., {Blaes}, O., {et~al.} 2021, Science, 373, 789

\bibitem[{{Chen} {et~al.}(2020){Chen}, {Wang}, {Deng}, {de Grijs}, {Yang}, \& {Tian}}]{ZTF_variable_catalog}
{Chen}, X., {Wang}, S., {Deng}, L., {et~al.} 2020, \apjs, 249, 18

\bibitem[{{Chen} {et~al.}(2024){Chen}, {Zhai}, {Liu}, {Guo}, {Peng}, {Li}, {Songsheng}, {Du}, {Hu}, \& {Wang}}]{ztf_crts_search}
{Chen}, Y.-J., {Zhai}, S., {Liu}, J.-R., {et~al.} 2024, \mnras, 527, 12154

\bibitem[{{Cocchiararo} {et~al.}(2024){Cocchiararo}, {Franchini}, {Lupi}, \& {Sesana}}]{Cocchiararo2024}
{Cocchiararo}, F., {Franchini}, A., {Lupi}, A., \& {Sesana}, A. 2024, \aap, 691, A250

\bibitem[{{Colpi}(2014)}]{MBH_path_to_coalescence}
{Colpi}, M. 2014, \ssr, 183, 189

\bibitem[{{Combi} {et~al.}(2022){Combi}, {Lopez Armengol}, {Campanelli}, {Noble}, {Avara}, {Krolik}, \& {Bowen}}]{SMBBH3}
{Combi}, L., {Lopez Armengol}, F.~G., {Campanelli}, M., {et~al.} 2022, \apj, 928, 187

\bibitem[{{Comerford} {et~al.}(2024){Comerford}, {Nevin}, {Negus}, {Barrows}, {Eracleous}, {M{\"u}ller-S{\'a}nchez}, {Roy}, {Stemo}, {Storchi-Bergmann}, \& {Wylezalek}}]{agn_fraction_mergers}
{Comerford}, J.~M., {Nevin}, R., {Negus}, J., {et~al.} 2024, \apj, 963, 53

\bibitem[{{D{\'a}lya} {et~al.}(2022){D{\'a}lya}, {D{\'\i}az}, {Bouchet}, {Frei}, {Jasche}, {Lavaux}, {Macas}, {Mukherjee}, {P{\'a}lfi}, {de Souza}, {Wandelt}, {Bilicki}, \& {Raffai}}]{glade+}
{D{\'a}lya}, G., {D{\'\i}az}, R., {Bouchet}, F.~R., {et~al.} 2022, \mnras, 514, 1403

\bibitem[{{d'Ascoli} {et~al.}(2018){d'Ascoli}, {Noble}, {Bowen}, {Campanelli}, {Krolik}, \& {Mewes}}]{simulation_MBH}
{d'Ascoli}, S., {Noble}, S.~C., {Bowen}, D.~B., {et~al.} 2018, \apj, 865, 140

\bibitem[{{De Rosa} {et~al.}(2019){De Rosa}, {Vignali}, {Bogdanovi{\'c}}, {Capelo}, {Charisi}, {Dotti}, {Husemann}, {Lusso}, {Mayer}, {Paragi}, {Runnoe}, {Sesana}, {Steinborn}, {Bianchi}, {Colpi}, {del Valle}, {Frey}, {Gab{\'a}nyi}, {Giustini}, {Guainazzi}, {Haiman}, {Herrera Ruiz}, {Herrero-Illana}, {Iwasawa}, {Komossa}, {Lena}, {Loiseau}, {Perez-Torres}, {Piconcelli}, \& {Volonteri}}]{Multi_messenger_SMBBH}
{De Rosa}, A., {Vignali}, C., {Bogdanovi{\'c}}, T., {et~al.} 2019, \nar, 86, 101525

\bibitem[{{Desvignes} {et~al.}(2016){Desvignes}, {Caballero}, {Lentati}, {Verbiest}, {Champion}, {Stappers}, {Janssen}, {Lazarus}, {Os{\l}owski}, {Babak}, {Bassa}, {Brem}, {Burgay}, {Cognard}, {Gair}, {Graikou}, {Guillemot}, {Hessels}, {Jessner}, {Jordan}, {Karuppusamy}, {Kramer}, {Lassus}, {Lazaridis}, {Lee}, {Liu}, {Lyne}, {McKee}, {Mingarelli}, {Perrodin}, {Petiteau}, {Possenti}, {Purver}, {Rosado}, {Sanidas}, {Sesana}, {Shaifullah}, {Smits}, {Taylor}, {Theureau}, {Tiburzi}, {van Haasteren}, \& {Vecchio}}]{EPTA}
{Desvignes}, G., {Caballero}, R.~N., {Lentati}, L., {et~al.} 2016, \mnras, 458, 3341

\bibitem[{{D'Orazio} {et~al.}(2015){D'Orazio}, {Haiman}, \& {Schiminovich}}]{doppler_boosting}
{D'Orazio}, D.~J., {Haiman}, Z., \& {Schiminovich}, D. 2015, \nat, 525, 351

\bibitem[{{Drake} {et~al.}(2009){Drake}, {Djorgovski}, {Mahabal}, {Beshore}, {Larson}, {Graham}, {Williams}, {Christensen}, {Catelan}, {Boattini}, {Gibbs}, {Hill}, \& {Kowalski}}]{CRTS}
{Drake}, A.~J., {Djorgovski}, S.~G., {Mahabal}, A., {et~al.} 2009, \apj, 696, 870

\bibitem[{{Duras} {et~al.}(2020){Duras}, {Bongiorno}, {Ricci}, {Piconcelli}, {Shankar}, {Lusso}, {Bianchi}, {Fiore}, {Maiolino}, {Marconi}, {Onori}, {Sani}, {Schneider}, {Vignali}, \& {La Franca}}]{bolomectric_correction_x_ray}
{Duras}, F., {Bongiorno}, A., {Ricci}, F., {et~al.} 2020, \aap, 636, A73

\bibitem[{D’Ascoli {et~al.}(2018)D’Ascoli, Noble, Bowen, Campanelli, Krolik, \& Mewes}]{bbh_emission_dascoli18}
D’Ascoli, S., Noble, S.~C., Bowen, D.~B., {et~al.} 2018, \apj, 865, 140

\bibitem[{{Ellison} {et~al.}(2013){Ellison}, {Mendel}, {Patton}, \& {Scudder}}]{post_merger_galaxies_properties}
{Ellison}, S.~L., {Mendel}, J.~T., {Patton}, D.~R., \& {Scudder}, J.~M. 2013, \mnras, 435, 3627

\bibitem[{{EPTA Collaboration} {et~al.}(2024){EPTA Collaboration}, {InPTA Collaboration}, {Antoniadis}, {Arumugam}, {Arumugam}, {Babak}, {Bagchi}, {Bak Nielsen}, {Bassa}, {Bathula}, {Berthereau}, {Bonetti}, {Bortolas}, {Brook}, {Burgay}, {Caballero}, {Chalumeau}, {Champion}, {Chanlaridis}, {Chen}, {Cognard}, {Dandapat}, {Deb}, {Desai}, {Desvignes}, {Dhanda-Batra}, {Dwivedi}, {Falxa}, {Ferdman}, {Franchini}, {Gair}, {Goncharov}, {Gopakumar}, {Graikou}, {Grie{\ss}meier}, {Gualandris}, {Guillemot}, {Guo}, {Gupta}, {Hisano}, {Hu}, {Iraci}, {Izquierdo-Villalba}, {Jang}, {Jawor}, {Janssen}, {Jessner}, {Joshi}, {Kareem}, {Karuppusamy}, {Keane}, {Keith}, {Kharbanda}, {Kikunaga}, {Kolhe}, {Kramer}, {Krishnakumar}, {Lackeos}, {Lee}, {Liu}, {Liu}, {Lyne}, {McKee}, {Maan}, {Main}, {Mickaliger}, {Ni{\c{t}}u}, {Nobleson}, {Paladi}, {Parthasarathy}, {Perera}, {Perrodin}, {Petiteau}, {Porayko}, {Possenti}, {Prabu}, {Quelquejay Leclere}, {Rana}, {Samajdar}, {Sanidas}, {Sesana}, {Shaifullah}, {Singha}, {Speri}, {Spiewak},
  {Srivastava}, {Stappers}, {Surnis}, {Susarla}, {Susobhanan}, {Takahashi}, {Tarafdar}, {Theureau}, {Tiburzi}, {van der Wateren}, {Vecchio}, {Venkatraman Krishnan}, {Verbiest}, {Wang}, {Wang}, {Wu}, {Auclair}, {Barausse}, {Caprini}, {Crisostomi}, {Fastidio}, {Khizriev}, {Middleton}, {Neronov}, {Postnov}, {Roper Pol}, {Semikoz}, {Smarra}, {Steer}, {Truant}, \& {Valtolina}}]{EPTA_detection}
{EPTA Collaboration}, {InPTA Collaboration}, {Antoniadis}, J., {et~al.} 2024, \aap, 685, A94

\bibitem[{Farris {et~al.}(2015{\natexlab{a}})Farris, Duffell, MacFadyen, \& Haiman}]{SimuBBH_inspiral_farris15}
Farris, B.~D., Duffell, P., MacFadyen, A.~I., \& Haiman, Z. 2015{\natexlab{a}}, \mnras, 447, L80

\bibitem[{Farris {et~al.}(2015{\natexlab{b}})Farris, Duffell, MacFadyen, \& Haiman}]{BBH_emission_lump_farris15}
Farris, B.~D., Duffell, P., MacFadyen, A.~I., \& Haiman, Z. 2015{\natexlab{b}}, \mnras, 446, L36

\bibitem[{{Foreman-Mackey} {et~al.}(2017){Foreman-Mackey}, {Agol}, {Ambikasaran}, \& {Angus}}]{celerite}
{Foreman-Mackey}, D., {Agol}, E., {Ambikasaran}, S., \& {Angus}, R. 2017, \apj, 154, 220

\bibitem[{{Foster} \& {Backer}(1990)}]{PTA_1990}
{Foster}, R.~S. \& {Backer}, D.~C. 1990, \apj, 361, 300

\bibitem[{{Fu} {et~al.}(2011){Fu}, {Zhang}, {Assef}, {Stockton}, {Myers}, {Yan}, {Djorgovski}, {Wrobel}, \& {Riechers}}]{dual_AGN}
{Fu}, H., {Zhang}, Z.-Y., {Assef}, R.~J., {et~al.} 2011, \apjl, 740, L44

\bibitem[{{Gao} {et~al.}(2020){Gao}, {Wang}, {Pearson}, {Gordon}, {Holwerda}, {Hopkins}, {Brown}, {Bland-Hawthorn}, \& {Owers}}]{AGN_mergers}
{Gao}, F., {Wang}, L., {Pearson}, W.~J., {et~al.} 2020, \aap, 637, A94

\bibitem[{{Gorbachev} {et~al.}(2024){Gorbachev}, {Butuzova}, {Nazarov}, \& {Zhovtan}}]{OJ_287_doubt}
{Gorbachev}, M.~A., {Butuzova}, M.~S., {Nazarov}, S.~V., \& {Zhovtan}, A.~V. 2024, Astroparticle Physics, 160, 102965

\bibitem[{{Goulding} {et~al.}(2018){Goulding}, {Greene}, {Bezanson}, {Greco}, {Johnson}, {Leauthaud}, {Matsuoka}, {Medezinski}, \& {Price-Whelan}}]{agn_mergers_3}
{Goulding}, A.~D., {Greene}, J.~E., {Bezanson}, R., {et~al.} 2018, \pasj, 70, S37

\bibitem[{{Graham} {et~al.}(2015){Graham}, {Djorgovski}, {Stern}, {Drake}, {Mahabal}, {Donalek}, {Glikman}, {Larson}, \& {Christensen}}]{graham}
{Graham}, M.~J., {Djorgovski}, S.~G., {Stern}, D., {et~al.} 2015, \mnras, 453, 1562

\bibitem[{{Haiman} {et~al.}(2009){Haiman}, {Kocsis}, \& {Menou}}]{residence_time_SMBH}
{Haiman}, Z., {Kocsis}, B., \& {Menou}, K. 2009, \apj, 700, 1952

\bibitem[{{Hopkins} {et~al.}(2008){Hopkins}, {Hernquist}, {Cox}, \& {Kere{\v{s}}}}]{merger_agn_trigger}
{Hopkins}, P.~F., {Hernquist}, L., {Cox}, T.~J., \& {Kere{\v{s}}}, D. 2008, \apjs, 175, 356

\bibitem[{{Husemann} {et~al.}(2018){Husemann}, {Worseck}, {Arrigoni Battaia}, \& {Shanks}}]{LBQS}
{Husemann}, B., {Worseck}, G., {Arrigoni Battaia}, F., \& {Shanks}, T. 2018, \aap, 610, L7

\bibitem[{{Ivezi{\'c}} {et~al.}(2019){Ivezi{\'c}}, {Kahn}, {Tyson}, {Abel}, {Acosta}, {Allsman}, {Alonso}, {AlSayyad}, {Anderson}, {Andrew}, {Angel}, {Angeli}, {Ansari}, {Antilogus}, {Araujo}, {Armstrong}, {Arndt}, {Astier}, {Aubourg}, {Auza}, {Axelrod}, {Bard}, {Barr}, {Barrau}, {Bartlett}, {Bauer}, {Bauman}, {Baumont}, {Bechtol}, {Bechtol}, {Becker}, {Becla}, {Beldica}, {Bellavia}, {Bianco}, {Biswas}, {Blanc}, {Blazek}, {Blandford}, {Bloom}, {Bogart}, {Bond}, {Booth}, {Borgland}, {Borne}, {Bosch}, {Boutigny}, {Brackett}, {Bradshaw}, {Brandt}, {Brown}, {Bullock}, {Burchat}, {Burke}, {Cagnoli}, {Calabrese}, {Callahan}, {Callen}, {Carlin}, {Carlson}, {Chandrasekharan}, {Charles-Emerson}, {Chesley}, {Cheu}, {Chiang}, {Chiang}, {Chirino}, {Chow}, {Ciardi}, {Claver}, {Cohen-Tanugi}, {Cockrum}, {Coles}, {Connolly}, {Cook}, {Cooray}, {Covey}, {Cribbs}, {Cui}, {Cutri}, {Daly}, {Daniel}, {Daruich}, {Daubard}, {Daues}, {Dawson}, {Delgado}, {Dellapenna}, {de Peyster}, {de Val-Borro}, {Digel}, {Doherty}, {Dubois},
  {Dubois-Felsmann}, {Durech}, {Economou}, {Eifler}, {Eracleous}, {Emmons}, {Fausti Neto}, {Ferguson}, {Figueroa}, {Fisher-Levine}, {Focke}, {Foss}, {Frank}, {Freemon}, {Gangler}, {Gawiser}, {Geary}, {Gee}, {Geha}, {Gessner}, {Gibson}, {Gilmore}, {Glanzman}, {Glick}, {Goldina}, {Goldstein}, {Goodenow}, {Graham}, {Gressler}, {Gris}, {Guy}, {Guyonnet}, {Haller}, {Harris}, {Hascall}, {Haupt}, {Hernandez}, {Herrmann}, {Hileman}, {Hoblitt}, {Hodgson}, {Hogan}, {Howard}, {Huang}, {Huffer}, {Ingraham}, {Innes}, {Jacoby}, {Jain}, {Jammes}, {Jee}, {Jenness}, {Jernigan}, {Jevremovi{\'c}}, {Johns}, {Johnson}, {Johnson}, {Jones}, {Juramy-Gilles}, {Juri{\'c}}, {Kalirai}, {Kallivayalil}, {Kalmbach}, {Kantor}, {Karst}, {Kasliwal}, {Kelly}, {Kessler}, {Kinnison}, {Kirkby}, {Knox}, {Kotov}, {Krabbendam}, {Krughoff}, {Kub{\'a}nek}, {Kuczewski}, {Kulkarni}, {Ku}, {Kurita}, {Lage}, {Lambert}, {Lange}, {Langton}, {Le Guillou}, {Levine}, {Liang}, {Lim}, {Lintott}, {Long}, {Lopez}, {Lotz}, {Lupton}, {Lust}, {MacArthur}, {Mahabal},
  {Mandelbaum}, {Markiewicz}, {Marsh}, {Marshall}, {Marshall}, {May}, {McKercher}, {McQueen}, {Meyers}, {Migliore}, {Miller}, {Mills}, {Miraval}, {Moeyens}, {Moolekamp}, {Monet}, {Moniez}, {Monkewitz}, {Montgomery}, {Morrison}, {Mueller}, {Muller}, {Mu{\~n}oz Arancibia}, {Neill}, {Newbry}, {Nief}, {Nomerotski}, {Nordby}, {O'Connor}, {Oliver}, {Olivier}, {Olsen}, {O'Mullane}, {Ortiz}, {Osier}, {Owen}, {Pain}, {Palecek}, {Parejko}, {Parsons}, {Pease}, {Peterson}, {Peterson}, {Petravick}, {Libby Petrick}, {Petry}, {Pierfederici}, {Pietrowicz}, {Pike}, {Pinto}, {Plante}, {Plate}, {Plutchak}, {Price}, {Prouza}, {Radeka}, {Rajagopal}, {Rasmussen}, {Regnault}, {Reil}, {Reiss}, {Reuter}, {Ridgway}, {Riot}, {Ritz}, {Robinson}, {Roby}, {Roodman}, {Rosing}, {Roucelle}, {Rumore}, {Russo}, {Saha}, {Sassolas}, {Schalk}, {Schellart}, {Schindler}, {Schmidt}, {Schneider}, {Schneider}, {Schoening}, {Schumacher}, {Schwamb}, {Sebag}, {Selvy}, {Sembroski}, {Seppala}, {Serio}, {Serrano}, {Shaw}, {Shipsey}, {Sick}, {Silvestri},
  {Slater}, {Smith}, {Smith}, {Sobhani}, {Soldahl}, {Storrie-Lombardi}, {Stover}, {Strauss}, {Street}, {Stubbs}, {Sullivan}, {Sweeney}, {Swinbank}, {Szalay}, {Takacs}, {Tether}, {Thaler}, {Thayer}, {Thomas}, {Thornton}, {Thukral}, {Tice}, {Trilling}, {Turri}, {Van Berg}, {Vanden Berk}, {Vetter}, {Virieux}, {Vucina}, {Wahl}, {Walkowicz}, {Walsh}, {Walter}, {Wang}, {Wang}, {Warner}, {Wiecha}, {Willman}, {Winters}, {Wittman}, {Wolff}, {Wood-Vasey}, {Wu}, {Xin}, {Yoachim}, \& {Zhan}}]{LSST}
{Ivezi{\'c}}, {\v{Z}}., {Kahn}, S.~M., {Tyson}, J.~A., {et~al.} 2019, \apj, 873, 111

\bibitem[{{Jeffreys}(1939)}]{Bayes_factor_scale}
{Jeffreys}, H. 1939, {Theory of Probability}

\bibitem[{{Kelley} {et~al.}(2019){Kelley}, {Haiman}, {Sesana}, \& {Hernquist}}]{2019MNRAS.485.1579K}
{Kelley}, L.~Z., {Haiman}, Z., {Sesana}, A., \& {Hernquist}, L. 2019, \mnras, 485, 1579

\bibitem[{{Kelly} {et~al.}(2009){Kelly}, {Bechtold}, \& {Siemiginowska}}]{DRW}
{Kelly}, B.~C., {Bechtold}, J., \& {Siemiginowska}, A. 2009, \apj, 698, 895

\bibitem[{{Komossa} {et~al.}(2003){Komossa}, {Burwitz}, {Hasinger}, {Predehl}, {Kaastra}, \& {Ikebe}}]{NGC6240}
{Komossa}, S., {Burwitz}, V., {Hasinger}, G., {et~al.} 2003, \apjl, 582, L15

\bibitem[{{Koss} {et~al.}(2011){Koss}, {Mushotzky}, {Treister}, {Veilleux}, {Vasudevan}, {Miller}, {Sanders}, {Schawinski}, \& {Trippe}}]{Mrk739}
{Koss}, M., {Mushotzky}, R., {Treister}, E., {et~al.} 2011, \apjl, 735, L42

\bibitem[{{Lackner} {et~al.}(2014){Lackner}, {Silverman}, {Salvato}, {Kampczyk}, {Kartaltepe}, {Sanders}, {Capak}, {Civano}, {Halliday}, {Ilbert}, {Jahnke}, {Koekemoer}, {Lee}, {Le F{\`e}vre}, {Liu}, {Scoville}, {Sheth}, \& {Toft}}]{agn_mergers_2}
{Lackner}, C.~N., {Silverman}, J.~D., {Salvato}, M., {et~al.} 2014, \apj, 148, 137

\bibitem[{{Li} {et~al.}(2014){Li}, {Wang}, {Hu}, {Du}, \& {Bai}}]{PyCALI}
{Li}, Y.-R., {Wang}, J.-M., {Hu}, C., {Du}, P., \& {Bai}, J.-M. 2014, \apjl, 786, L6

\bibitem[{{MacLeod} {et~al.}(2010){MacLeod}, {Ivezic}, {Kozlowski}, {Kochanek}, {de Vries}, {Sesar}, {Becker}, {Brooks}, \& {Gibson}}]{tau_qso_1}
{MacLeod}, C., {Ivezic}, Z., {Kozlowski}, S., {et~al.} 2010, in American Astronomical Society Meeting Abstracts, Vol. 215, American Astronomical Society Meeting Abstracts \#215, 433.26

\bibitem[{{Masci} {et~al.}(2019){Masci}, {Laher}, {Rusholme}, {Shupe}, {Groom}, {Surace}, {Jackson}, {Monkewitz}, {Beck}, {Flynn}, {Terek}, {Landry}, {Hacopians}, {Desai}, {Howell}, {Brooke}, {Imel}, {Wachter}, {Ye}, {Lin}, {Cenko}, {Cunningham}, {Rebbapragada}, {Bue}, {Miller}, {Mahabal}, {Bellm}, {Patterson}, {Juri{\'c}}, {Golkhou}, {Ofek}, {Walters}, {Graham}, {Kasliwal}, {Dekany}, {Kupfer}, {Burdge}, {Cannella}, {Barlow}, {Van Sistine}, {Giomi}, {Fremling}, {Blagorodnova}, {Levitan}, {Riddle}, {Smith}, {Helou}, {Prince}, \& {Kulkarni}}]{ZTF}
{Masci}, F.~J., {Laher}, R.~R., {Rusholme}, B., {et~al.} 2019, \pasp, 131, 018003

\bibitem[{{Merloni} {et~al.}(2024){Merloni}, {Lamer}, {Liu}, {Ramos-Ceja}, {Brunner}, {Bulbul}, {Dennerl}, {Doroshenko}, {Freyberg}, {Friedrich}, {Gatuzz}, {Georgakakis}, {Haberl}, {Igo}, {Kreykenbohm}, {Liu}, {Maitra}, {Malyali}, {Mayer}, {Nandra}, {Predehl}, {Robrade}, {Salvato}, {Sanders}, {Stewart}, {Tub{\'\i}n-Arenas}, {Weber}, {Wilms}, {Arcodia}, {Artis}, {Aschersleben}, {Avakyan}, {Aydar}, {Bahar}, {Balzer}, {Becker}, {Berger}, {Boller}, {Bornemann}, {Br{\"u}ggen}, {Brusa}, {Buchner}, {Burwitz}, {Camilloni}, {Clerc}, {Comparat}, {Coutinho}, {Czesla}, {Dannhauer}, {Dauner}, {Dauser}, {Dietl}, {Dolag}, {Dwelly}, {Egg}, {Ehl}, {Freund}, {Friedrich}, {Gaida}, {Garrel}, {Ghirardini}, {Gokus}, {Gr{\"u}nwald}, {Grandis}, {Grotova}, {Gruen}, {Gueguen}, {H{\"a}mmerich}, {Hamaus}, {Hasinger}, {Haubner}, {Homan}, {Ider Chitham}, {Joseph}, {Joyce}, {K{\"o}nig}, {Kaltenbrunner}, {Khokhriakova}, {Kink}, {Kirsch}, {Kluge}, {Knies}, {Krippendorf}, {Krumpe}, {Kurpas}, {Li}, {Liu}, {Locatelli}, {Lorenz}, {M{\"u}ller},
  {Magaudda}, {Mannes}, {McCall}, {Meidinger}, {Michailidis}, {Migkas}, {Mu{\~n}oz-Giraldo}, {Musiimenta}, {Nguyen-Dang}, {Ni}, {Olechowska}, {Ota}, {Pacaud}, {Pasini}, {Perinati}, {Pires}, {Pommranz}, {Ponti}, {Poppenhaeger}, {P{\"u}hlhofer}, {Rau}, {Reh}, {Reiprich}, {Roster}, {Saeedi}, {Santangelo}, {Sasaki}, {Schmitt}, {Schneider}, {Schrabback}, {Schuster}, {Schwope}, {Seppi}, {Serim}, {Shreeram}, {Sokolova-Lapa}, {Starck}, {Stelzer}, {Stierhof}, {Suleimanov}, {Tenzer}, {Traulsen}, {Tr{\"u}mper}, {Tsuge}, {Urrutia}, {Veronica}, {Waddell}, {Willer}, {Wolf}, {Yeung}, {Zainab}, {Zangrandi}, {Zhang}, {Zhang}, \& {Zheng}}]{erosita_cat}
{Merloni}, A., {Lamer}, G., {Liu}, T., {et~al.} 2024, \aap, 682, A34

\bibitem[{{Mignon-Risse} {et~al.}(2023){Mignon-Risse}, {Varniere}, \& {Casse}}]{lump}
{Mignon-Risse}, R., {Varniere}, P., \& {Casse}, F. 2023, \mnras [\eprint[arXiv]{2301.06566}]

\bibitem[{Mignon-Risse {et~al.}(2024)Mignon-Risse, Varniere, \& Casse}]{lump_modulation_MR24}
Mignon-Risse, R., Varniere, P., \& Casse, F. 2024, submitted to Monthly Notices of the Royal Astronomical Society

\bibitem[{{Nandra} {et~al.}(2013){Nandra}, {Barret}, {Barcons}, {Fabian}, {den Herder}, {Piro}, {Watson}, {Adami}, {Aird}, {Afonso}, {Alexander}, {Argiroffi}, {Amati}, {Arnaud}, {Atteia}, {Audard}, {Badenes}, {Ballet}, {Ballo}, {Bamba}, {Bhardwaj}, {Stefano Battistelli}, {Becker}, {De Becker}, {Behar}, {Bianchi}, {Biffi}, {B{\^\i}rzan}, {Bocchino}, {Bogdanov}, {Boirin}, {Boller}, {Borgani}, {Borm}, {Bouch{\'e}}, {Bourdin}, {Bower}, {Braito}, {Branchini}, {Branduardi-Raymont}, {Bregman}, {Brenneman}, {Brightman}, {Br{\"u}ggen}, {Buchner}, {Bulbul}, {Brusa}, {Bursa}, {Caccianiga}, {Cackett}, {Campana}, {Cappelluti}, {Cappi}, {Carrera}, {Ceballos}, {Christensen}, {Chu}, {Churazov}, {Clerc}, {Corbel}, {Corral}, {Comastri}, {Costantini}, {Croston}, {Dadina}, {D'Ai}, {Decourchelle}, {Della Ceca}, {Dennerl}, {Dolag}, {Done}, {Dovciak}, {Drake}, {Eckert}, {Edge}, {Ettori}, {Ezoe}, {Feigelson}, {Fender}, {Feruglio}, {Finoguenov}, {Fiore}, {Galeazzi}, {Gallagher}, {Gandhi}, {Gaspari}, {Gastaldello}, {Georgakakis},
  {Georgantopoulos}, {Gilfanov}, {Gitti}, {Gladstone}, {Goosmann}, {Gosset}, {Grosso}, {Guedel}, {Guerrero}, {Haberl}, {Hardcastle}, {Heinz}, {Alonso Herrero}, {Herv{\'e}}, {Holmstrom}, {Iwasawa}, {Jonker}, {Kaastra}, {Kara}, {Karas}, {Kastner}, {King}, {Kosenko}, {Koutroumpa}, {Kraft}, {Kreykenbohm}, {Lallement}, {Lanzuisi}, {Lee}, {Lemoine-Goumard}, {Lobban}, {Lodato}, {Lovisari}, {Lotti}, {McCharthy}, {McNamara}, {Maggio}, {Maiolino}, {De Marco}, {de Martino}, {Mateos}, {Matt}, {Maughan}, {Mazzotta}, {Mendez}, {Merloni}, {Micela}, {Miceli}, {Mignani}, {Miller}, {Miniutti}, {Molendi}, {Montez}, {Moretti}, {Motch}, {Naz{\'e}}, {Nevalainen}, {Nicastro}, {Nulsen}, {Ohashi}, {O'Brien}, {Osborne}, {Oskinova}, {Pacaud}, {Paerels}, {Page}, {Papadakis}, {Pareschi}, {Petre}, {Petrucci}, {Piconcelli}, {Pillitteri}, {Pinto}, {de Plaa}, {Pointecouteau}, {Ponman}, {Ponti}, {Porquet}, {Pounds}, {Pratt}, {Predehl}, {Proga}, {Psaltis}, {Rafferty}, {Ramos-Ceja}, {Ranalli}, {Rasia}, {Rau}, {Rauw}, {Rea}, {Read}, {Reeves},
  {Reiprich}, {Renaud}, {Reynolds}, {Risaliti}, {Rodriguez}, {Rodriguez Hidalgo}, {Roncarelli}, {Rosario}, {Rossetti}, {Rozanska}, {Rovilos}, {Salvaterra}, {Salvato}, {Di Salvo}, {Sanders}, {Sanz-Forcada}, {Schawinski}, {Schaye}, {Schwope}, {Sciortino}, {Severgnini}, {Shankar}, {Sijacki}, {Sim}, {Schmid}, {Smith}, {Steiner}, {Stelzer}, {Stewart}, {Strohmayer}, {Str{\"u}der}, {Sun}, {Takei}, {Tatischeff}, {Tiengo}, {Tombesi}, {Trinchieri}, {Tsuru}, {Ud-Doula}, {Ursino}, {Valencic}, {Vanzella}, {Vaughan}, {Vignali}, {Vink}, {Vito}, {Volonteri}, {Wang}, {Webb}, {Willingale}, {Wilms}, {Wise}, {Worrall}, {Young}, {Zampieri}, {In't Zand}, {Zane}, {Zezas}, {Zhang}, \& {Zhuravleva}}]{Athena}
{Nandra}, K., {Barret}, D., {Barcons}, X., {et~al.} 2013, arXiv e-prints, arXiv:1306.2307

\bibitem[{{Netzer}(2015)}]{AGN_unification}
{Netzer}, H. 2015, \araa, 53, 365

\bibitem[{{Noble} {et~al.}(2012){Noble}, {Mundim}, {Nakano}, {Krolik}, {Campanelli}, {Zlochower}, \& {Yunes}}]{Noble12}
{Noble}, S.~C., {Mundim}, B.~C., {Nakano}, H., {et~al.} 2012, \apj, 755, 51

\bibitem[{{O'Neill} {et~al.}(2022){O'Neill}, {Kiehlmann}, {Readhead}, {Aller}, {Blandford}, {Liodakis}, {Lister}, {Mr{\'o}z}, {O'Dea}, {Pearson}, {Ravi}, {Vallisneri}, {Cleary}, {Graham}, {Grainge}, {Hodges}, {Hovatta}, {L{\"a}hteenm{\"a}ki}, {Lamb}, {Lazio}, {Max-Moerbeck}, {Pavlidou}, {Prince}, {Reeves}, {Tornikoski}, {Vergara de la Parra}, \& {Zensus}}]{GLS_significance}
{O'Neill}, S., {Kiehlmann}, S., {Readhead}, A.~C.~S., {et~al.} 2022, \apjl, 926, L35

\bibitem[{{Papaloizou} \& {Pringle}(1977)}]{tidal_torques_close_binary_systems}
{Papaloizou}, J. \& {Pringle}, J.~E. 1977, \mnras, 181, 441

\bibitem[{{Peters}(1964)}]{time_to_merge}
{Peters}, P.~C. 1964, Physical Review, 136, 1224

\bibitem[{{Rakshit} {et~al.}(2020){Rakshit}, {Stalin}, \& {Kotilainen}}]{Masse_candidates}
{Rakshit}, S., {Stalin}, C.~S., \& {Kotilainen}, J. 2020, \apjs, 249, 17

\bibitem[{{Reardon} {et~al.}(2023){Reardon}, {Zic}, {Shannon}, {Hobbs}, {Bailes}, {Di Marco}, {Kapur}, {Rogers}, {Thrane}, {Askew}, {Bhat}, {Cameron}, {Cury{\l}o}, {Coles}, {Dai}, {Goncharov}, {Kerr}, {Kulkarni}, {Levin}, {Lower}, {Manchester}, {Mandow}, {Miles}, {Nathan}, {Os{\l}owski}, {Russell}, {Spiewak}, {Zhang}, \& {Zhu}}]{PTA_GW_background}
{Reardon}, D.~J., {Zic}, A., {Shannon}, R.~M., {et~al.} 2023, \apjl, 951, L6

\bibitem[{Roedig {et~al.}(2014)Roedig, Krolik, \& Miller}]{bbh_sed_notch_roedig14}
Roedig, C., Krolik, J.~H., \& Miller, M.~C. 2014, \apj, 785, 115

\bibitem[{{Runnoe} {et~al.}(2017){Runnoe}, {Eracleous}, {Pennell}, {Mathes}, {Boroson}, {Sigur{\dh}sson}, {Bogdanovi{\'c}}, {Halpern}, {Liu}, \& {Brown}}]{spectroscopic_mbbh_candidates}
{Runnoe}, J.~C., {Eracleous}, M., {Pennell}, A., {et~al.} 2017, \mnras, 468, 1683

\bibitem[{{Satyapal} {et~al.}(2014){Satyapal}, {Ellison}, {McAlpine}, {Hickox}, {Patton}, \& {Mendel}}]{galaxy_pairs_agn}
{Satyapal}, S., {Ellison}, S.~L., {McAlpine}, W., {et~al.} 2014, \mnras, 441, 1297

\bibitem[{{Savonije} {et~al.}(1994){Savonije}, {Papaloizou}, \& {Lin}}]{m2_modes_accretion_disk}
{Savonije}, G.~J., {Papaloizou}, J.~C.~B., \& {Lin}, D.~N.~C. 1994, \mnras, 268, 13

\bibitem[{{Shi} \& {Krolik}(2015)}]{SMBBH2}
{Shi}, J.-M. \& {Krolik}, J.~H. 2015, \apj, 807, 131

\bibitem[{Shi {et~al.}(2012)Shi, Krolik, Lubow, \& Hawley}]{Simu_BBH_lump_shi12}
Shi, J.-M., Krolik, J.~H., Lubow, S.~H., \& Hawley, J.~F. 2012, \apj, 749, 118

\bibitem[{{Shi} {et~al.}(2012){Shi}, {Krolik}, {Lubow}, \& {Hawley}}]{Shi12}
{Shi}, J.-M., {Krolik}, J.~H., {Lubow}, S.~H., \& {Hawley}, J.~F. 2012, \apj, 749, 118

\bibitem[{{Sillanpaa} {et~al.}(1988){Sillanpaa}, {Haarala}, {Valtonen}, {Sundelius}, \& {Byrd}}]{OJ287}
{Sillanpaa}, A., {Haarala}, S., {Valtonen}, M.~J., {Sundelius}, B., \& {Byrd}, G.~G. 1988, \apj, 325, 628

\bibitem[{Tang {et~al.}(2018)Tang, Haiman, \& MacFadyen}]{tang_late_2018}
Tang, Y., Haiman, Z., \& MacFadyen, A. 2018, \mnras, 476, 2249, arXiv: 1801.02266

\bibitem[{{Timmer} \& {K{\"o}nig}(1995)}]{lightcurves_simulations}
{Timmer}, J. \& {K{\"o}nig}, M. 1995, \aap, 300, 707

\bibitem[{{Trindade Falc{\~a}o} {et~al.}(2024){Trindade Falc{\~a}o}, {Turner}, {Kraemer}, {Reeves}, {Braito}, {Schmitt}, \& {Feuillet}}]{2024ApJ...972..185}
{Trindade Falc{\~a}o}, A., {Turner}, T.~J., {Kraemer}, S.~B., {et~al.} 2024, \apj, 972, 185

\bibitem[{{Vanden Berk} {et~al.}(2004){Vanden Berk}, {Wilhite}, {Kron}, {Anderson}, {Brunner}, {Hall}, {Ivezi{\'c}}, {Richards}, {Schneider}, {York}, {Brinkmann}, {Lamb}, {Nichol}, \& {Schlegel}}]{quasar_variability}
{Vanden Berk}, D.~E., {Wilhite}, B.~C., {Kron}, R.~G., {et~al.} 2004, \apj, 601, 692

\bibitem[{{VanderPlas}(2018)}]{understanding_LS}
{VanderPlas}, J.~T. 2018, \apjs, 236, 16

\bibitem[{{Vaughan} {et~al.}(2016){Vaughan}, {Uttley}, {Markowitz}, {Huppenkothen}, {Middleton}, {Alston}, {Scargle}, \& {Farr}}]{False_periodicities}
{Vaughan}, S., {Uttley}, P., {Markowitz}, A.~G., {et~al.} 2016, \mnras, 461, 3145

\bibitem[{{Verbiest} {et~al.}(2016){Verbiest}, {Lentati}, {Hobbs}, {van Haasteren}, {Demorest}, {Janssen}, {Wang}, {Desvignes}, {Caballero}, {Keith}, {Champion}, {Arzoumanian}, {Babak}, {Bassa}, {Bhat}, {Brazier}, {Brem}, {Burgay}, {Burke-Spolaor}, {Chamberlin}, {Chatterjee}, {Christy}, {Cognard}, {Cordes}, {Dai}, {Dolch}, {Ellis}, {Ferdman}, {Fonseca}, {Gair}, {Garver-Daniels}, {Gentile}, {Gonzalez}, {Graikou}, {Guillemot}, {Hessels}, {Jones}, {Karuppusamy}, {Kerr}, {Kramer}, {Lam}, {Lasky}, {Lassus}, {Lazarus}, {Lazio}, {Lee}, {Levin}, {Liu}, {Lynch}, {Lyne}, {Mckee}, {McLaughlin}, {McWilliams}, {Madison}, {Manchester}, {Mingarelli}, {Nice}, {Os{\l}owski}, {Palliyaguru}, {Pennucci}, {Perera}, {Perrodin}, {Possenti}, {Petiteau}, {Ransom}, {Reardon}, {Rosado}, {Sanidas}, {Sesana}, {Shaifullah}, {Shannon}, {Siemens}, {Simon}, {Smits}, {Spiewak}, {Stairs}, {Stappers}, {Stinebring}, {Stovall}, {Swiggum}, {Taylor}, {Theureau}, {Tiburzi}, {Toomey}, {Vallisneri}, {van Straten}, {Vecchio}, {Wang}, {Wen}, {You},
  {Zhu}, \& {Zhu}}]{iPTA}
{Verbiest}, J.~P.~W., {Lentati}, L., {Hobbs}, G., {et~al.} 2016, \mnras, 458, 1267

\bibitem[{{Voggel} {et~al.}(2022){Voggel}, {Seth}, {Baumgardt}, {Husemann}, {Neumayer}, {Hilker}, {Pechetti}, {Mieske}, {Dumont}, \& {Georgiev}}]{NGC7727}
{Voggel}, K.~T., {Seth}, A.~C., {Baumgardt}, H., {et~al.} 2022, \aap, 658, A152

\bibitem[{{Volonteri} {et~al.}(2003){Volonteri}, {Madau}, \& {Haardt}}]{model_hierarchique_mergers}
{Volonteri}, M., {Madau}, P., \& {Haardt}, F. 2003, \apj, 593, 661

\bibitem[{{Volonteri} {et~al.}(2009){Volonteri}, {Miller}, \& {Dotti}}]{MBBHs_low_redshift_expectations}
{Volonteri}, M., {Miller}, J.~M., \& {Dotti}, M. 2009, \apjl, 703, L86

\bibitem[{{Westernacher-Schneider} {et~al.}(2022){Westernacher-Schneider}, {Zrake}, {MacFadyen}, \& {Haiman}}]{Westernacher22}
{Westernacher-Schneider}, J.~R., {Zrake}, J., {MacFadyen}, A., \& {Haiman}, Z. 2022, \prd, 106, 103010

\bibitem[{{Xu} {et~al.}(2023){Xu}, {Chen}, {Guo}, {Jiang}, {Wang}, {Xu}, {Xue}, {Nicolas Caballero}, {Yuan}, {Xu}, {Wang}, {Hao}, {Luo}, {Lee}, {Han}, {Jiang}, {Shen}, {Wang}, {Wang}, {Xu}, {Wu}, {Manchester}, {Qian}, {Guan}, {Huang}, {Sun}, \& {Zhu}}]{2023RAA....23g5024X}
{Xu}, H., {Chen}, S., {Guo}, Y., {et~al.} 2023, Research in Astronomy and Astrophysics, 23, 075024

\bibitem[{{Yu}(2002)}]{2002MNRAS.331..935Y}
{Yu}, Q. 2002, \mnras, 331, 935

\bibitem[{Zechmeister \& Kürster(2009)}]{GLS}
Zechmeister, M. \& Kürster, M. 2009, \aap, 496, 577–584

\end{thebibliography}
\begin{appendix}
\onecolumn
\section{36 Candidates CRTS/ZTF intercalibrated lightcurves} \label{appendix:Appendix_A_lightcurves}
We show below the CRTS (black) and ZTF (red) lightcurves of the 36 potential candidates we found. Dashed green line represents the best fitted sine function. Orange envelope is the 3$\sigma$ best fitted sine function.

\begin{figure*}[!h]
    \centering
    \begin{subfigure}[t]{0.45\textwidth}
        \includegraphics[height=2.1in]{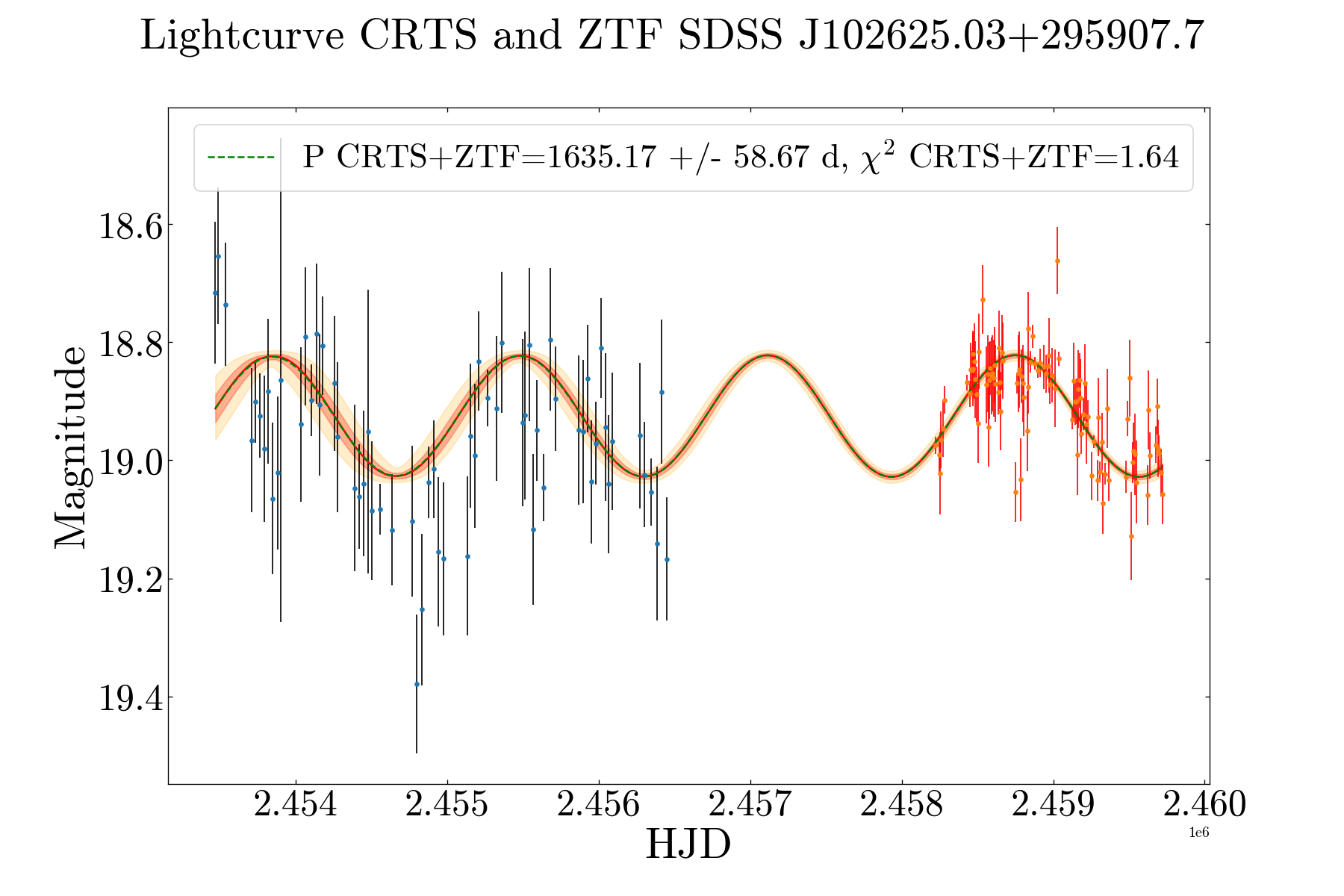}
    \end{subfigure}
    \begin{subfigure}[t]{0.45\textwidth}
        \includegraphics[height=2.1in]{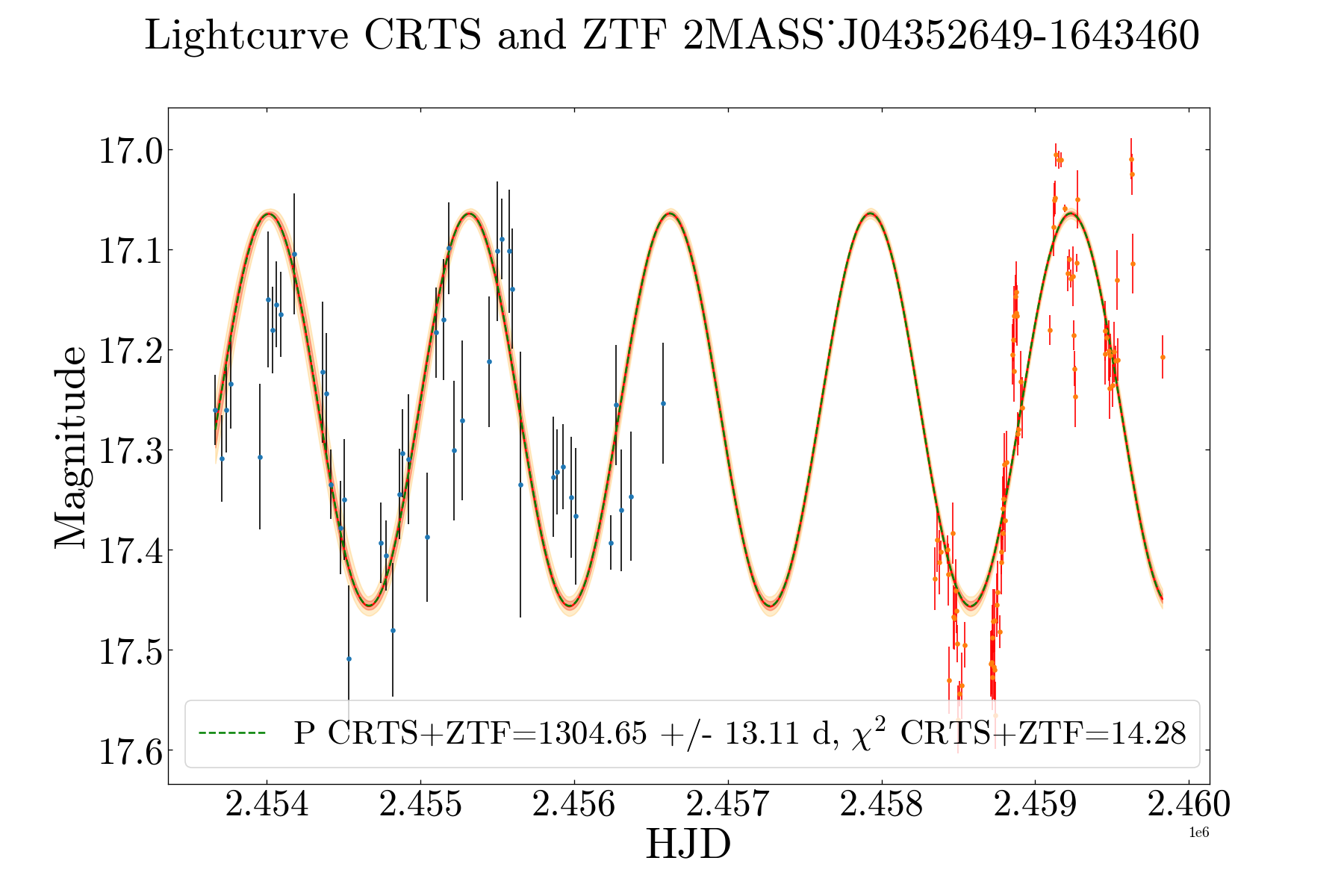}
    \end{subfigure}
\end{figure*}

\begin{figure*}[!h]
    \centering
    \begin{subfigure}[t]{0.45\textwidth}
        \centering
        \includegraphics[height=2.1in]{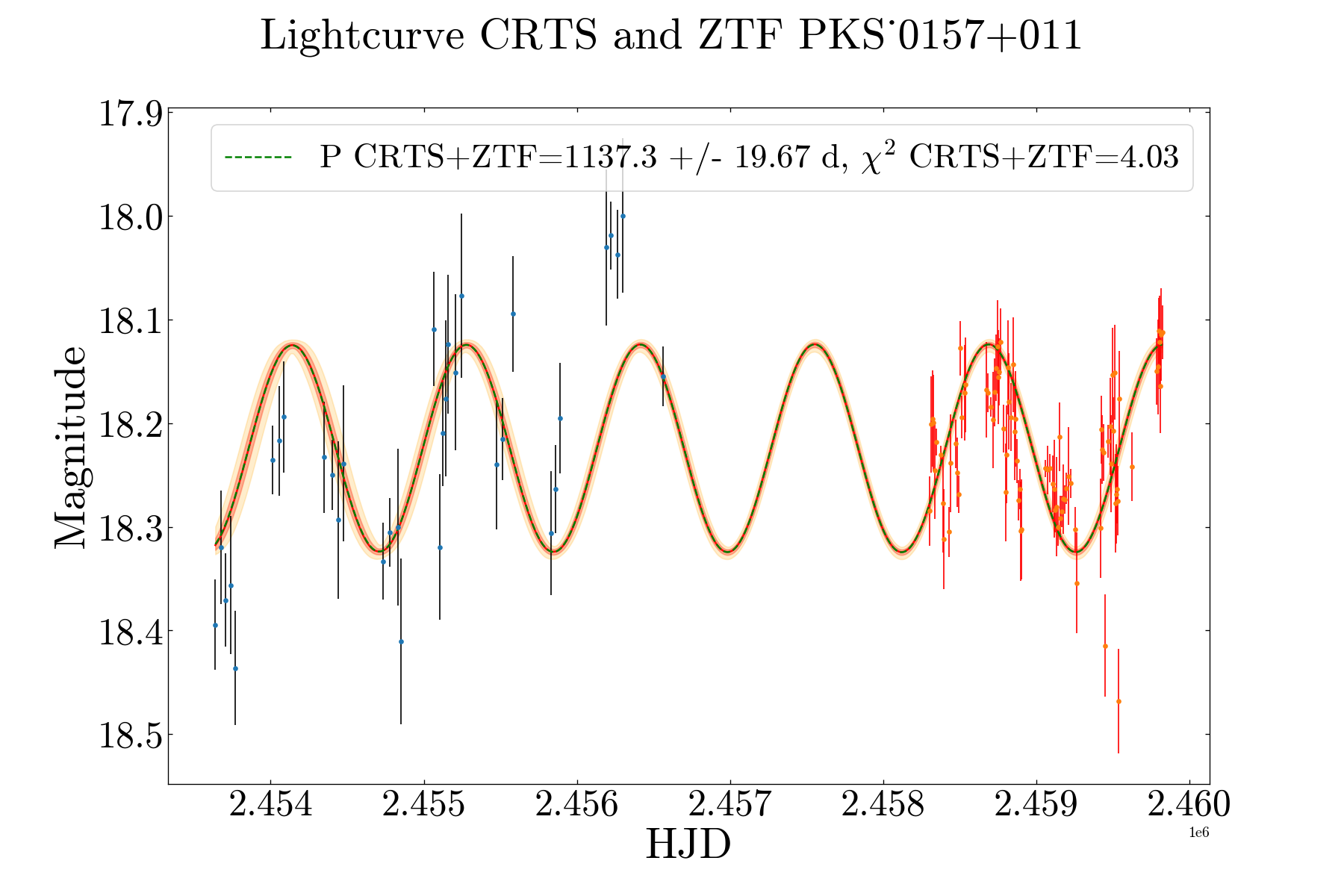}
    \end{subfigure}
    \begin{subfigure}[t]{0.45\textwidth}
        \centering
        \includegraphics[height=2.1in]{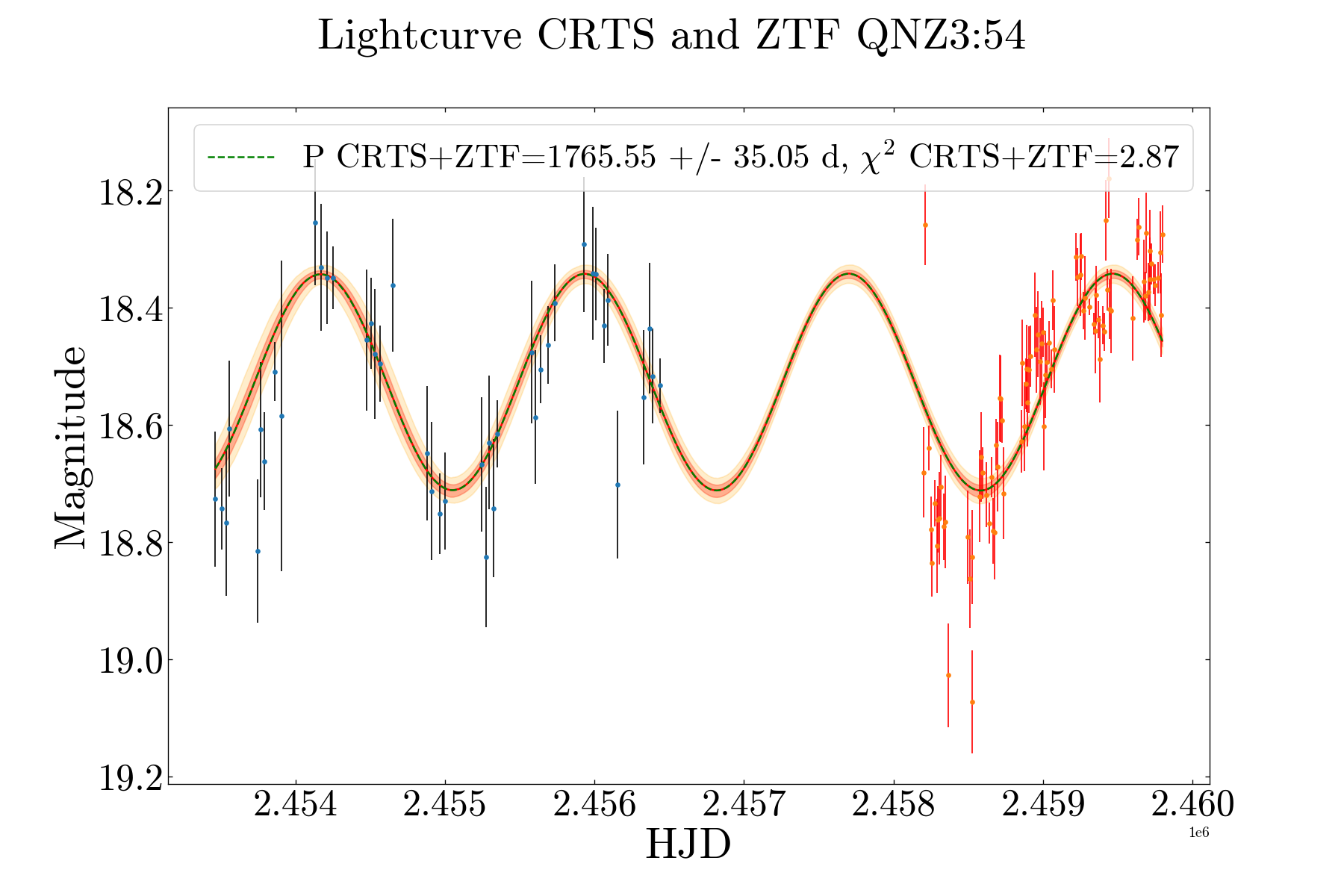}
    \end{subfigure}
\end{figure*}

\begin{figure*}[!h]
    \centering
    \begin{subfigure}[t]{0.45\textwidth}
        \centering
        \includegraphics[height=2.1in]{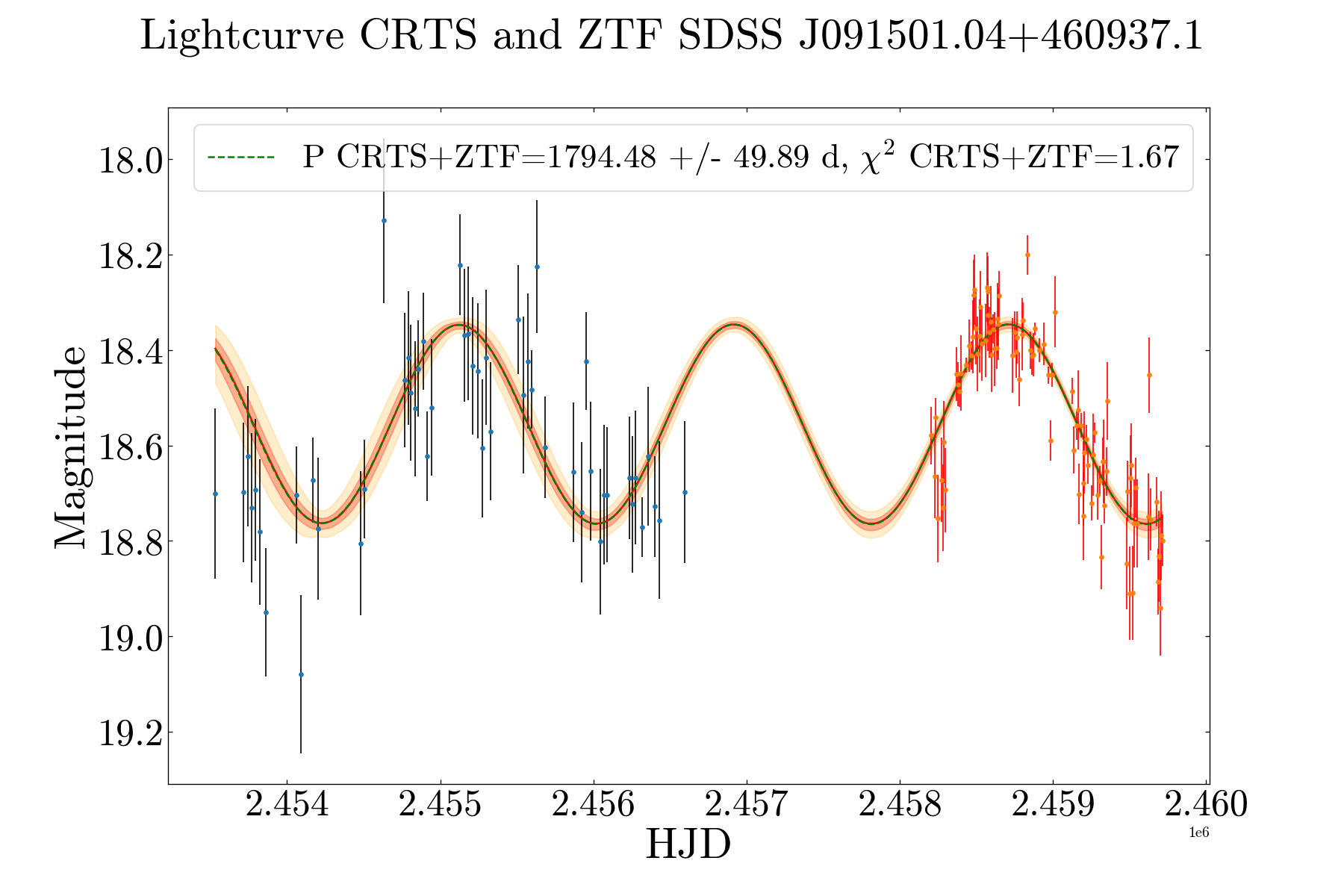}
    \end{subfigure}  
    \begin{subfigure}[t]{0.45\textwidth}
        \centering
        \includegraphics[height=2.1in]{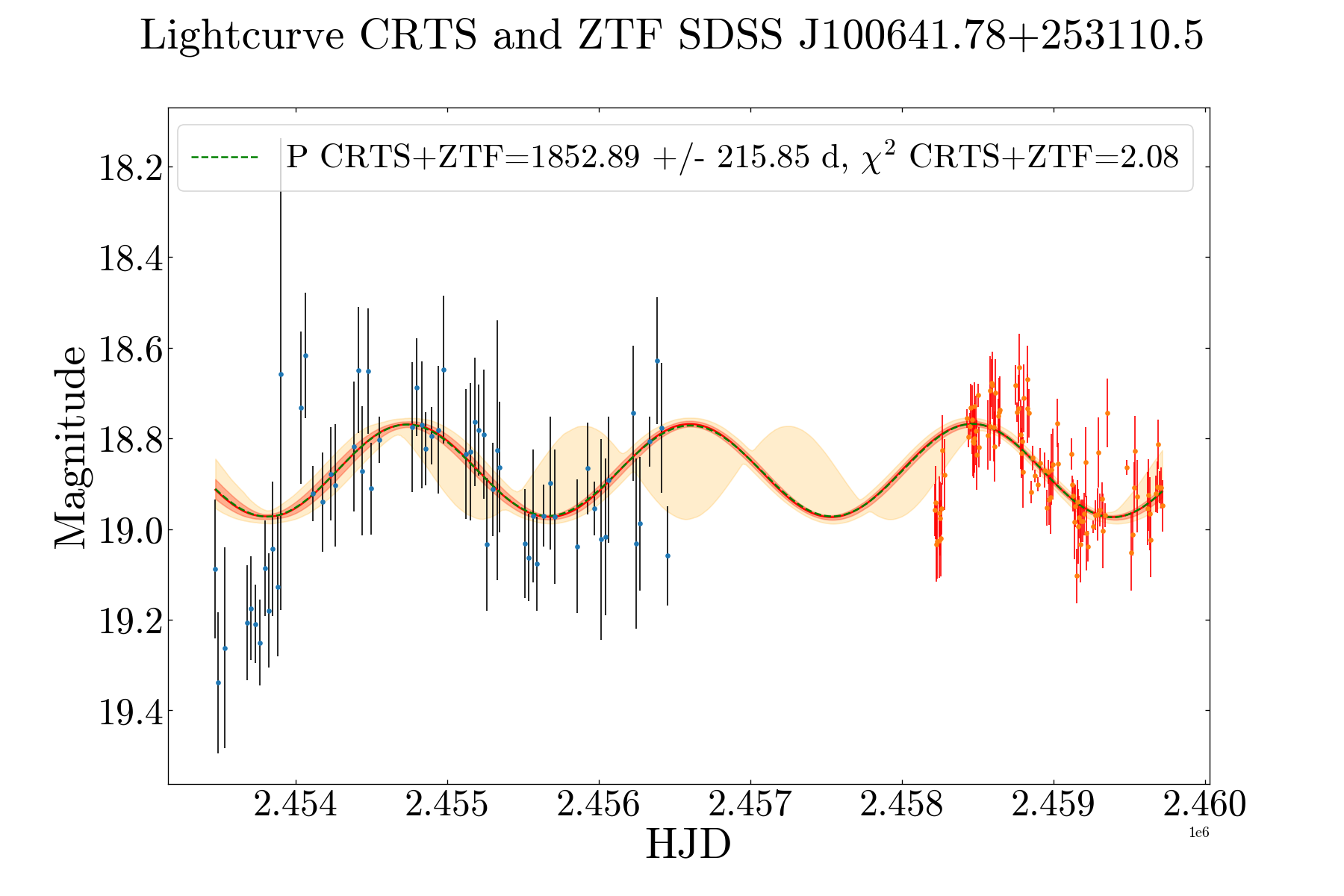}
    \end{subfigure}
\end{figure*}

\begin{figure*}[!h]
    \centering
    \begin{subfigure}[t]{0.45\textwidth}
        \centering
        \includegraphics[height=2.1in]{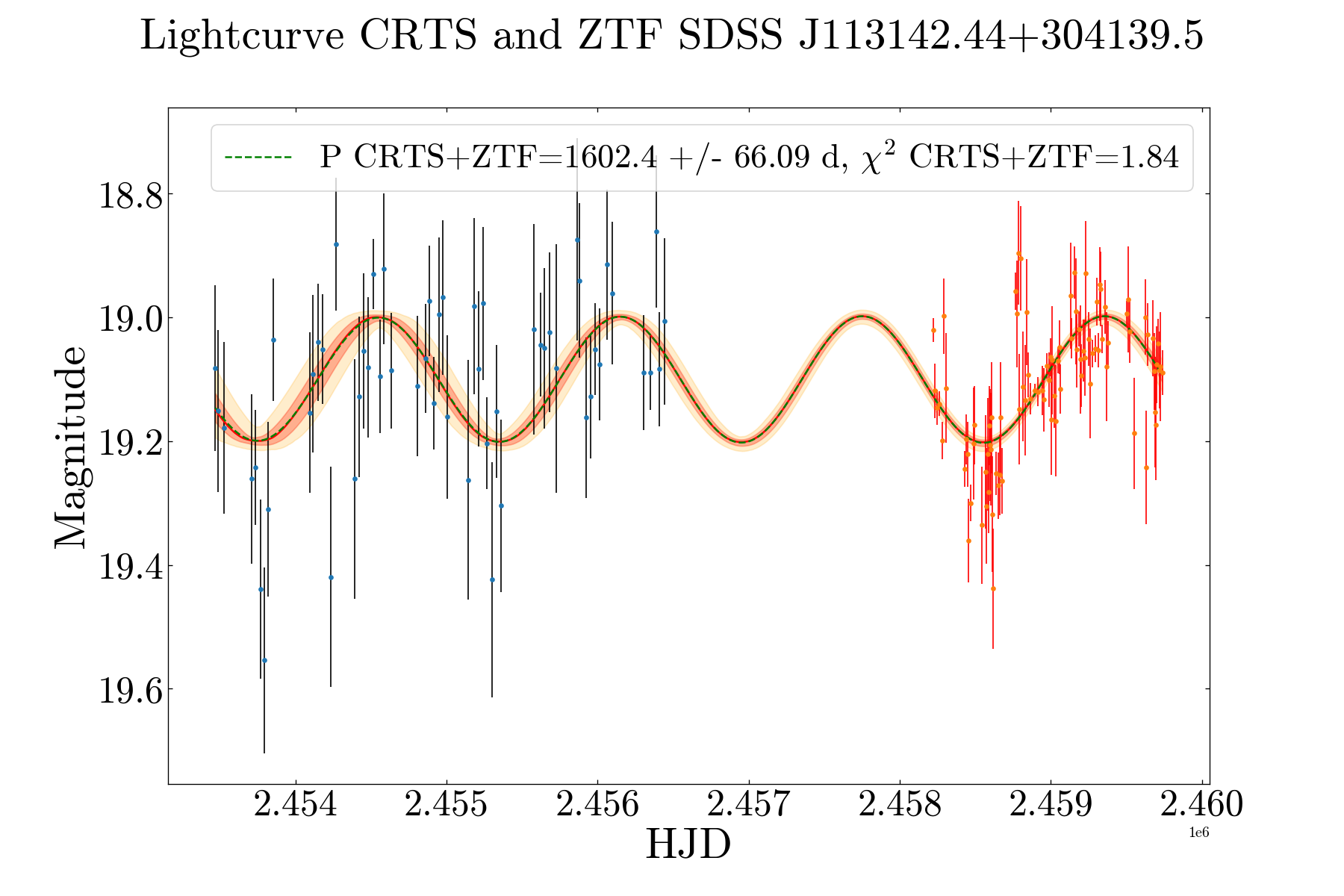}
    \end{subfigure} 
    \begin{subfigure}[t]{0.45\textwidth}
        \centering
        \includegraphics[height=2.1in]{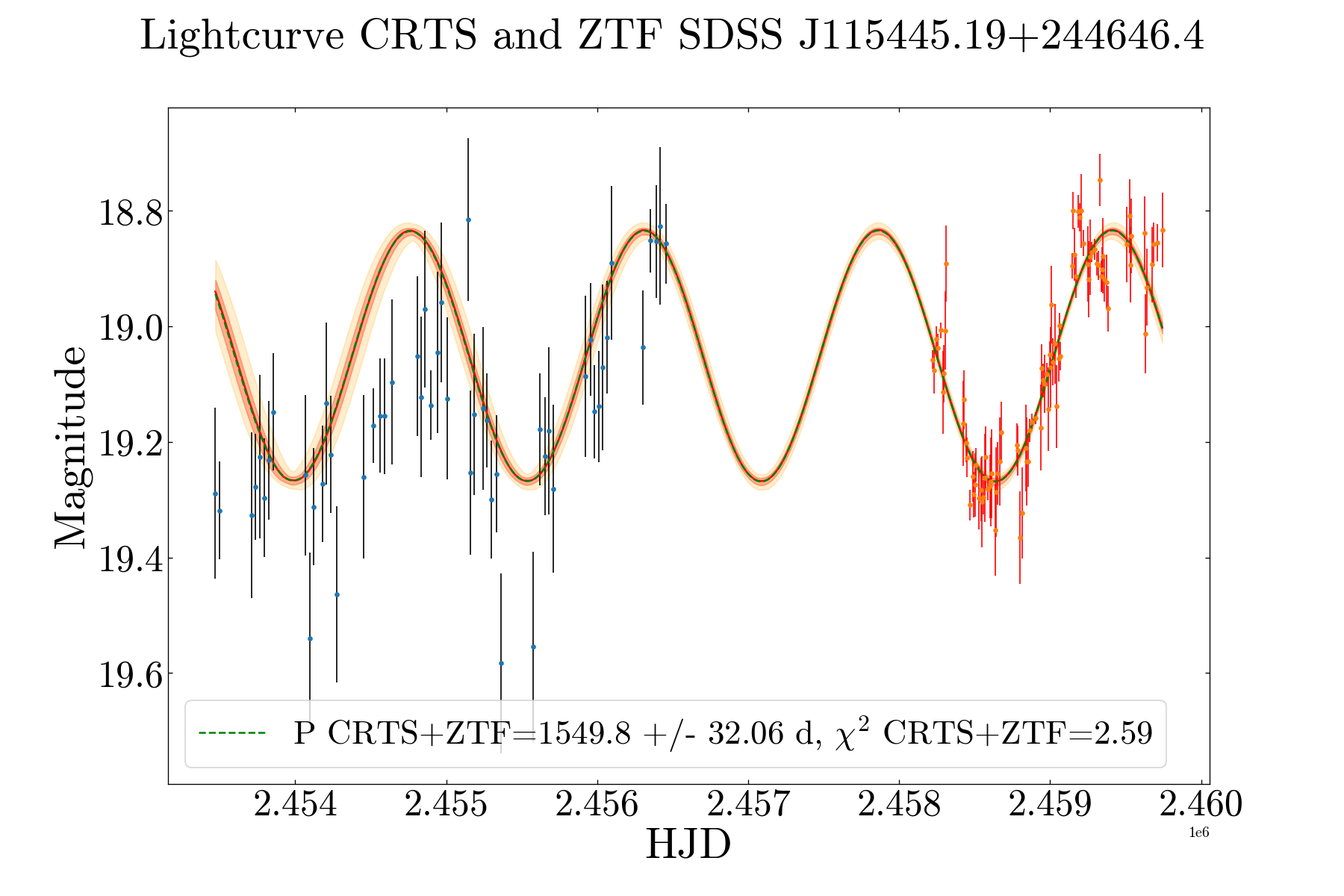}
    \end{subfigure}
\end{figure*}

\begin{figure*}[!h]
    \centering
    \begin{subfigure}[t]{0.45\textwidth}
        \centering
        \includegraphics[height=2.1in]{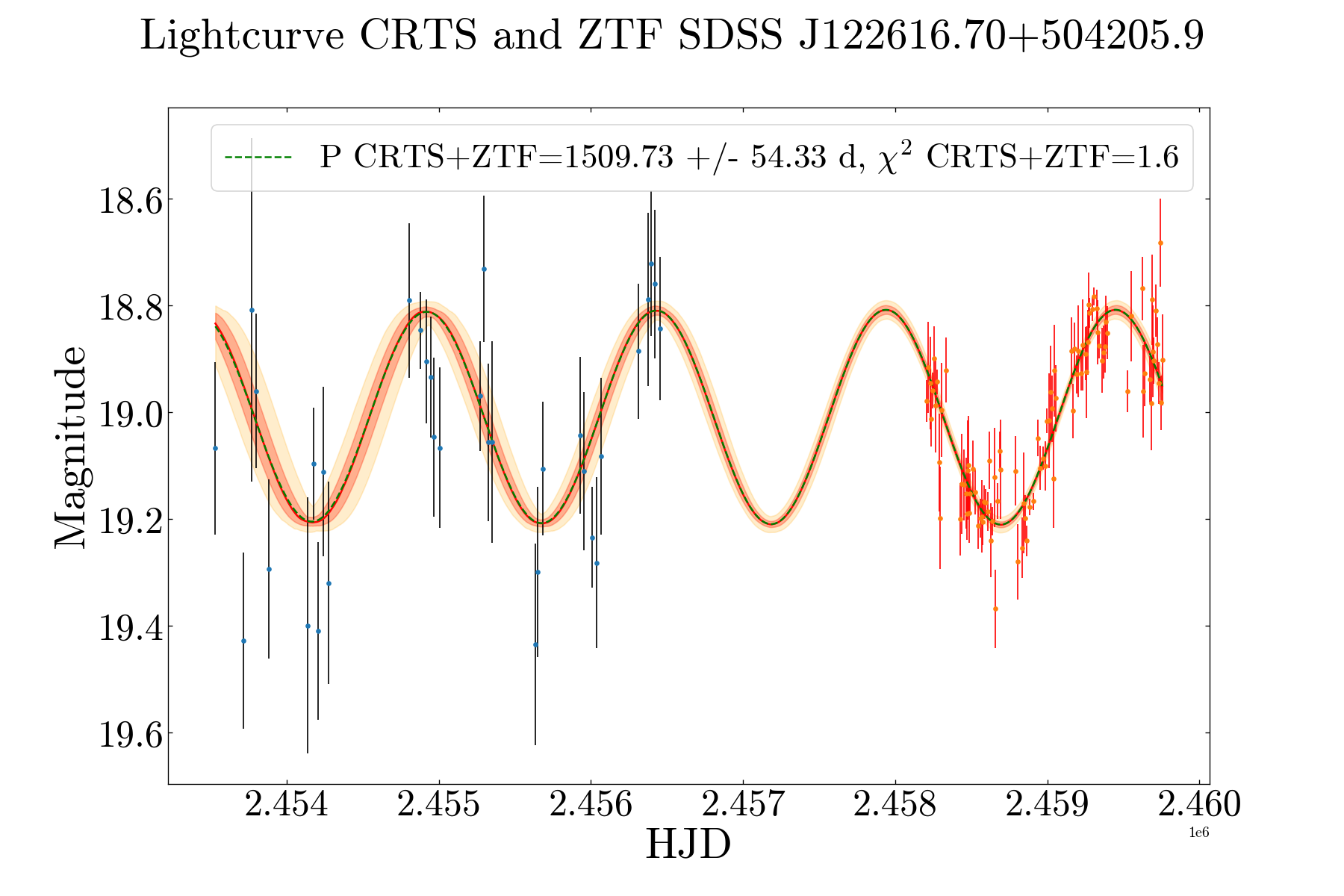}
    \end{subfigure} 
    \begin{subfigure}[t]{0.45\textwidth}
        \centering
        \includegraphics[height=2.1in]{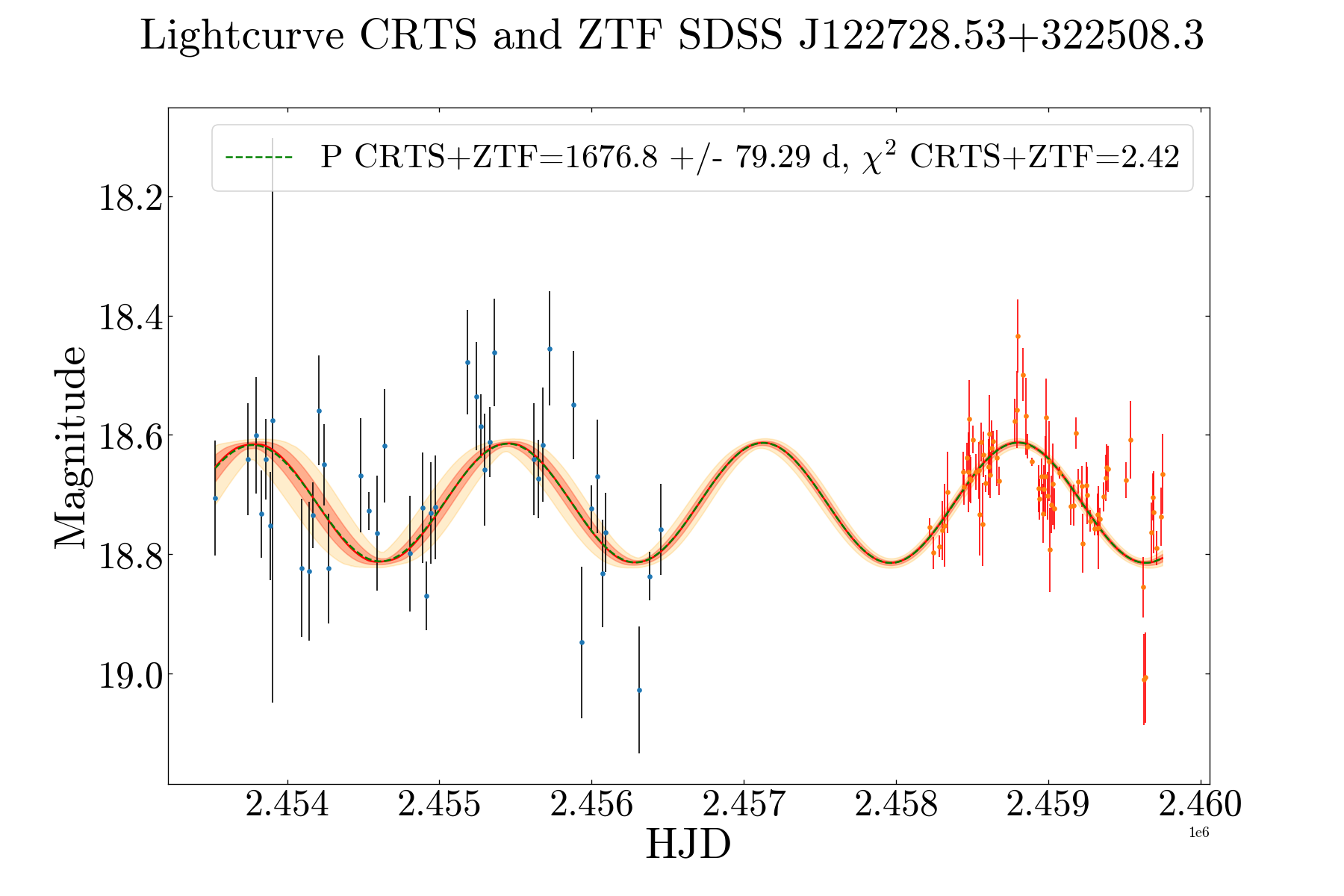}
    \end{subfigure}
\end{figure*}

\begin{figure*}[!h]
    \centering
    \begin{subfigure}[t]{0.45\textwidth}
        \centering
        \includegraphics[height=2.1in]{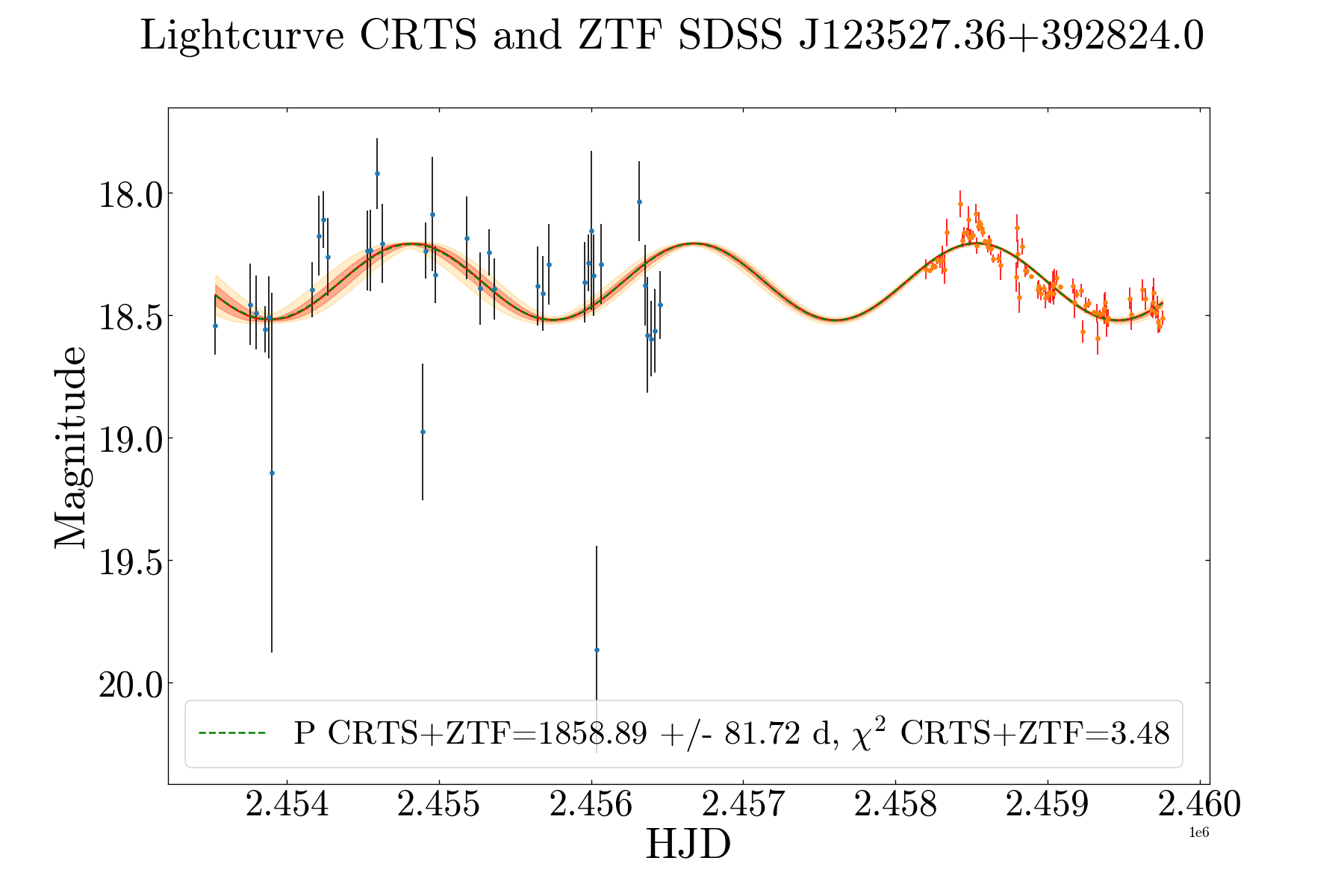}
    \end{subfigure}  
    \begin{subfigure}[t]{0.45\textwidth}
        \centering
        \includegraphics[height=2.1in]{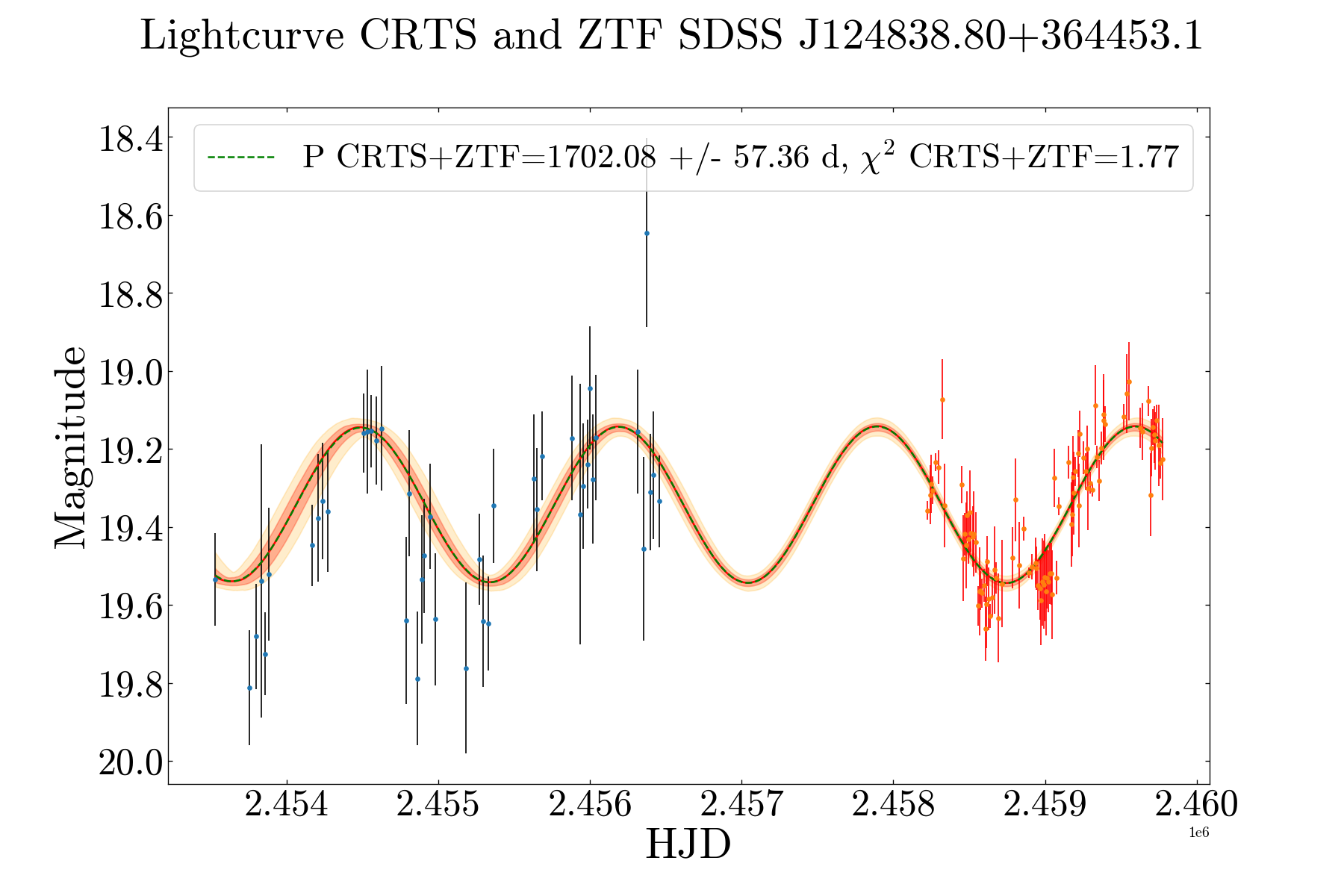}
    \end{subfigure}
\end{figure*}

\begin{figure*}[!h]
    \centering
    \begin{subfigure}[t]{0.45\textwidth}
        \centering
        \includegraphics[height=2.1in]{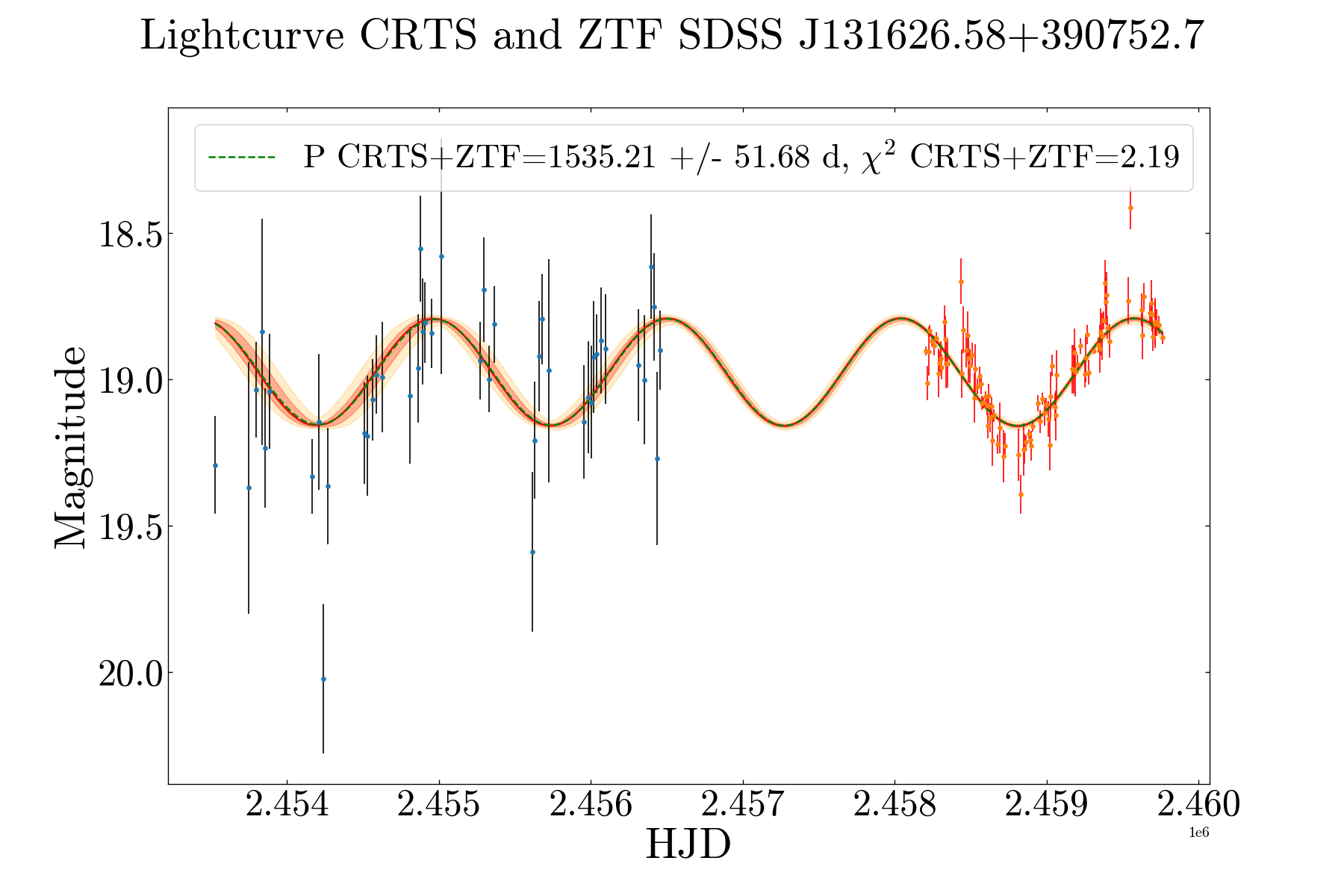}
    \end{subfigure}  
    \begin{subfigure}[t]{0.45\textwidth}
        \centering
        \includegraphics[height=2.1in]{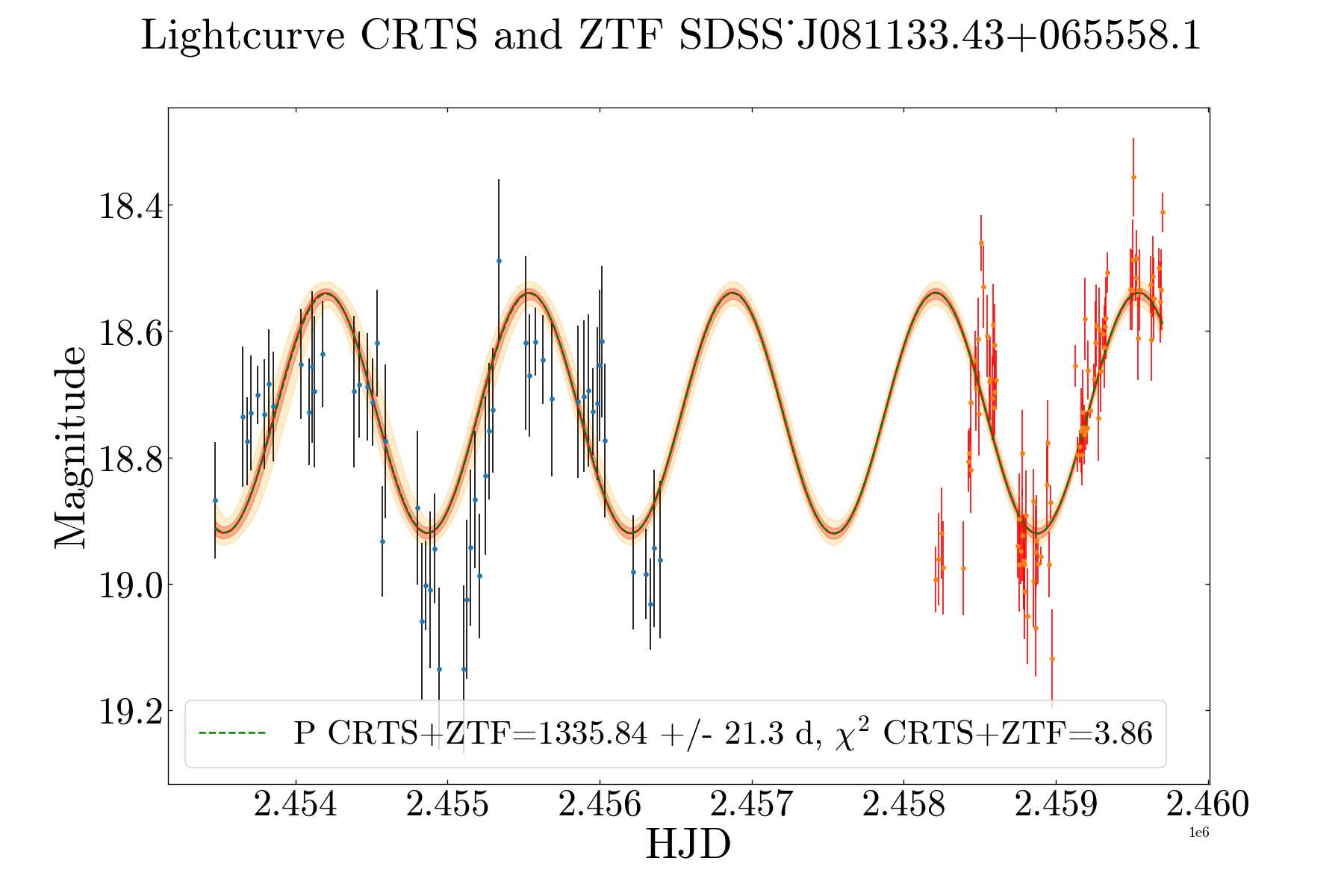}
    \end{subfigure}
\end{figure*}

\begin{figure*}[!h]
    \centering
    \begin{subfigure}[t]{0.45\textwidth}
        \centering
        \includegraphics[height=2.1in]{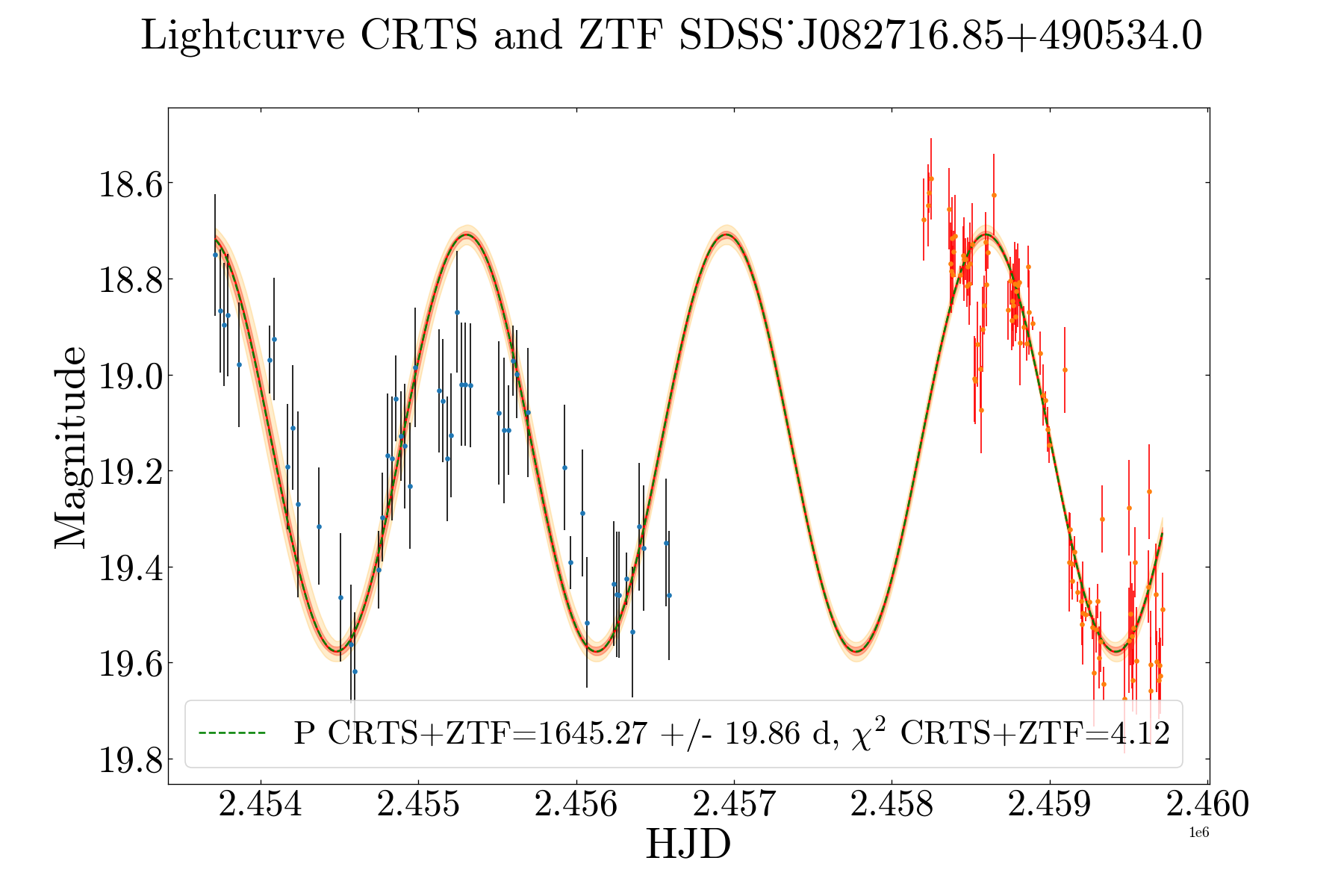}
    \end{subfigure} 
    \begin{subfigure}[t]{0.45\textwidth}
        \centering
        \includegraphics[height=2.1in]{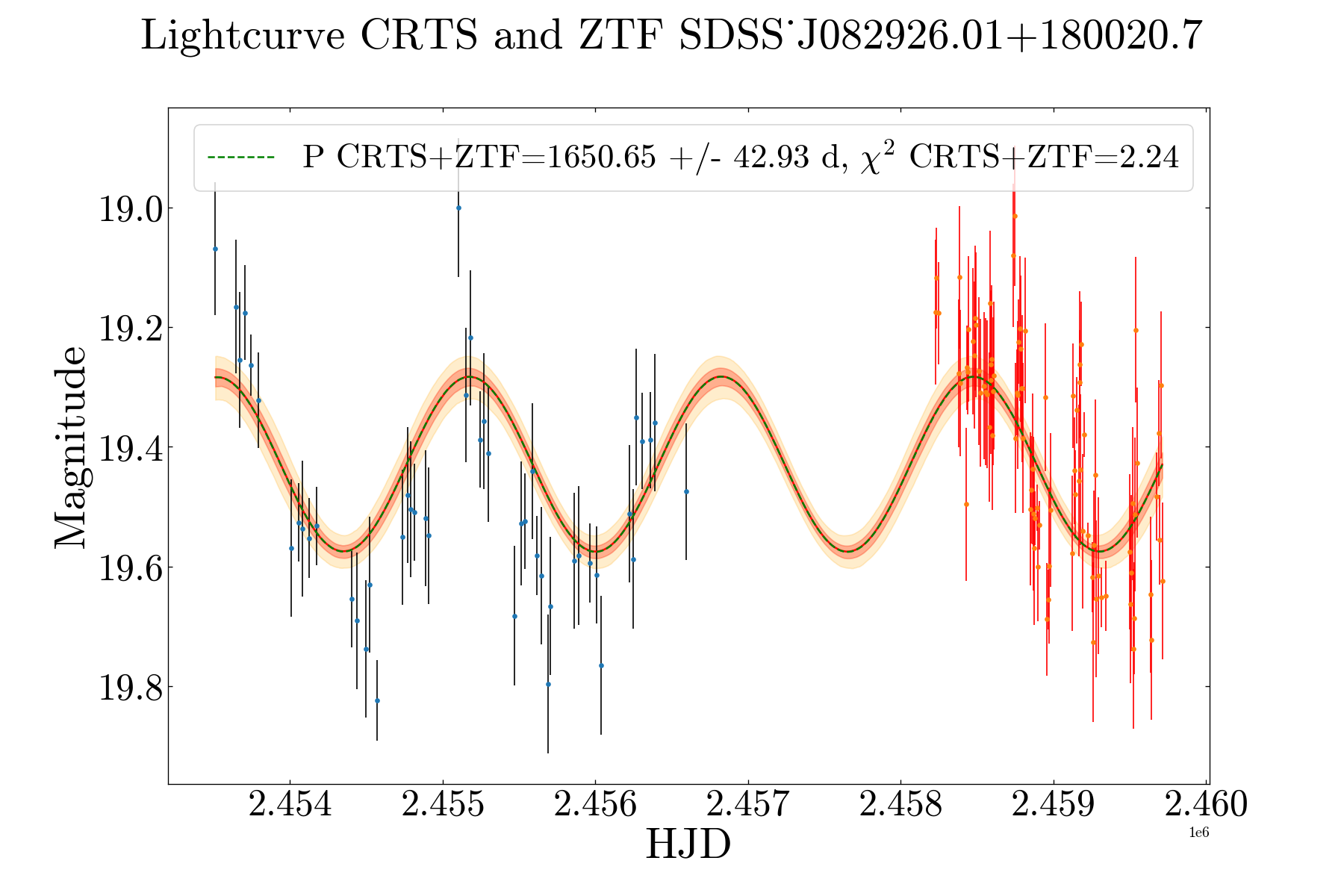}
    \end{subfigure}
\end{figure*}

\begin{figure*}[!h]
    \centering
    \begin{subfigure}[t]{0.45\textwidth}
        \centering
        \includegraphics[height=2.1in]{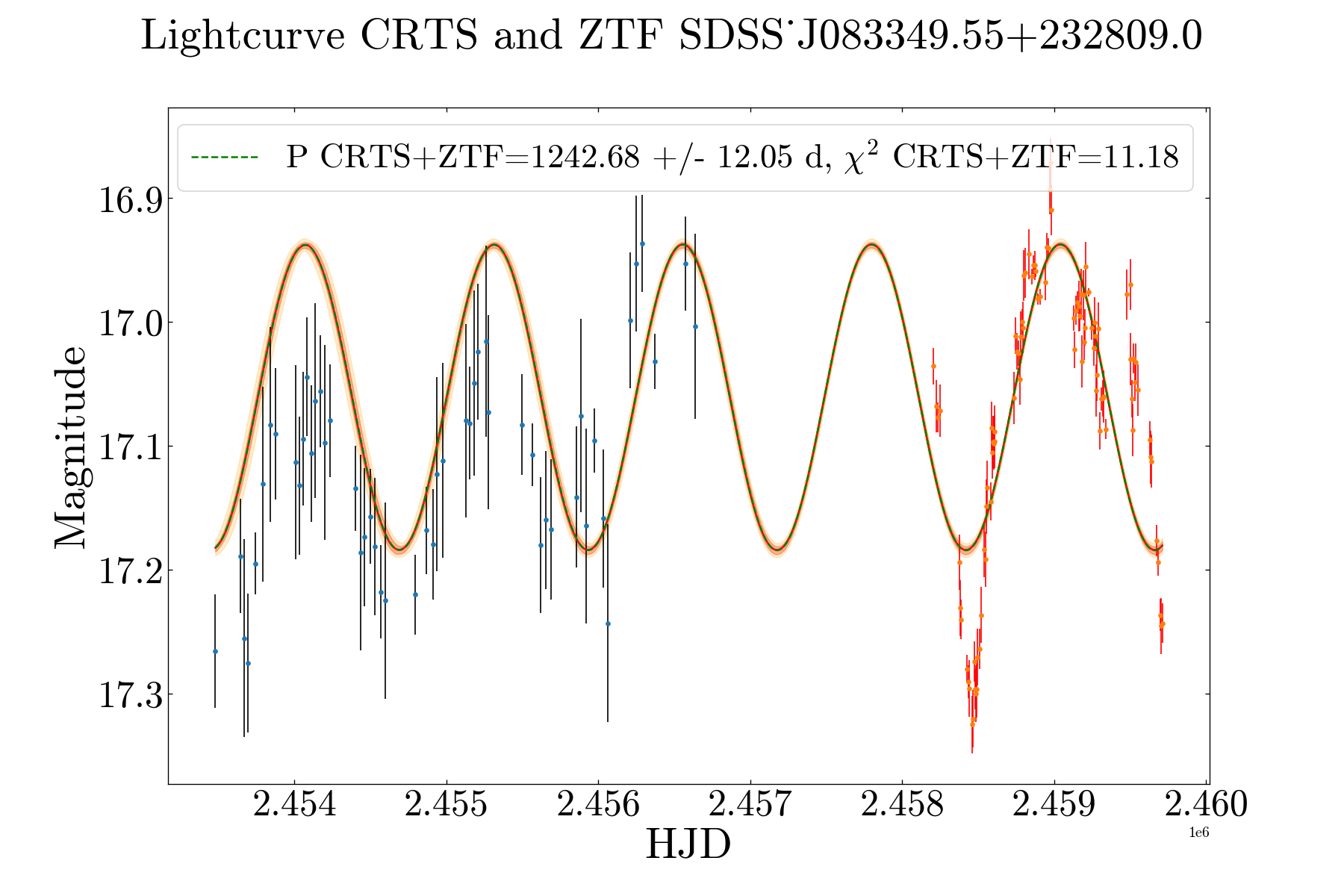}
    \end{subfigure} 
    \begin{subfigure}[t]{0.45\textwidth}
        \centering
        \includegraphics[height=2.1in]{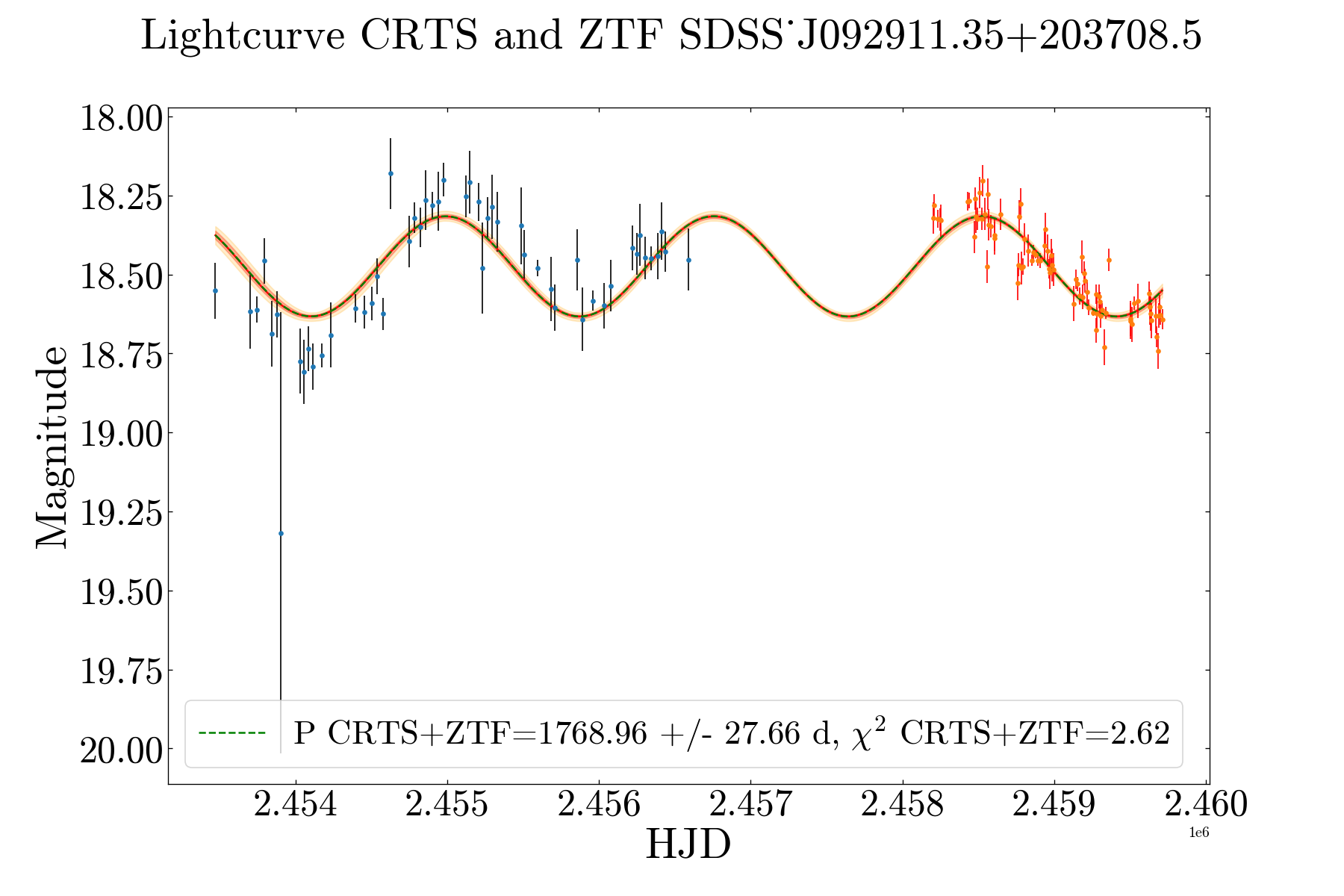}
    \end{subfigure}
\end{figure*}

\begin{figure*}[!h]
    \centering
    \begin{subfigure}[t]{0.45\textwidth}
        \centering
        \includegraphics[height=2.1in]{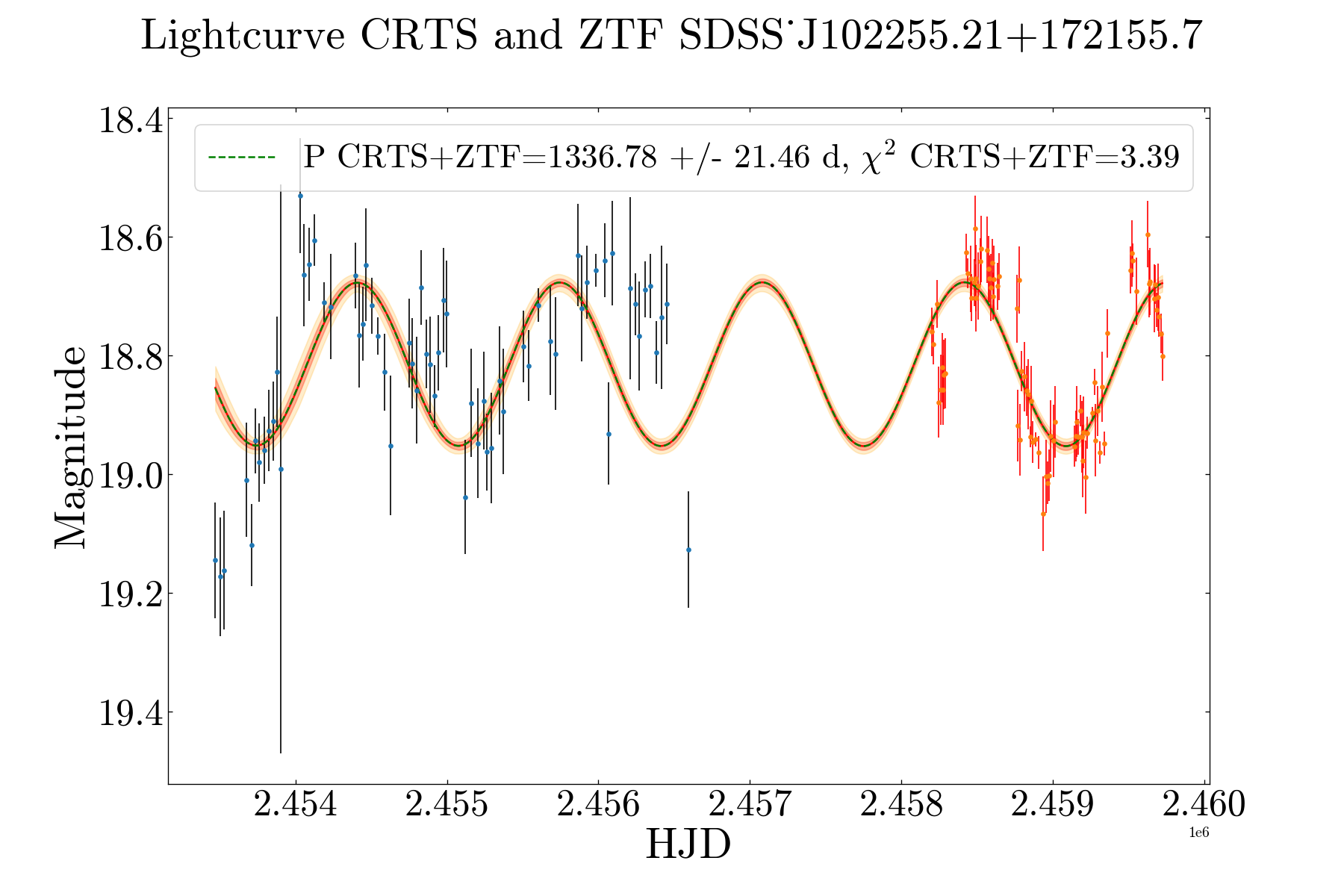}
    \end{subfigure}
    \begin{subfigure}[t]{0.45\textwidth}
        \centering
        \includegraphics[height=2.1in]{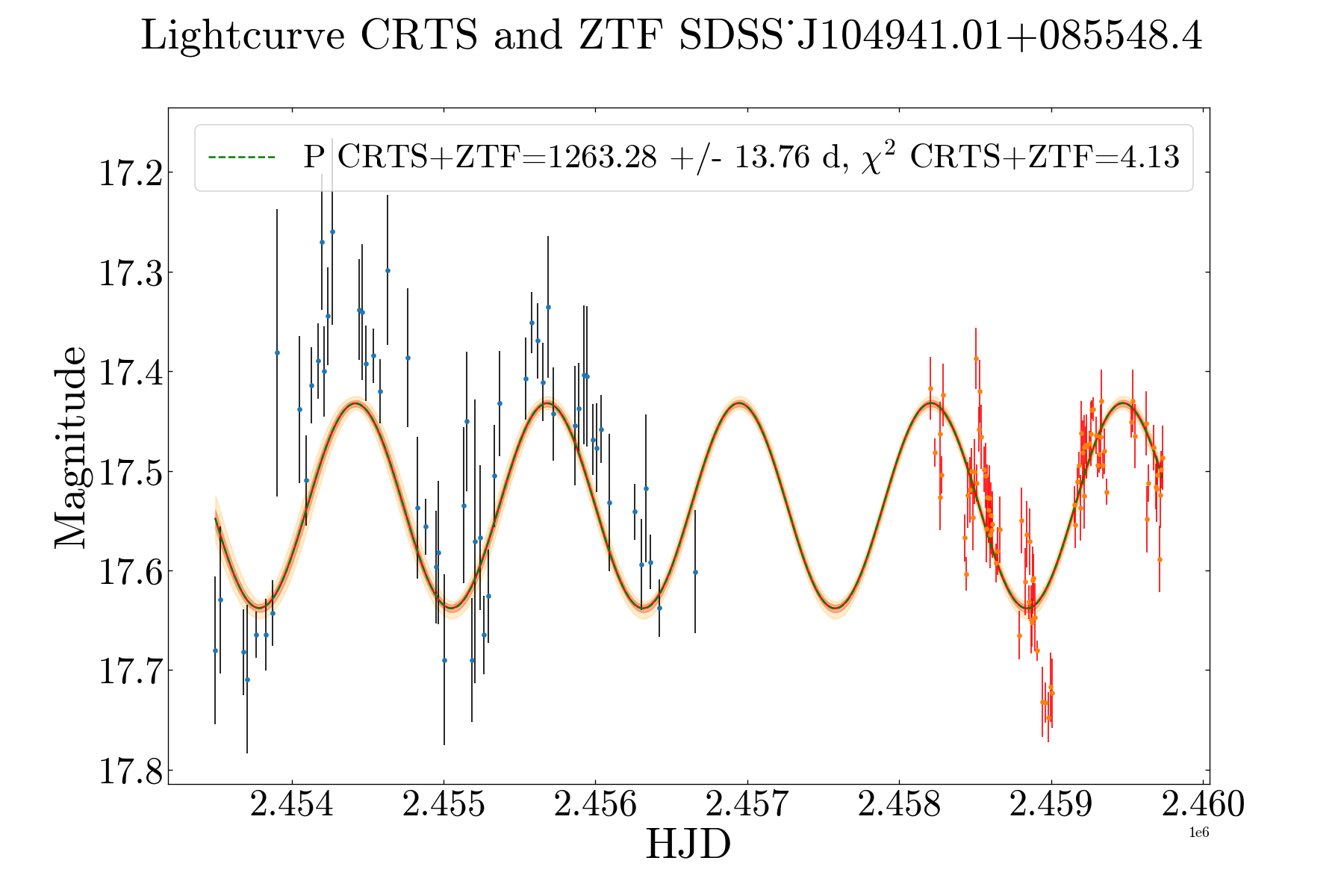}
    \end{subfigure}
\end{figure*}

\begin{figure*}[!h]
    \centering
    \begin{subfigure}[t]{0.45\textwidth}
        \centering
        \includegraphics[height=2.1in]{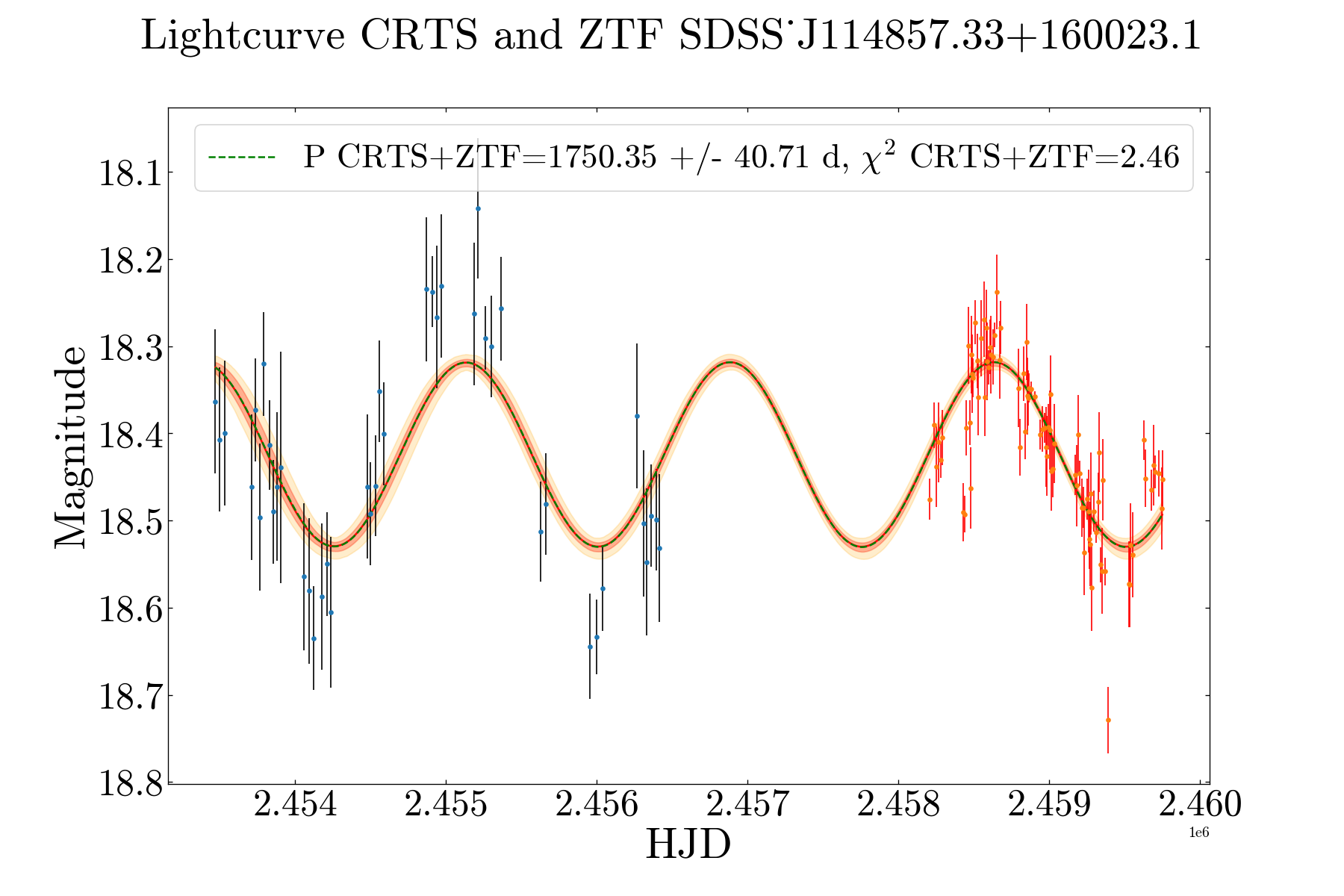}
    \end{subfigure}
    \begin{subfigure}[t]{0.45\textwidth}
        \centering
        \includegraphics[height=2.1in]{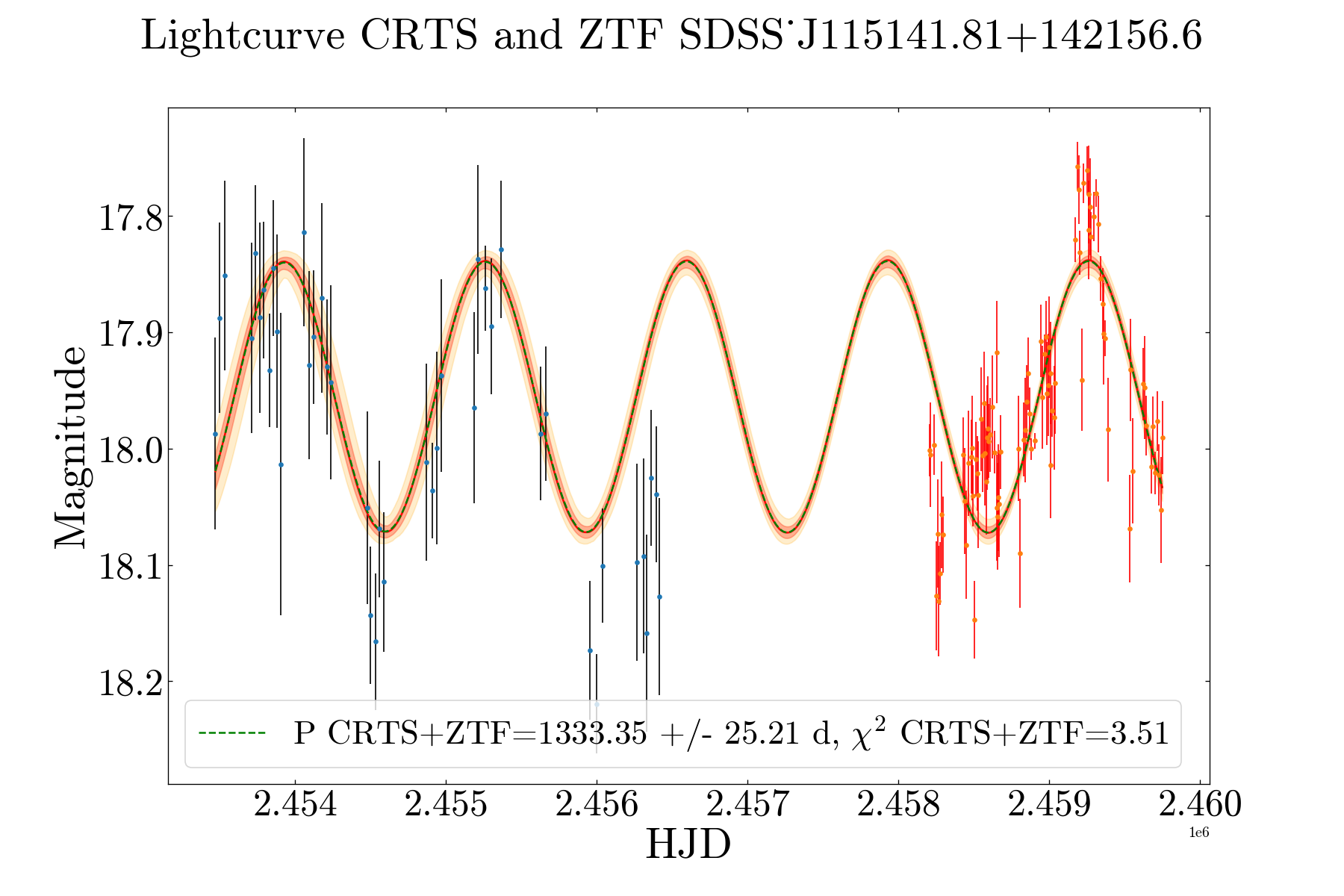}
    \end{subfigure}
\end{figure*}

\begin{figure*}[!h]
    \centering
    \begin{subfigure}[t]{0.45\textwidth}
        \centering
        \includegraphics[height=2.1in]{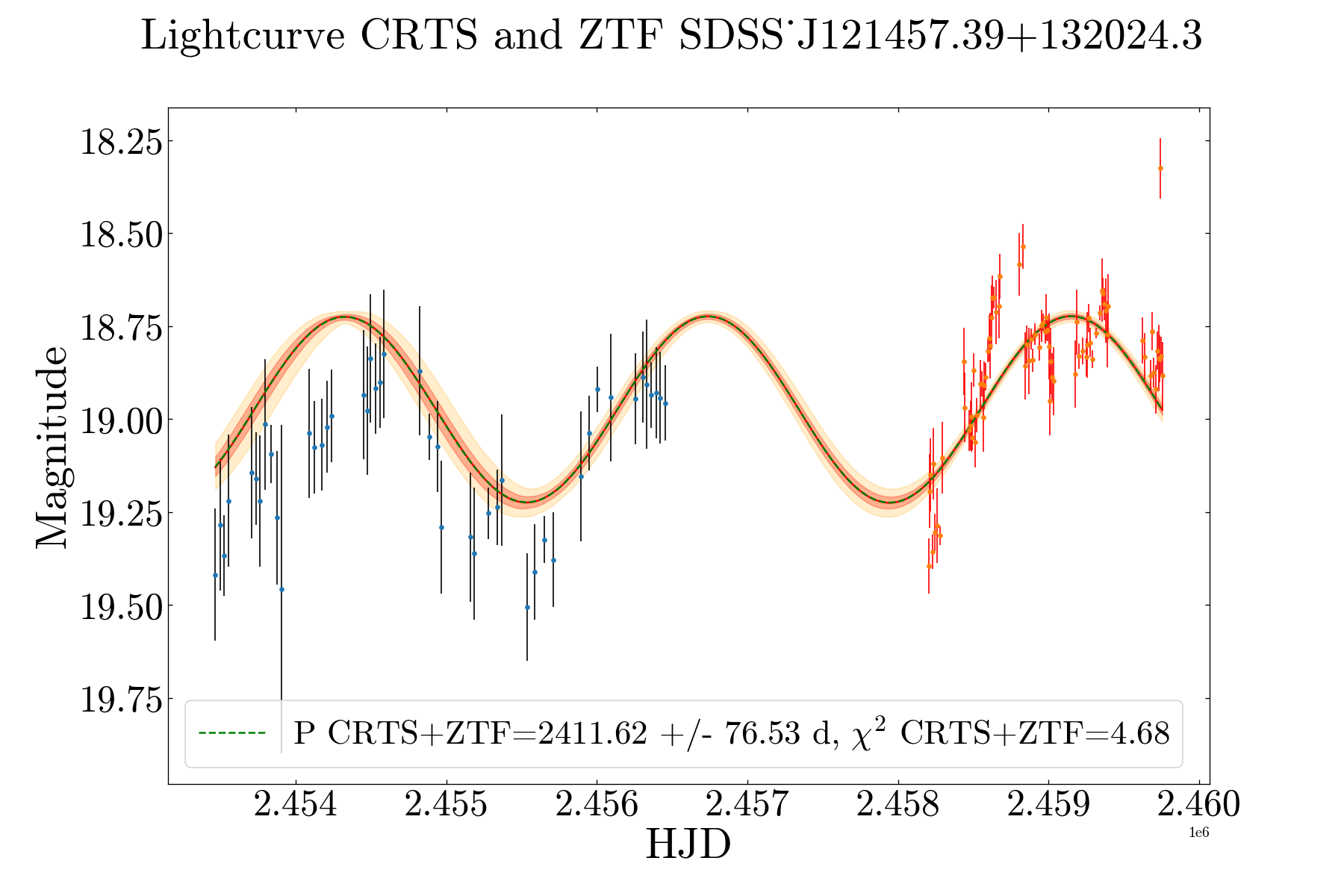}
    \end{subfigure}
    \begin{subfigure}[t]{0.45\textwidth}
        \centering
        \includegraphics[height=2.1in]{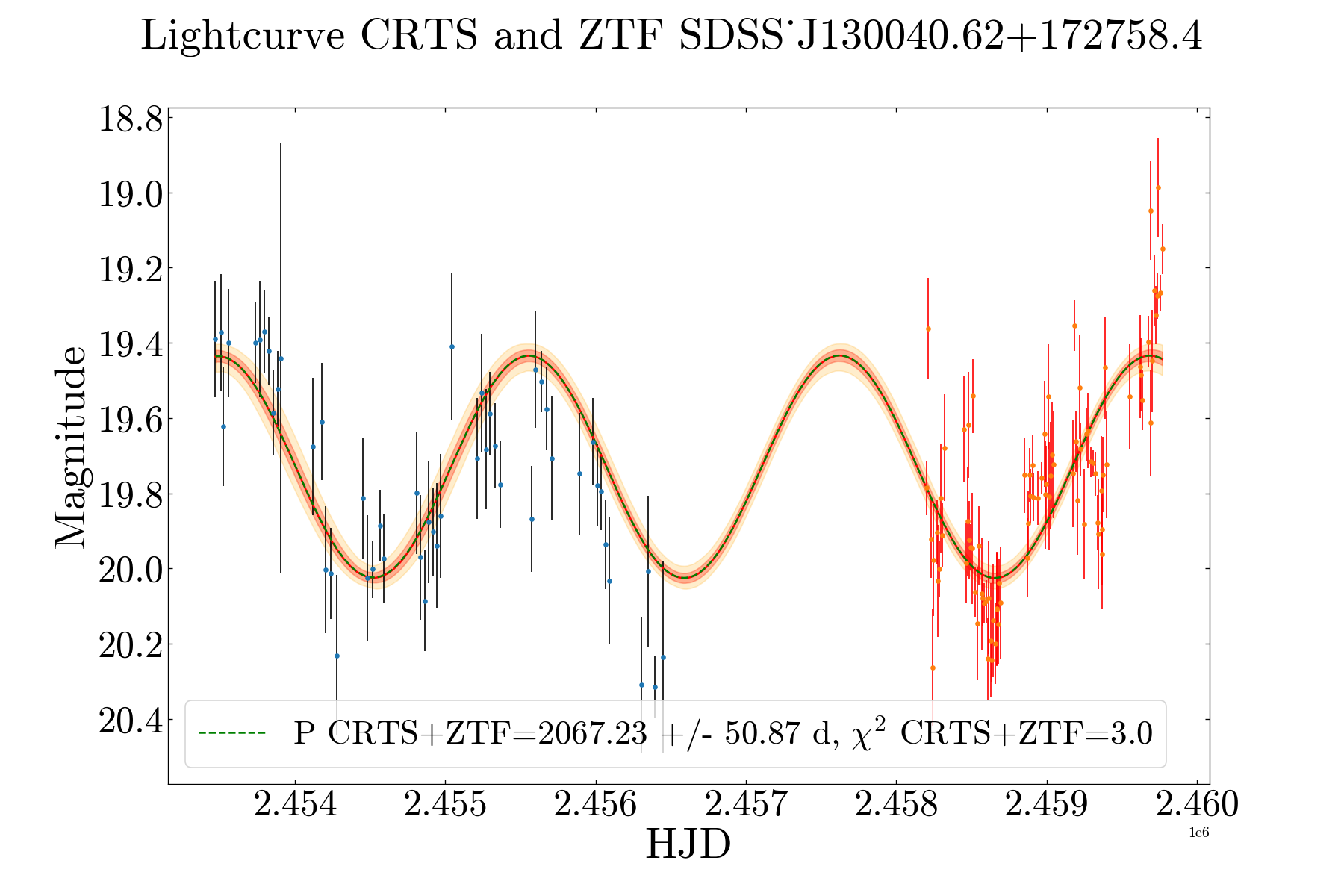}
    \end{subfigure}
\end{figure*}

\begin{figure*}[!h]
    \centering
    \begin{subfigure}[t]{0.45\textwidth}
        \centering
        \includegraphics[height=2.1in]{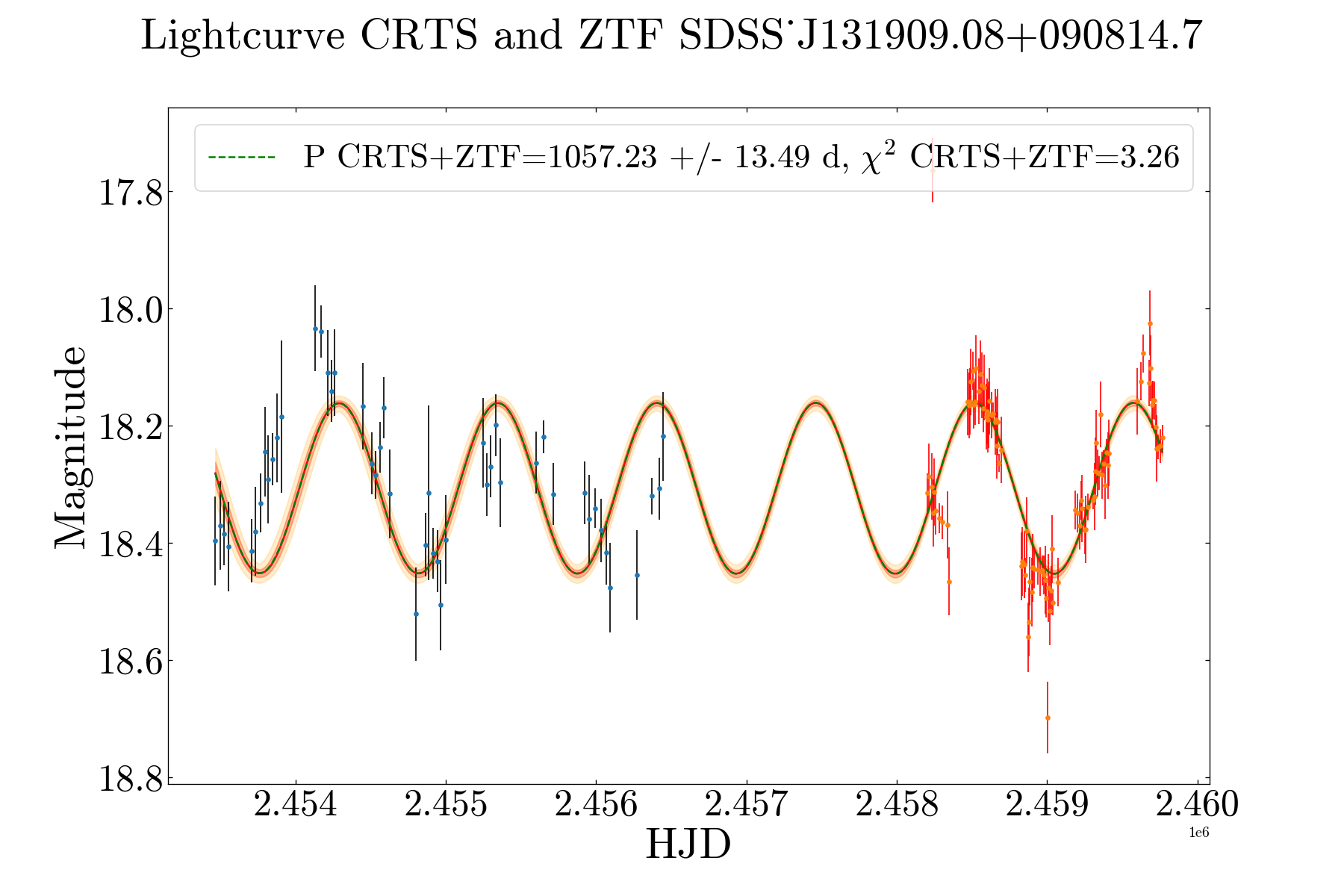}
    \end{subfigure}
    \begin{subfigure}[t]{0.45\textwidth}
        \centering
        \includegraphics[height=2.1in]{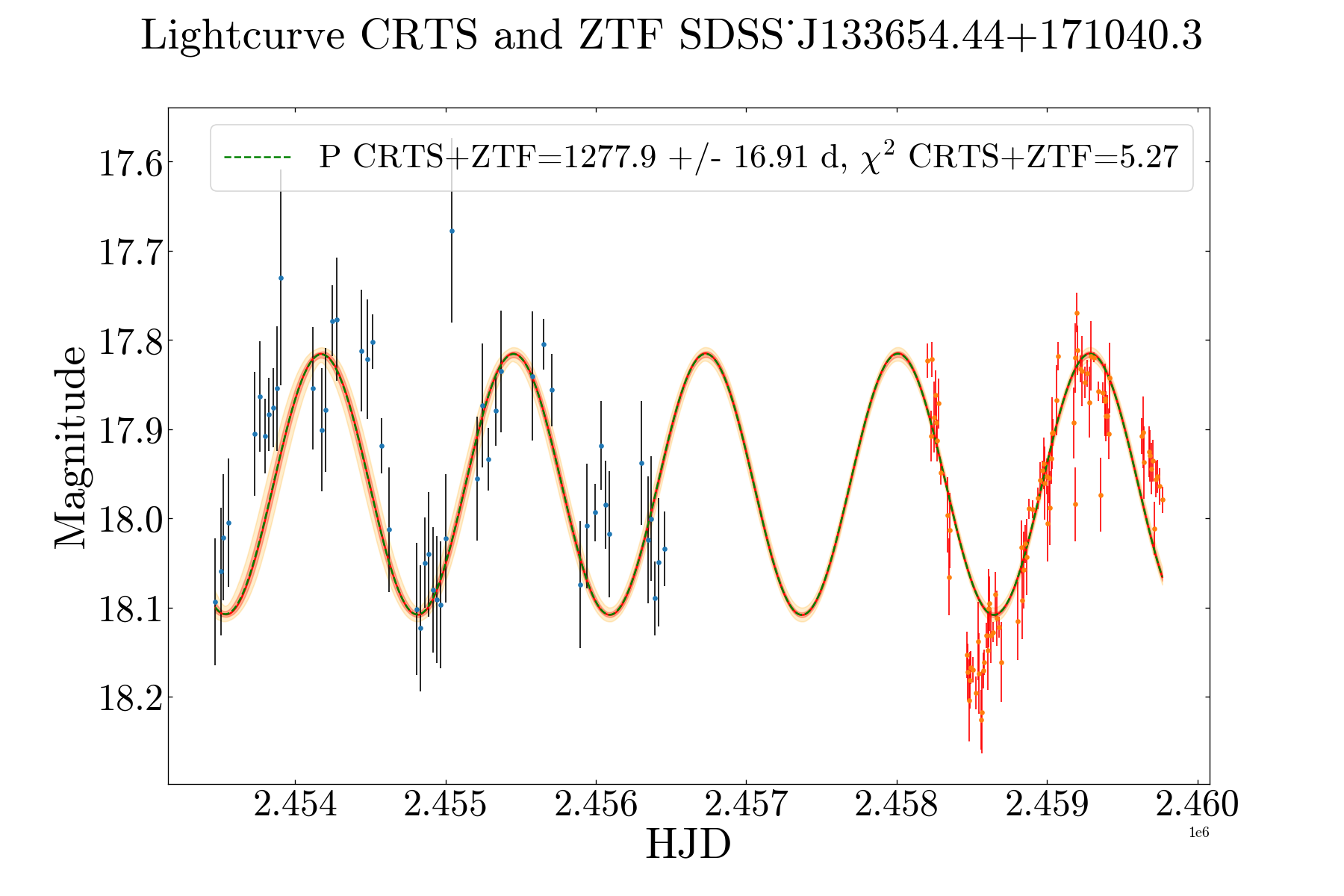}
    \end{subfigure}
\end{figure*}

\begin{figure*}[!h]
    \centering
    \begin{subfigure}[t]{0.45\textwidth}
        \centering
        \includegraphics[height=2.1in]{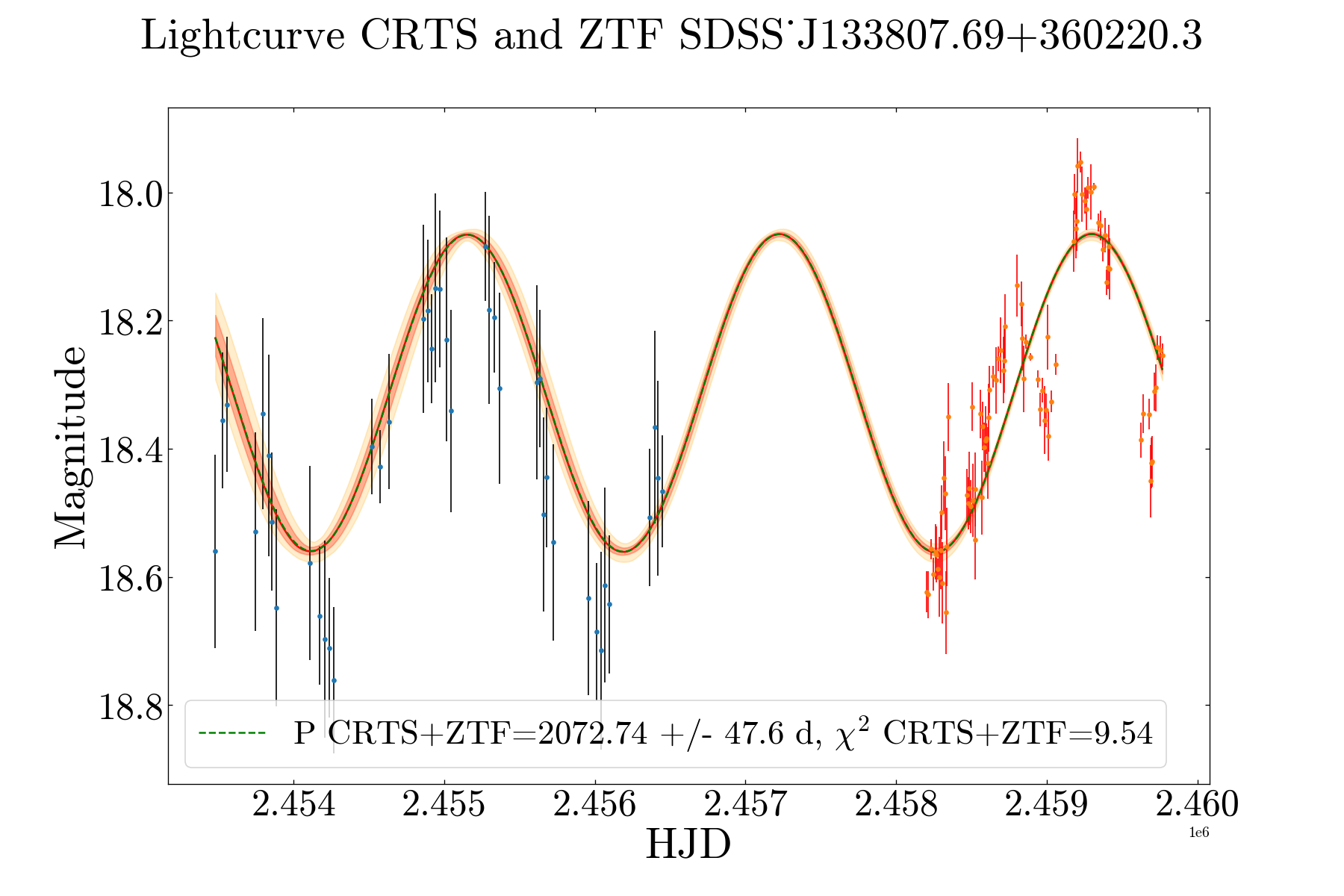}
    \end{subfigure} 
    \begin{subfigure}[t]{0.45\textwidth}
        \centering
        \includegraphics[height=2.1in]{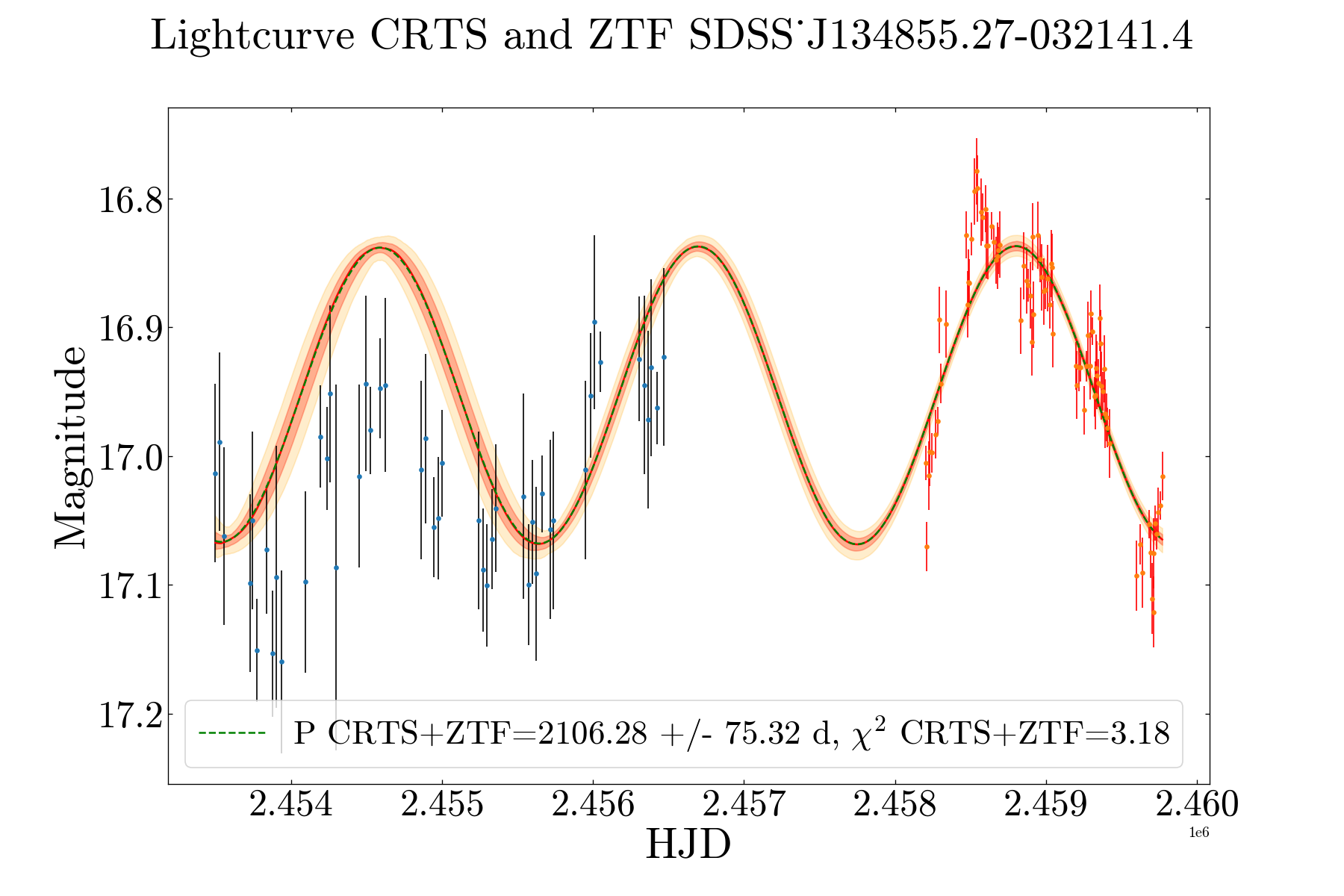}
    \end{subfigure}
\end{figure*}

\begin{figure*}[!h]
    \centering
    \begin{subfigure}[t]{0.45\textwidth}
        \centering
        \includegraphics[height=2.1in]{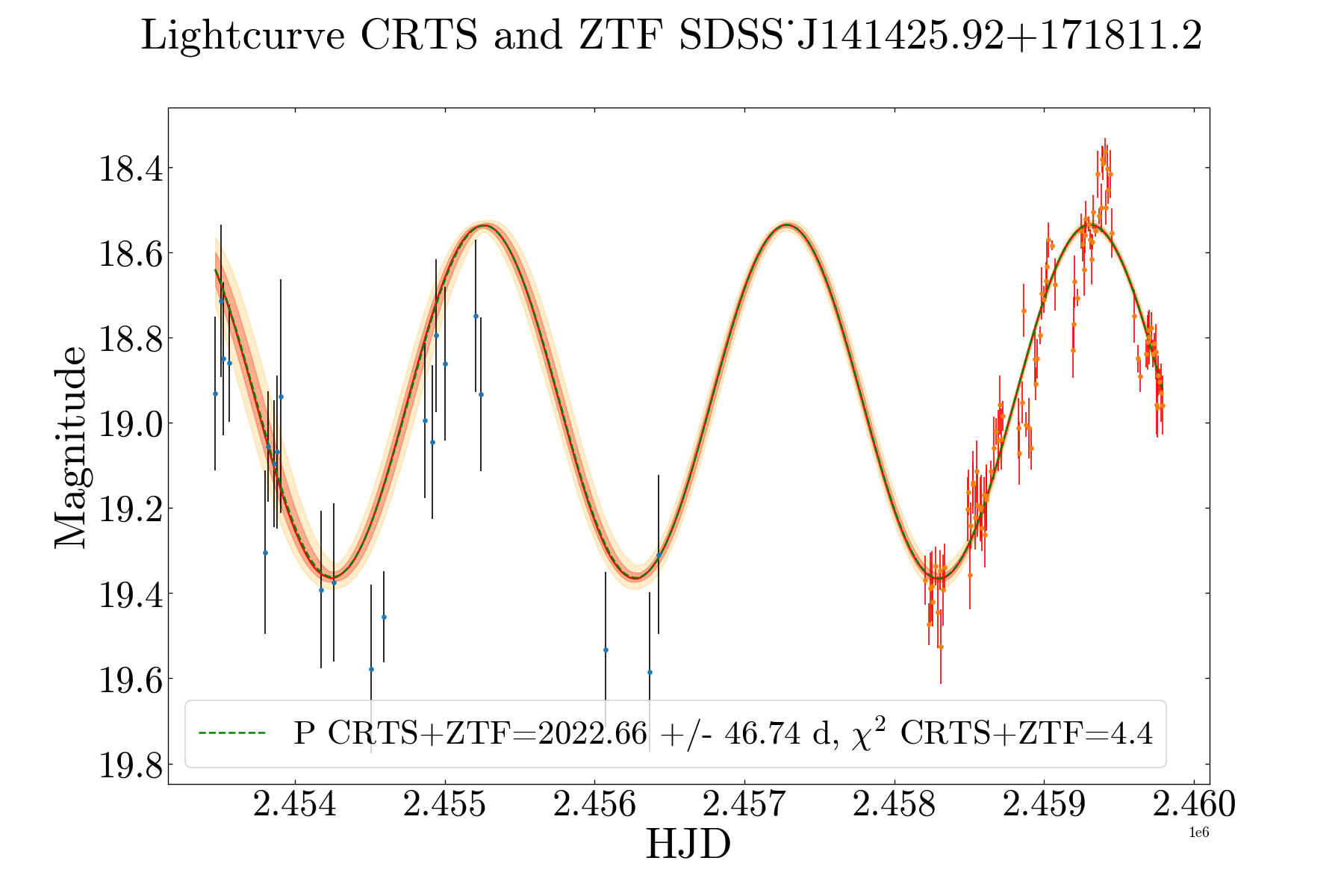}
    \end{subfigure}  
    \begin{subfigure}[t]{0.45\textwidth}
        \centering
        \includegraphics[height=2.1in]{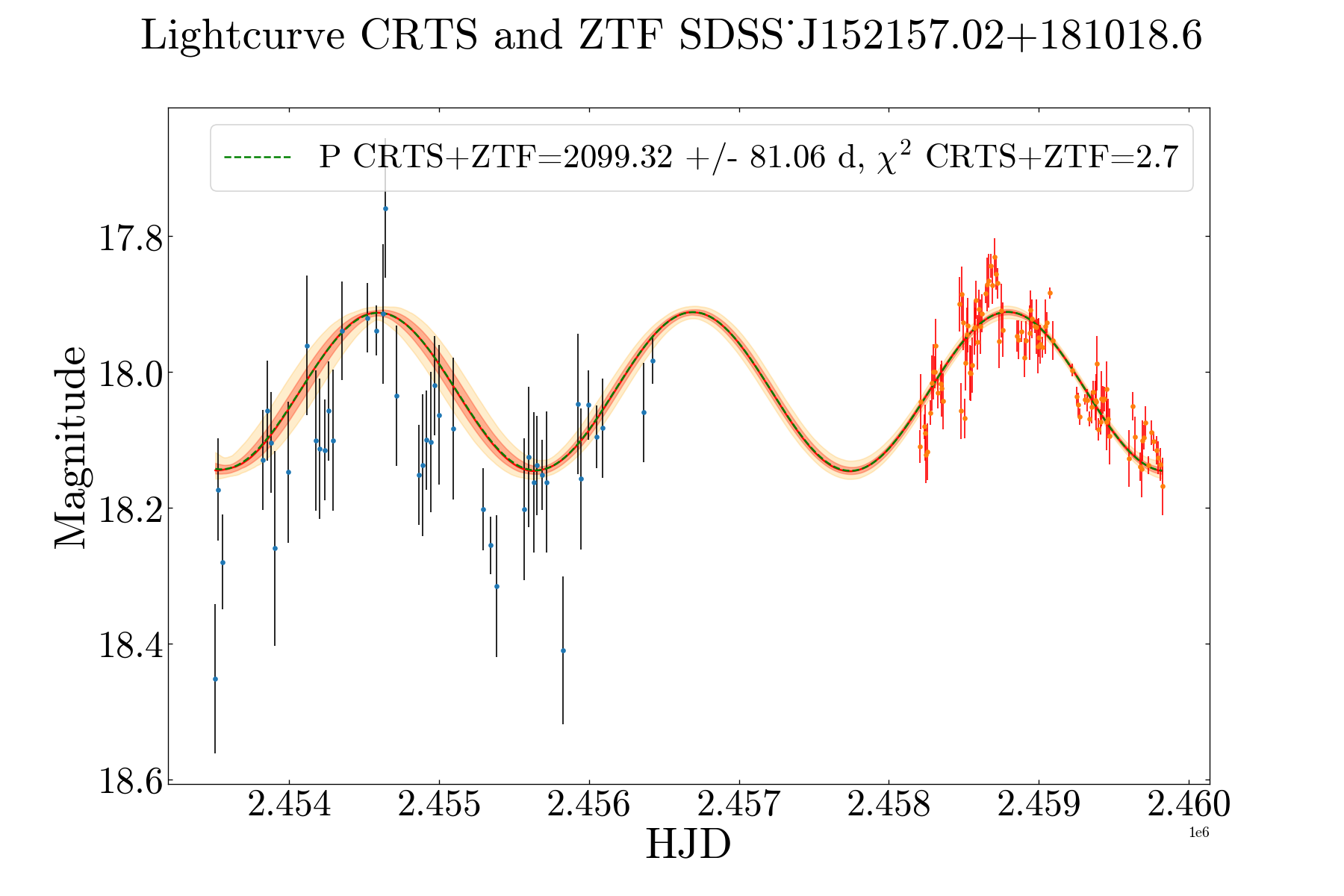}
    \end{subfigure}
\end{figure*}

\begin{figure*}[!h]
    \centering
    \begin{subfigure}[t]{0.45\textwidth}
        \centering
        \includegraphics[height=2.1in]{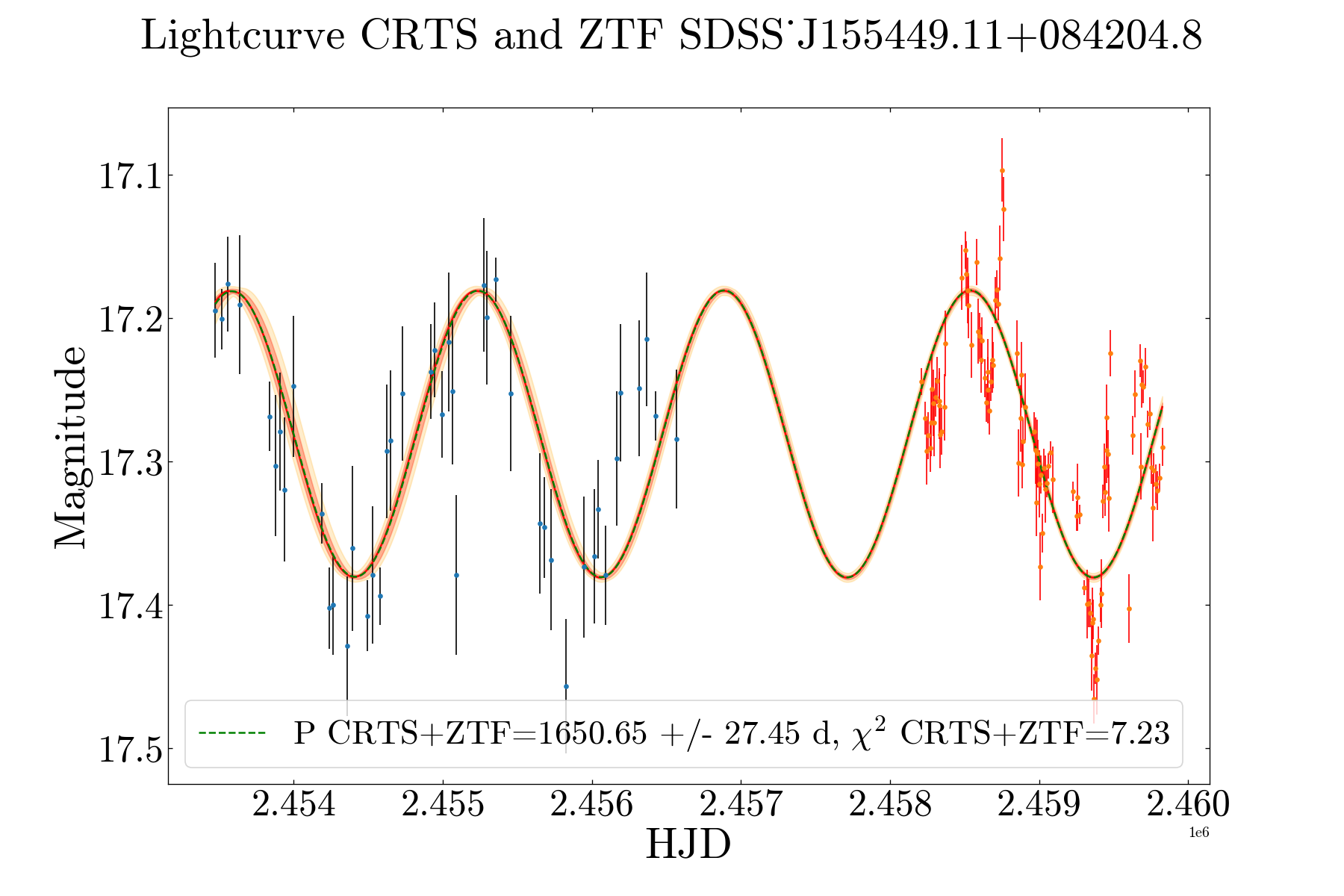}
    \end{subfigure} 
    \begin{subfigure}[t]{0.45\textwidth}
        \centering
        \includegraphics[height=2.1in]{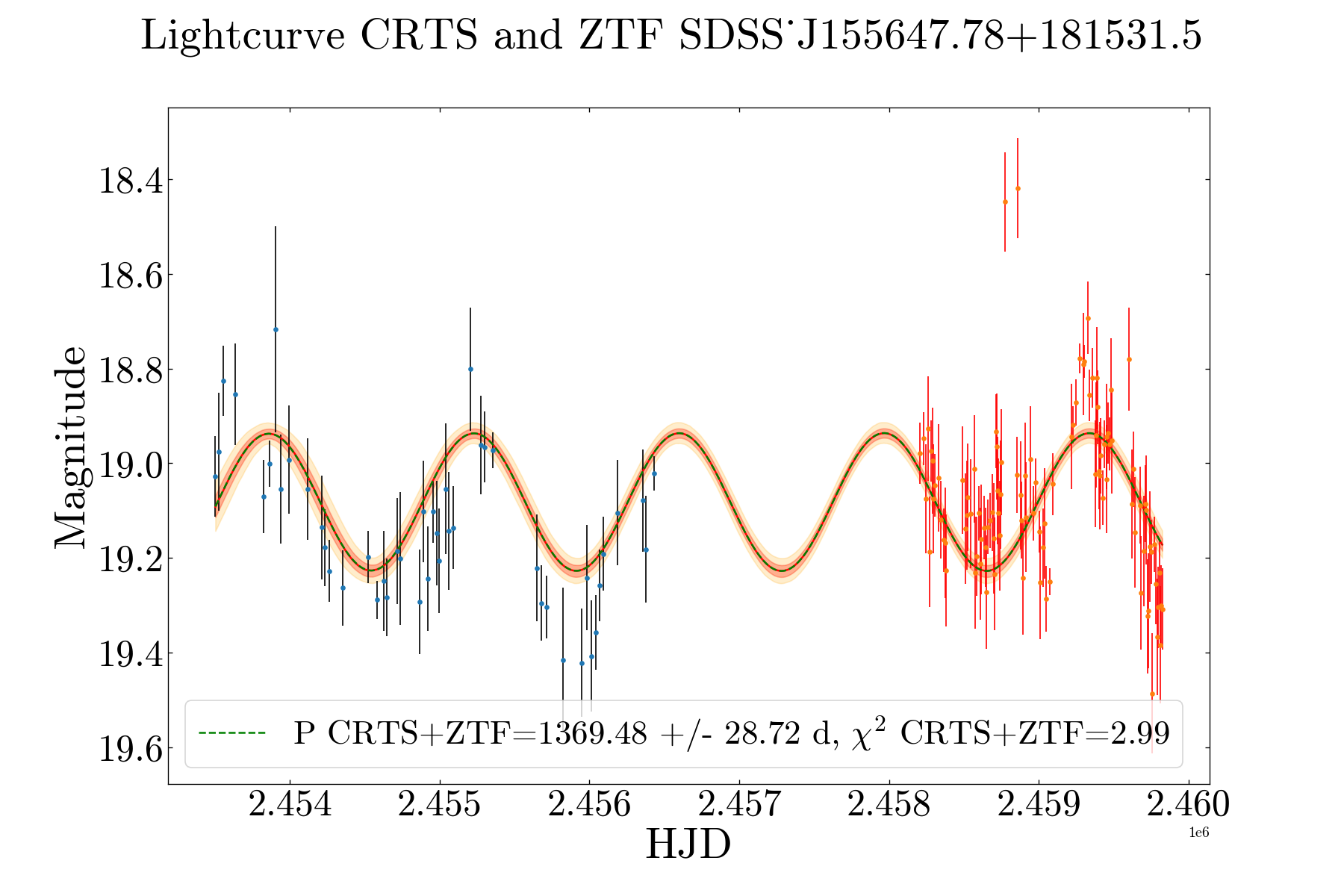}
    \end{subfigure}
\end{figure*}

\begin{figure*}[!h]
    \centering
    \begin{subfigure}[t]{0.45\textwidth}
        \centering
        \includegraphics[height=2.1in]{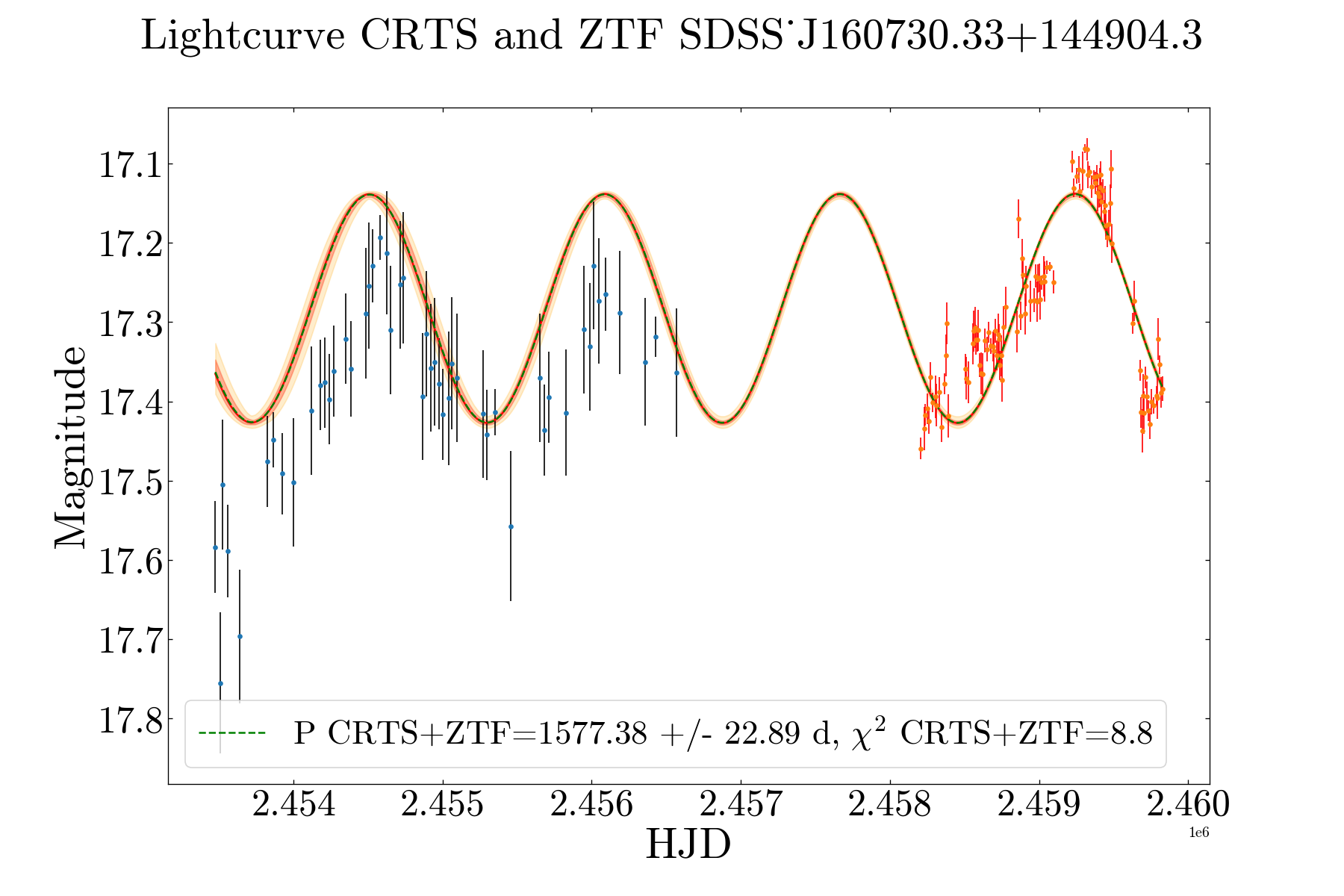}
    \end{subfigure} 
    \begin{subfigure}[t]{0.45\textwidth}
        \centering
        \includegraphics[height=2.1in]{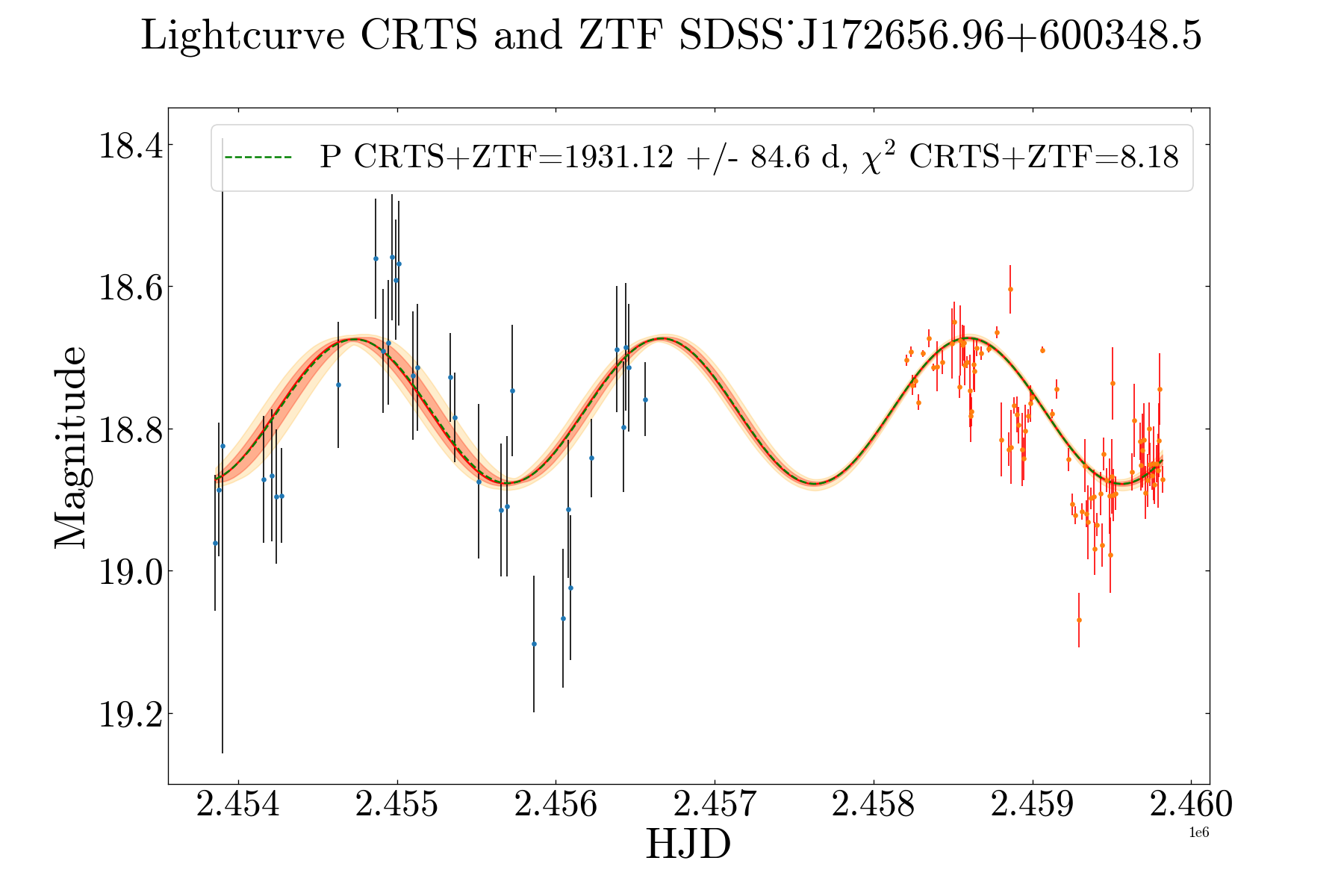}
    \end{subfigure}
\end{figure*}

\begin{figure*}[!h]
    \centering
    \begin{subfigure}[t]{0.45\textwidth}
        \centering
        \includegraphics[height=2.1in]{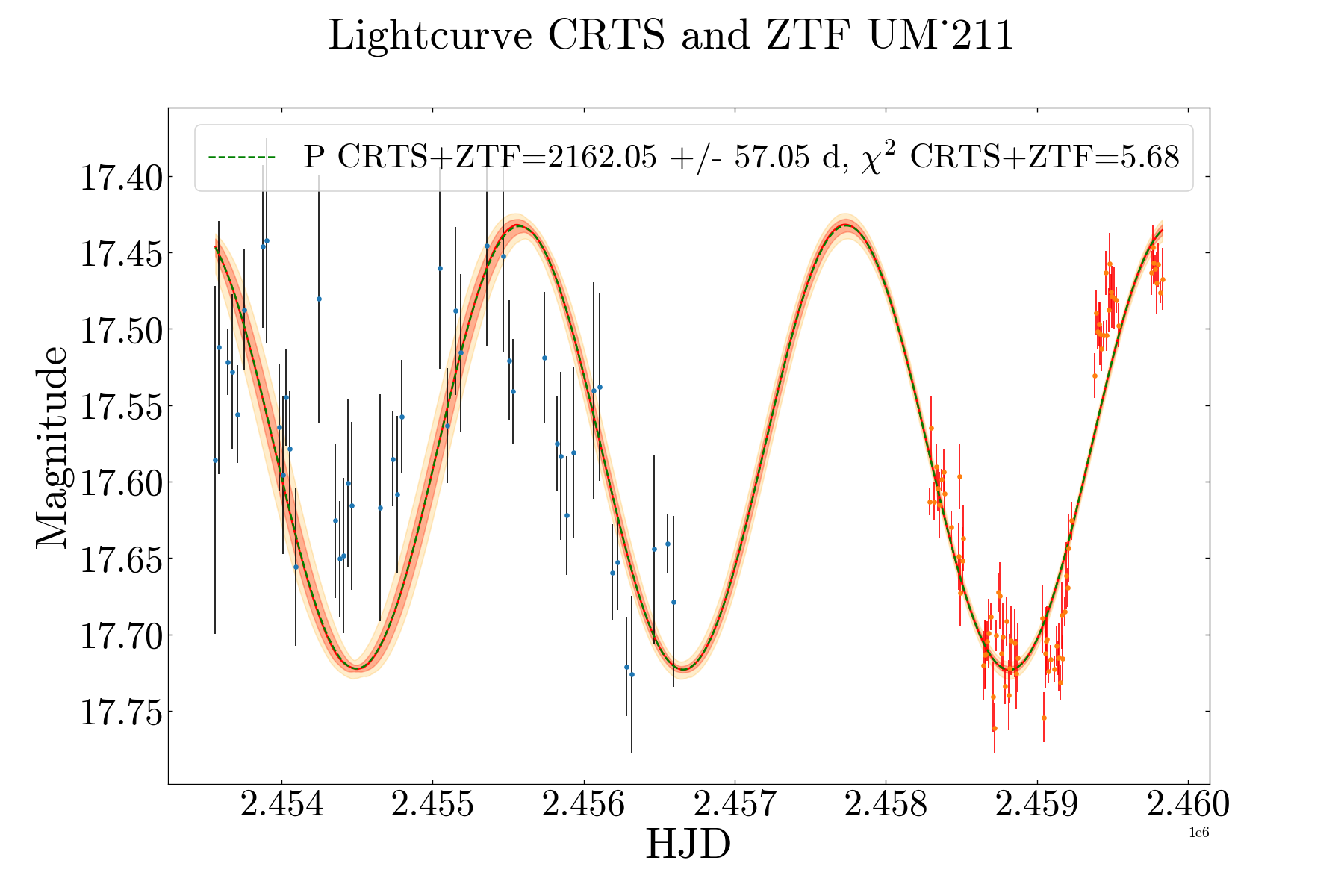}
    \end{subfigure} 
    \begin{subfigure}[t]{0.45\textwidth}
        \centering
        \includegraphics[height=2.1in]{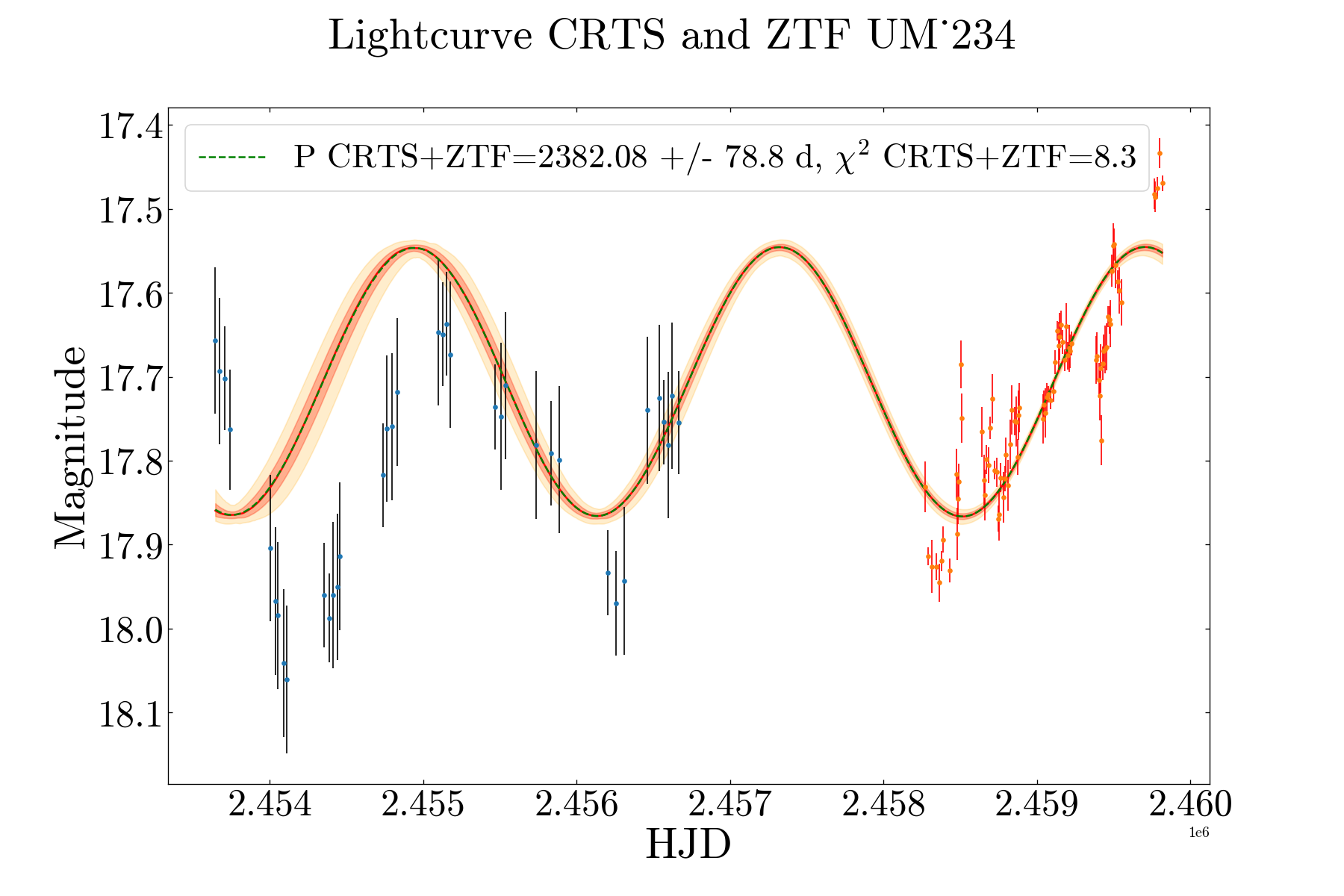}
    \end{subfigure}
\end{figure*}

\clearpage

\section{Results Table}
\addtolength{\tabcolsep}{-0.4em}
{\renewcommand{\arraystretch}{1.1}
\begin{table*}[!h]
\caption{36 MBBH Candidates results}
\begin{tabular*}{1.0\textwidth}{@{\extracolsep{\fill}}ccccccccccc}
\hline
\begin{tabular}[c]{@{}c@{}} (1) \\ Identifier \end{tabular}   & \begin{tabular}[c]{@{}c@{}} (2) \\ Fitted \\ Period \\ (days)\end{tabular} & \begin{tabular}[c]{@{}c@{}} (3) \\ $\chi^{2}_{\nu}$  \\ (d.o.f) \end{tabular} & \begin{tabular}[c]{@{}c@{}} (4) \\ G.L.S.\\ Period / \it{p}-value \\ (days)\end{tabular}  & \begin{tabular}[c]{@{}c@{}} (5) \\ $A_{r}$ \\ (mag)\end{tabular} & \begin{tabular}[c]{@{}c@{}} (6) \\ $A_{g}$ \\ (mag)\end{tabular}  & \begin{tabular}[c]{@{}c@{}} (7) \\ Estimated \\ separation \\ (mpc)\end{tabular} & \begin{tabular}[c]{@{}c@{}} (8) \\ $z$ \end{tabular} & \begin{tabular}[c]{@{}c@{}} (9) \\ Log M \\ (M$_{\odot}$)\end{tabular} & \begin{tabular}[c]{@{}c@{}} (10) \\  P$_{r}$ \\ (days)  \end{tabular} &  \begin{tabular}[c]{@{}c@{}} (11) \\ R  \end{tabular} \\ 
\hline
\,
J141425+171811 & 2017 $\pm$ 43 & 4.4 (112) & 2017 / 0.06 & 0.25 $\pm$ 0.01 & 0.40 $\pm$ 0.02 & 6.0 & 0.41 & 8.1 $\pm$ 0.4 & 1994 $\pm$ 180 & 7.84 \\
QNZ3:54 & 1765 $\pm$ 34 & 2.9 (132) & 1765 / 0.09 & 0.15 $\pm$ 0.01 & 0.18 $\pm$ 0.02 & 8.3 & 1.4 & 9.1 $\pm$ 0.1 & 1724 $\pm$ 123 & 3.49 \\
J133654+171040 & 1278 $\pm$ 16   &  5.3 (159) &  1278 / 0.002  & 0.13 $\pm$ 0.01 & 0.15 $\pm$ 0.01 &             7.6                      & 1.23   & 9.2 $\pm$ 0.1   & 1262 $\pm$ 62 & 3.12   \\
J122616+504205  & 1508 $\pm$ 55  & 1.6  (192)     &  1508 / 0.003    &  0.17 $\pm$ 0.02 & 0.21 $\pm$ 0.02  &          2.9                      & 0.75  & 7.6 $\pm$ 1.2 & 1515 $\pm$ 110 & 1.95 \\
J092911+203708  & 1768 $\pm$ 27  & 2.6 (140)  & 1768 / 0.007 &  0.17 $\pm$ 0.02  & 0.18 $\pm$ 0.02  &              10.9                      & 1.84 &  9.6 $\pm$ 0.2 & 1755 $\pm$ 167 & 1.36  \\
J152157+181018 & 2093 $\pm$ 78 & 2.7 (151) & 2092 / 0.19 & 0.12 $\pm$ 0.01 & 0.12 $\pm$ 0.01 & 4.6 & 0.73 & 7.9 $\pm$ 1.3 & 2186 $\pm$ 238 & 1.11 \\
J133807+360220 & 2073 $\pm$ 42 & 9.5 (128) & 2072 / 0.32 & 0.19 $\pm$ 0.01 & 0.25 $\pm$ 0.01 & 9.9 & 1.19 & 9.1 $\pm$ 0.1 & 2037 $\pm$ 253 & 0.65 \\
J114857+160023  & 1748 $\pm$ 40 &  2.5 (144) & 1749 / 0.144 &  0.10 $\pm$ 0.01 & 0.11 $\pm$ 0.01 &     12.8                     & 1.22  & 9.6 $\pm$ 0.2  & 1774 $\pm$ 162 & 0.57 \\
J130040+172758 & 2068 $\pm$ 53 & 3.0 (136) & 2068 / 0.16 & 0.27 $\pm$ 0.03 & 0.29 $\pm$ 0.03 & 9.4 & 0.86 & 8.9 $\pm$ 0.1 & 2129 $\pm$ 407 & 0.40 \\
J131626+390752  & 1532 $\pm$ 50    & 2.2  (217)      &   1531 / 0.07  &   0.17 $\pm$ 0.02 & 0.18 $\pm$ 0.02   &         5.1                    & 2.42   &  8.9 $\pm$ 0.1 & 1529 $\pm$ 136  & 0.27  \\
J124838+364453  & 1700 $\pm$ 50     & 1.8  (206) &  1699 / 0.07 &   0.20 $\pm$ 0.03   & 0.24 $\pm$ 0.03  &            8.7                        & 2.44   & 9.5 $\pm$ 0.2 & 1704 $\pm$ 154 & 0.26   \\
J082716+490534 & 1646 $\pm$ 20 & 4.1 (138) & 1645 / 0.001 & 0.41 $\pm$ 0.02 & 0.37 $\pm$ 0.02 & 10.1 & 0.68 & 9.1 $\pm$ 0.2 & 1632 $\pm$ 227 & 0.25 \\
J172656+600348 & 1924 $\pm$ 66 & 8.2 (109) & 1925 / 0.66 & 0.1 $\pm$ 0.01 & 0.1 $\pm$ 0.01 & 13.6 & 0.99 & 9.5 $\pm$ 0.7 & 1813 $\pm$ 129 & 0.22 \\ 
J160730+144904 & 1577 $\pm$ 22 & 8.8 (160) & 1576 / 0.08 & 0.1 $\pm$ 0.01 & 0.14 $\pm$ 0.01 & 11.0 & 1.8 & 9.7 $\pm$ 0.1 & 1584 $\pm$ 177 & 0.18 \\
J104941+085548 & 1263 $\pm$ 15 & 4.1 (171) & 1263 / 0.12 & 0.10 $\pm$ 0.01 & 0.10 $\pm$ 0.01 & 8.3 & 1.18 & 9.3 $\pm$ 0.1 & 1291 $\pm$ 91 & 0.18 \\
J131909+090814 & 1057 $\pm$ 13 & 3.3 (144) & 1057 / 0.0001 & 0.13 $\pm$ 0.01 & 0.15 $\pm$ 0.01 & 4.8 & 0.88 & 8.6 $\pm$ 0.3 & 1379 $\pm$ 96 & 0.17 \\
UM211 & 2162 $\pm$ 54 & 5.7 (123) & 2165 / 0.08 & 0.12 $\pm$ 0.01 & 0.14 $\pm$ 0.01 &   & 1.99 & & 2207 $\pm$ 324 & 0.17 \\
J121457+132024 & 2405 $\pm$ 73 & 4.7 (139) & 2404 / 0.44 & 0.16 $\pm$ 0.02 & 0.17 $\pm$ 0.02 & 12.6 & 1.49 & 9.4 $\pm$ 0.2 & 2394 $\pm$ 688 & 0.15 \\
J091501+460937  & 1793 $\pm$ 51   &  1.7   (190)     &   1792 / 0.04 &   0.16 $\pm$ 0.03 & 0.21 $\pm$ 0.03  &          6.6                     & 1.77 &  8.9 $\pm$ 0.1  & 1825 $\pm$ 535  & 0.14   \\
J115445+244646  & 1548 $\pm$ 34    &  2.6 (200)   &  1548 / 0.007 &  0.22 $\pm$ 0.02  & 0.23 $\pm$ 0.02   &           11.7                         & 0.8   &  9.4 $\pm$ 0.3 & 1310 $\pm$ 190 & 0.10  \\
J082926+180020 & 1652 $\pm$ 45 & 2.2 (141) & 1651 / 0.34 & 0.15 $\pm$ 0.02 & 0.17 $\pm$ 0.02 & 6.1 & 0.81 & 8.5 $\pm$ 0.1 & 1639 $\pm$ 333 & 0.06 \\
J083349+232809 & 1243 $\pm$ 15 & 11.2 (182) & 1243 / 0.30 & 0.10 $\pm$ 0.01 & 0.12 $\pm$ 0.01 & 8.3 & 1.15 & 9.3 $\pm$ 0.2 & 1270 $\pm$ 103 & 0.06 \\
J134855-032141 & 2101 $\pm$ 72 &  3.2 (136) & 2100 / 0.09 & 0.10 $\pm$ 0.01 & 0.10 $\pm$ 0.01 & 12.5 & 2.1 & 9.7 $\pm$ 0.1 & 2112 $\pm$ 495 & 0.05 \\
J102255+172155 & 1336 $\pm$ 22 & 3.4 (164) & 1336 / 0.05 & 0.11 $\pm$ 0.01 & 0.13 $\pm$ 0.02 & 5.6 & 1.06 & 8.7 $\pm$ 0.3 & 1929 $\pm$ 393 & 0.05 \\
UM234 & 2375 $\pm$ 86 & 8.3 (110) & 2323 / 0.32 & 0.15 $\pm$ 0.01 & 0.15 $\pm$ 0.01 & 12.7 & 0.73 & 9.1 $\pm$ 0.1 & 1696 $\pm$ 273 & 0.04 \\
J155449+084204 & 1647 $\pm$ 25 & 7.2 (155) & 1646 / 0.29 & 0.10 $\pm$ 0.01 & 0.10 $\pm$ 0.01 & 7.1 & 0.79 & 8.7 $\pm$ 0.2 & 1679 $\pm$ 129 & 0.04 \\
J081133+065558 & 1335 $\pm$ 22 & 3.9 (146) & 1335 / 0.19 & 0.15 $\pm$ 0.01 & 0.19 $\pm$ 0.02 & 9.1 & 1.27 & 9.4 $\pm$ 0.1 & 1385 $\pm$ 225 & 0.03 \\
J100641+253110  & 1851 $\pm$ 182     & 2.1 (200) &  1857 / 0.46  &  0.11 $\pm$ 0.03 & 0.15 $\pm$ 0.03 &               10.9                         & 1.13 &  9.3 $\pm$ 0.5 & 2053 $\pm$ 491 & 0.03  \\
J115141+142156 & 1332 $\pm$ 26 & 3.5 (134) & 1331 / 0.13 & 0.11 $\pm$ 0.01 & 0.12 $\pm$ 0.01 & 6.7 & 1.00 & 8.9 $\pm$ 0.2 & 1310 $\pm$ 191 & 0.02 \\
J155647+181531 & 1370 $\pm$ 29 & 3.0 (153) & 1370 / 0.24 & 0.1 $\pm$ 0.01 & 0.15 $\pm$ 0.02 & 7.4 & 1.50 & 9.2 $\pm$ 0.1 & 1392 $\pm$ 176 & 0.02 \\
J123527+392824  & 1863 $\pm$ 77   & 3.5  (203)    &   1861 / 0.13  &  0.16  $\pm$ 0.01 & 0.18 $\pm$ 0.01  &          6.2                   & 2.15   & 8.9 $\pm$ 0.1 & 2008 $\pm$ 412 & 0.014  \\
J113142+304139  & 1604 $\pm$ 71     & 1.8  (216)  &  1603 / 0.41  &  0.10 $\pm$ 0.02 & 0.11 $\pm$ 0.02  &        7.9                       & 2.34   & 9.4 $\pm$ 0.3 & 1609 $\pm$ 161 & 0.012  \\
J102625+295907  & 1634 $\pm$ 63      &   1.6  (213)  &  1633 / 0.15 &  0.10 $\pm$ 0.01 & 0.11 $\pm$ 0.01  &            8.4                   & 3.39   & 9.7 $\pm$ 0.1  & 1667 $\pm$ 300 & 0.01 \\
PKS 0157+011  & 1127  $\pm$ 26 & 4.0  (165)   &  1127 / 0.53 & 0.05 $\pm$ 0.01 &  0.06 $\pm$ 0.01   &          5.7                      & 1.16  & 8.9 $\pm$ 0.1  & 1137 $\pm$ 239 & 0.008 \\
J043526-164346 & 1304 $\pm$ 12 & 14.3 (127) & 1304 / 0.10 & 0.17 $\pm$ 0.01 & 0.18 $\pm$ 0.01 & & 0.01 &  & 1300 $\pm$ 250 & 0.008 \\
J122728+322508  & 1655  $\pm$ 86  & 2.4  (189)   &  1652 / 0.48 & 0.06 $\pm$ 0.01 &  0.06 $\pm$ 0.01  &          5.3                     & 0.75  & 8.3 $\pm$ 0.3 & 1694 $\pm$ 284  & 0.007  \\
\hline
\end{tabular*}
\tablefoot{36 good massive black hole binary candidates showing a similar periodicity in both CRTS and ZTF observations. Columns show : (1) : The identifier of the sources. (2) : The fitted periodicity to combined CRTS and ZTF data and the 3$\sigma$ error. (3) : The reduced chi-squared value of the fitted periodicity to combined CRTS and ZTF data (and the number of degrees of freedom). (4) : The period corresponding to the maximum peak in the Generalized Lomb-Scargle periodogram and the p-value computed using the method described in \ref{sec: GLS} (5) :  The fitted amplitude of the ZTF periodicity in the r-band and the 3$\sigma$ error. (6) : The fitted amplitude of the ZTF modulation in the g-band and the 3$\sigma$ error. (7) : The estimated separation of the sources calculated from the periodicity corrected with redshift and mass estimate  (assuming it corresponds to the total MBBH mass and that the two black holes are of similar mass). (8) : the redshift $z$ of the candidates. (9) : Logarithm of the mass estimated for the central black hole(s) \citep{Masse_candidates}. (10) : The period estimated using sinusoidal variability modulated by a red noise model. Uncertainties are given in a 3$\sigma$ confidence interval. Also, \cite{understanding_LS} explains that Gaussian error bars should be avoided in the context of a periodogram analysis, so uncertainties of identified Lomb-Scargle periods are not specified. (11) Bayes ratio comparing the DRW+Sine model to the DRW model.}
\label{Tab:Table1}
\end{table*}
}
\clearpage

\section{20 First rows of the additional candidates catalogue}
{\renewcommand{\arraystretch}{1.3}
\begin{table*}[!h]
\caption{Additional candidates catalogue}
\hspace{-0.5cm}
\begin{tabular*}{1.03\textwidth}{@{\extracolsep{\fill}}llllllllll}
    \hline
        oid & ra & dec & Redshift & Fitted Period & Error Period & Log(M) & Error Log(M) & $\chi^2_\nu$ ZTF & CRTS obs \\ \hline
        722214100002451 & 246.183 & 43.994 & 1.93 & 1517 & 55 & 9.24 & 0.02 & 1.08 & No \\ \hline
        762207400010289 & 253.448 & 46.594 & 2.29 & 1547 & 650 & 8.1 & 0.14 & 1.23 & Yes \\ \hline
        796215300002648 & 254.941 & 57.525 & 2.05 & 1354 & 45 & ~ & ~ & 0.92 & No \\ \hline
        721210400000714 & 237.562 & 41.418 & 1.89 & 1527 & 720 & 8.81 & 0.05 & 1.21 & Yes \\ \hline
        760213300014840 & 237.592 & 49.784 & 2.09 & 1545 & 722 & 8.69 & 0.07 & 1.02 & Yes \\ \hline
        792206200005745 & 214.018 & 54.177 & 1.48 & 1773 & 27 & 9.17 & 0.02 & 1.16 & No \\ \hline
        792210400012079 & 216.131 & 55.414 & 2.36 & 1477 & 696 & 9.31 & 0.05 & 1.17 & Yes \\ \hline
        793210300003169 & 225.462 & 55.544 & 1.15 & 1512 & 613 & 9.28 & 0.03 & 0.96 & Yes \\ \hline
        793213100021821 & 228.540 & 57.820 & 2.33 & 817 & 25 & 9.67 & 0.17 & 1.01 & No \\ \hline
        793216100016751 & 235.480 & 54.179 & 0.85 & 788 & 18 & 8.66 & 0.06 & 1.22 & No \\ \hline
        795211400020541 & 231.904 & 59.202 & 0.93 & 1283 & 509 & 9.83 & 0.23 & 0.95 & Yes \\ \hline
        823203400004287 & 238.373 & 61.827 & 2.44 & 1538 & 620 & 9.64 & 0.04 & 1.26 & Yes \\ \hline
        823205300004979 & 234.299 & 61.807 & 2.65 & 1534 & 398 & 9.34 & 0.07 & 0.88 & Yes \\ \hline
        678211200026561 & 239.653 & 35.532 & 2.09 & 1530 & 475 & 8.87 & 0.02 & 1.08 & Yes \\ \hline
        678213400006806 & 244.586 & 31.964 & 0.19 & 1520 & 289 & ~ & ~ & 1.26 & Yes \\ \hline
        720206400001109 & 229.529 & 42.999 & 2.04 & 1495 & 632 & 9.21 & 0.11 & 1.08 & Yes \\ \hline
        760216400003842 & 243.705 & 44.680 & 0.46 & 1470 & 26 & ~ & ~ & 1.06 & No \\ \hline
        792209300011535 & 215.639 & 55.088 & 1.02 & 1227 & 33 & 8.72 & 0.06 & 1.28 & No \\ \hline
        793206400000506 & 226.619 & 58.033 & 0.92 & 1527 & 592 & 8.06 & 0.08 & 1.22 & Yes \\ \hline
\end{tabular*}
\tablefoot{First 20 lines of the catalogue of weak MBBH candidates. Columns in the catalogue show from left to right, the ZTF object ID (oid), the right ascension (R.A.) and declination (Dec.) of the sources in degrees, their redshift z, their identified period and 1 $\sigma$ error in days, their Log(M) and 1 $\sigma$ error in M$_{\odot}$, the reduced chi square $\chi^{2}_{\nu}$ and finally if the object has CRTS observations. The whole catalogue will be available online.}
\end{table*}
}
\end{appendix}
\end{document}